\documentclass[12pt,a4paper]{article} 
\pdfoutput=1
\usepackage{style}
\usepackage{rotating}
\usepackage{lscape}
\usepackage{multirow}

\usepackage[utf8]{inputenc}
\usepackage[T1]{fontenc}
\usepackage{mathrsfs}
\usepackage{amsmath}
\usepackage{amssymb}
\usepackage{amsthm}
\usepackage{psfrag}
\usepackage{graphicx}
\usepackage{hyperref}
\usepackage{color}
\usepackage{bm}
\usepackage{graphicx}

\usepackage{bbm}
\usepackage{scalerel}
\usepackage{upgreek}

\newcommand{\be}{\begin{equation}}
\newcommand{\ee}{\end{equation}}
\newcommand{\ba}{\begin{align*}}
\newcommand{\ea}{\end{align*}}
\newcommand{\beqa}{\begin{eqnarray}}
\newcommand{\eeqa}{\end{eqnarray}}
\newcommand{\bseq}{\begin{subequations}}
\newcommand{\eseq}{\end{subequations}}

\def\d{\partial}

\def\l{\left(}
\def\r{\right)}
\def\k{{\bf k}}
\def\x{{\bf x}}
\def\tree{{\rm tree}}
\def\oneloop{{\rm 1\text{-}loop}}

\def\ctr{{\rm ctr}}
\def\Apr{{\cal A}_{\rm asp}}
\def\Actr{{\cal A}^{\rm ctr}}

\def\P{{\cal P}}

\def\H{{\cal H}}
\def\ksc{k_{\rm sc}}

\renewcommand\a{\alpha}
\renewcommand\b{\beta}
\def\e{{\rm e}}

\def\D{\mathcal{D}}
\def\ch{{\rm ch\,}}
\def\sh{{\rm sh\,}}

\def\kin{{\rm kin}}
\def\pot{{\rm pot}}
\def\foc{\varpi}

\newcommand\Mpc{\,\text{Mpc}}
\newcommand\Mpch{\,\text{Mpc}/h}

\newcommand{\kNL}{k_{\rm NL}}

\newcommand\vk{\varkappa}

\def\Xint#1{\mathchoice
	{\XXint\displaystyle\textstyle{#1}}%
	{\XXint\textstyle\scriptstyle{#1}}%
	{\XXint\scriptstyle\scriptscriptstyle{#1}}%
	{\XXint\scriptscriptstyle\scriptscriptstyle{#1}}%
	\!\int}
\def\XXint#1#2#3{{\setbox0=\hbox{$#1{#2#3}{\int}$}
		\vcenter{\hbox{$#2#3$}}\kern-.5\wd0}}

\def\dashint{\Xint-}
\newcommand*{\paral}{\stretchrel*{\parallel}{\perp}}

\relax

\title{
    Renormalizing one-point probability distribution function for
    cosmological counts in cells  
}

\author[a,b,c]{Anton Chudaykin\footnote{\texttt{chudayka@mcmaster.ca}}}
\author[d,e]{Mikhail M. Ivanov\footnote{\texttt{ivanov99@mit.edu}}}
\author[a,b]{Sergey Sibiryakov\footnote{\texttt{ssibiryakov@perimeterinstitute.ca}}}

\affiliation[a]{Department of Physics \& Astronomy, McMaster University,\\ 
	1280 Main Street West, Hamilton, ON L8S 4M1, Canada}
\affiliation[b]{Perimeter Institute for Theoretical Physics, Waterloo,
  Ontario, N2L 2Y5, Canada} 
\affiliation[c]{Institute for Nuclear Research of the
	Russian Academy of Sciences, \\ 
	\normalsize \it  60th October Anniversary Prospect, 7a, 117312
	Moscow, Russia}
  \affiliation[d]{Center for Theoretical Physics, Massachusetts Institute of Technology, \\  Cambridge, MA 02139, USA}
\affiliation[e]{School of Natural Sciences, Institute for Advanced Study, \\
	1 Einstein Drive, Princeton, NJ 08540, USA}

\abstract{
	We study the one-point probability 
	distribution function (PDF) for matter density averaged over
        spherical cells. The leading part to the PDF is defined by spherical collapse dynamics, whereas the next-to-leading
        part 
        comes from the integration over fluctuations around the
        saddle-point solution. The latter calculation receives sizable
        contributions from short modes and must be
        renormalized. We propose a new approach to renormalization by
        modeling the effective
        stress-energy tensor for short perturbations. The model
        contains three free parameters. Two of them are related to the
        counterterms in the one-loop matter power spectrum and
        bispectrum, one more parameterizes their redshift dependence.
        This relation can be used to impose priors in fitting the
        model to the PDF
        data. 
 We confront the model with 
the results of high-resolution N-body
        simulations and find excellent agreement for cell radii
        $r_*\geq 10\Mpc/h$ at all redshifts down to $z=0$. Discrepancies at a few
        per cent level are detected at low redshifts for $r_*\leq
        10\Mpc/h$ and are associated with two-loop corrections to the
        model.

\begin{flushright}
{\it Dedicated to the memory of Valery Rubakov, great scientist and
  teacher.}
\end{flushright}
}

\begin{document}
	
	\begin{flushright}
		INR-TH-2022-028
	\end{flushright}
	
	\maketitle

\section{Introduction and Summary}
\label{sec:intro}

Motivated by the rapid accumulation of observational data, there has
been a lot of progress in the analytic modeling of the
Large-Scale-Structure (LSS) of the Universe. To a large extent, this progress is associated with the development of the cosmological perturbation
theory. The latter is built on a diagrammatic expansion quite similar
to that in perturbative quantum field theory. Various cosmological
correlators are obtained as a sum of the leading tree-level
contribution and corrections ordered by the number of loops in the
diagram \cite{Bernardeau:2001qr}. An important step was the
realization that the loop contributions must be renormalized because
they involve unphysical integration over modes with short wavelengths
which are beyond the perturbative regime. This renormalization is
systematically carried out within the framework of effective field
theory (EFT) where the contribution of short modes is accounted for by an
effective stress tensor added to the equations of motion for the long
modes
\cite{Baumann:2010tm,Carrasco:2012cv,Pajer:2013jj,Baldauf:2014qfa,
Abolhasani:2015mra,Cabass:2022avo,Ivanov:2022mrd}.   
The effect of short-scale dynamics on the LSS correlators is thereby
captured by a few free parameters which are to be fixed from the
data. The EFT approach, combined with other developments, such as
resummation of large bulk flows 
\cite{Senatore:2014via,Baldauf:2015xfa,Blas:2016sfa,Ivanov:2018gjr}
and FFTLog computation of the loop integrals \cite{Simonovic:2017mhp},
has been shown to provide an efficient and accurate method for
calculation of the LSS power spectrum \cite{Nishimichi:2020tvu} and
bispectrum \cite{Steele:2020tak,Philcox:2022frc}. It has been
implemented in numerical codes \cite{DAmico:2020kxu,Chudaykin:2020aoj,Chen:2020zjt}.
Recently, it  has been successfully used to extract cosmological parameters from
the data \cite{Ivanov:2019pdj,DAmico:2019fhj,Ivanov:2019hqk,Colas:2019ret,Philcox:2020vvt,Chudaykin:2020ghx,DAmico:2020tty,Ivanov:2021zmi,Chen:2021wdi,White:2021yvw,Philcox:2021kcw,Chen:2022jzq,Cabass:2022wjy,Cabass:2022ymb,Cabass:2022oap,Chudaykin:2022nru},
including analyses
with controlled theoretical uncertainty
\cite{Baldauf:2016sjb,Chudaykin:2019ock,Chudaykin:2020hbf}.  

In this paper we go beyond the perturbation theory and consider
non-perturbative one-point probability distribution function (PDF),
also known as counts-in-cells (CiC) statistics. To define it, one
separates the density field in spherical cells of radius $r_*$ in
position space and takes the average value of the density inside each
cell.\footnote{ In the case of discrete tracers, one counts the number
  of objects inside each cell.}
The one-point PDF is then the probability of finding a given average
density. It is estimated as the number of cells with this density divided by
the total number of cells. 
Note that the deviation of this averaged density from the mean density
of the universe does not need to be small, so evaluation of the PDF
requires a non-perturbative treatment. 

CiC statistics was introduced by E.~Hubble \cite{1934ApJ....79....8H}
and has been measured in galaxy surveys
\cite{2dFGRS:2004amf,Hurtado-Gil:2017dbm,Repp:2020kfd}. Related
one-point statistics of the matter field have been used in the analysis of 
weak
lensing maps
\cite{DES:2016wzy,DES:2017eav,Burger:2022lwh}. 
The PDF is known to contain information from all $n$-point correlation
functions in a compressed way \cite{Bernardeau:2001qr} and is also sensitive to rare fluctuations. Thus, it is a promising probe of LSS, complementary to the perturbative
observables. Combining the PDF with the matter power spectrum can
significantly enhance the constraining power on the standard
cosmological parameters and the neutrino mass \cite{Uhlemann:2019gni},
as well as on the extended models of gravity and dark energy
\cite{Cataneo:2021xlx,Gough:2021hlr}. The PDF and its generalizations
\cite{Jamieson:2020wxf} also provide a promising tool for probing primordial
non-Gaussianity \cite{Matarrese:2000iz,Friedrich:2019byw} and for
analyzing future 21cm intensity mapping data \cite{Leicht:2018ryx}.

There are several theoretical approaches to PDF modeling. 
Pioneering studies of the counts-in-cell statistics used perturbative methods \cite{Bernardeau:1992zw,Bernardeau:1993ac}. They were extended beyond perturbation theory in Refs.~\cite{Valageas:2001zr,Valageas:2001td} which show that the most probable configuration producing a given averaged density inside a spherical cell respects the symmetry of the problem, i.e. it is given by the spherical collapse.
These results were generalized to several concentric cells \cite{Bernardeau:2013dua} and interpreted in the context of 
Large Deviation Principle \cite{Bernardeau:2015khs}.
Ref.~\cite{Uhlemann:2015npz} suggested applying the logarithmic transformation of the density field to improve the accuracy of the method. This approach uses the non-linear density variance as an input and has proven to be successful in describing the results of high-resolution simulations, see Refs.~\cite{Uhlemann:2017tex,Uhlemann:2019gni,Boyle:2020bqn,Friedrich:2021xff,Cataneo:2021xlx}.
An alternative approach to the counts-in-cells statistics was developed in Refs.~\cite{1970Doroshkevich,Hui:1999ku,Sheth:1998ew,BetancortRijo:2001ge}
based on the Lagrangian-space description of LSS.
This method involves the relation between the initial (Lagrangian) PDF and the physical (Eulerian) PDF that was addressed in Refs.~\cite{Lam:2007qw,Paranjape:2011wa,Musso:2012qk}.
Ref.~\cite{Pajer:2017ulp} discusses application of this method to $(1+1)$-dimensional toy model and convergence of the perturbative series for PDF. 

Here we pursue the path-integral
approach developed in \cite{Ivanov:2018lcg} following earlier works
\cite{Valageas:2001zr,Valageas:2001td}. This approach uses the linear
variance of the density field within the cell as an expansion
parameter and leads to representation of the PDF in the form similar
to the semiclassical expansion of quantum field theory. The PDF
factorizes into the leading-order exponential part and a
prefactor. The former is given by the spherically symmetric
saddle-point configuration that is found exactly using the dynamics of
spherical collapse. On the other hand, the next-to-leading prefactor
results from integration over fluctuations around the saddle-point
solution. 
A numerical procedure for calculating it was implemented in an open-source \texttt{Python} code 
\texttt{AsPy} \cite{AsPy}.

The prefactor can be thought of as an one-loop contribution
in the background of the saddle-point configuration. It receives
sizable contribution from short --- or `ultraviolet'
(UV) --- modes and must be renormalized with an appropriate
counterterm. Unlike in perturbation theory, this procedure cannot be
carried out relying only on the symmetries of the long-distance
dynamics. Indeed, these symmetries allow the counterterm to be an
arbitrary function of the cell density contrast, thereby destroying
the predictive power. Another way to see the problem is to
recall that the PDF aggregates the information about all correlation
functions. Within the EFT of LSS, the coefficients of their
counterterms are independent, implying that the PDF counterterm contains an infinite number of parameters. In an attempt to overcome
this difficulty Ref.~\cite{Ivanov:2018lcg} proposed two {\it ad hoc}
counterterm models motivated by the structure of the UV contribution
into the aspherical prefactor. The models  
agree with each other at moderate density contrasts, but 
deviate by $30\%$ at the tails of the distribution which was interpreted as theoretical uncertainty of the method.  

The goal of this paper is to refine the counterterm model and put it on firm physical grounds. 
To this aim, we revisit the calculation of the aspherical prefactor
and isolate the contribution of the UV modes.
We 
find that it is
described by an effective stress tensor consisting of two physically
distinct parts, which we refer to as {\it kinetic} and {\it potential}. One corresponds to the pressure of the short
modes, and the other to their gravitational pull. We propose that
this stress tensor should be renormalized by the counterterm with the
same structure, but with additional factors in front
of the two contributions controlled by the shell-crossing scale in the background solution. The key physical assumption behind this
proposal is that the counterterm is dominated by the modes in a
relatively narrow range of wavenumbers which experience
the shell crossing, but are not yet virialized. 

Approximating the power spectrum at the relevant wavenumbers by
a power-law, we can fix the dependence of these factors on time and density. This leaves
the model with three free parameters --- the coefficients in front of
the kinetic and the potential parts of the counterterm 
and the exponent of the power law describing their time and density dependence.
These parameters must be fitted from the data. We show that certain 
combinations of these parameters can be matched to the EFT
counterterms in the one-loop power spectrum and bispectrum. While the
former can be reliably measured from the power spectrum of the 
real or N-body data,
measuring the latter requires measurement of the bispectrum and
is challenging. 
Thus, we suggest to restrict the freedom of the PDF
model by imposing priors following from the value and the redshift
dependence of the
power-spectrum counterterm.  

We validate our theoretical framework using N-body data from
high-resolution Farpoint simulation which evolves $12288^3$ particles 
in a $1\,({\rm Gpc}/h)^3$ volume \cite{HACC:2021sgt}. We find that the
model describes the data within their statistical uncertainty for
density contrasts $0.1\leq 1+\delta_*\leq 10$, cell radii\footnote{The
  parameter $h\approx 0.7$ is defined through the 
  present-day value of the Hubble constant, ${H_0=100\,h~{\rm
    km}/({\rm s}\cdot {\rm Mpc})}$.}  
${r_*=15\,{\rm Mpc}/h}$ and $10\,{\rm Mpc}/h$ and up to redshift
zero. For smaller cell radii $r_*=7.5\,{\rm Mpc}/h$ and $5\,{\rm
  Mpc}/h$ we observe small (a few per cent), but significant,
discrepancies between the model and the data. We explain that these
discrepancies cannot arise from a deficiency of the counterterm
model, but represent the first detection of the two-loop corrections
to the PDF.

The paper is organized as follows. In Sec.~\ref{sec:theory} we review
the approach to the PDF developed in \cite{Ivanov:2018lcg}, with
the focus on the UV contribution to the prefactor.  
In Sec.~\ref{sec:rel} we derive the model-independent relations
between the PDF counterterm and the counterterms of the $n$-point
correlators. In Sec.~\ref{sec:ctr} we propose our model for the
counterterm stress tensor and compute the resulting counterterm for
the PDF prefactor. Section~\ref{sec:result} contains comparison to the 
N-body data. We conclude in
Sec.~\ref{sec:con}. 

A number of appendices complement the main text. 
Appendix~\ref{app:notations} summarizes our conventions. 
In appendix \ref{app:path} we discuss the PDF restricted to spherical
configurations and derive its analytic form. 
Appendix~\ref{sec:log} presents a simple phenomenological fit to the
aspherical prefactor and contrasts it with the full theoretical
model. 
Appendix \ref{app:Apr} summarizes the dynamical equations governing
the density and velocity perturbations. 
In appendix \ref{app:WKB} we discuss the 
Wentzel–Kramers–Brillouin (WKB) approximation for high multipoles.
In appendix \ref{app:ksc} we explore alternative definitions of shell-crossing
scale in the non-trivial background of the spherical collapse
solution. 
Appendix \ref{app:gamma} explores the sensitivity of the
power-spectrum counterterm to the cosmological parameters.
Appendix \ref{app:ZA} contains discussion of transients in the N-body PDF
due to inaccuracy of the Zeldovich 
initial conditions.
In appendix \ref{app:comp} we investigate 
the possibility of fitting PDF with reduced sets of counterterm parameters.

\section{One-Point Probability Distribution Function}
\label{sec:theory}

\subsection{Saddle-point expansion}
\label{sec:semi}

We consider the density contrast averaged over a spherical cell of
radius $r_*$,\footnote{See appendix~\ref{app:notations} for the notations
  and conventions.} 
\be
\label{eq:wf}
\bar \delta_W =  \int \frac{d^3x}{r_*^3} \,\tilde{W}(r/r_*)\, 
\delta(\x)=\int_\k W(kr_*) \delta(\k)\,,
\ee
where $\delta({\bf x})\equiv \delta\rho({\bf x})/\rho_{\rm univ}$, with
$\rho_{\rm univ}$ being the average matter density of the Universe. 
$\tilde 
W(r/r_*)$ is a window function in position space and $W(kr_*)$ denotes
its Fourier counterpart. Generally, $\tilde W$ can be any
spherically symmetric function normalized to unity, $\int d^3z\, \tilde
W=1$.  
It is commonly chosen to be top-hat in the literature,
\be
\label{thx}
\tilde{W}_{\rm th}(r/r_*)=\frac{3}{4\pi}
\Theta_{\rm H}\left(1-\frac{r}{r_*}\right)~~~~~ \Longleftrightarrow~~~~~
W_{\rm th}(kr_*)= \frac{3j_1(kr_*)}{kr_*}\,,
\ee
where $\Theta_{\rm H}$ stands for the Heaviside theta-function.
As explained in \cite{Ivanov:2018lcg}, this choice greatly simplifies
the analysis and we adopt it throughout this work. 

The initial conditions for the density perturbations are assumed to be
adiabatic and Gaussian, so that 
their statistical properties are entirely described by the two-point
correlator,
\be 
\langle\delta_{\rm in}(\k)\delta_{\rm in}(\k')\rangle = 
(2\pi)^3\delta_{\rm D}(\k+\k')\, g^2(z_{\rm in})\,P(k)\,,
\ee
where $\delta_{\rm D}$ is the Dirac
delta-function. Here $P(k)$ is the linear power spectrum at $z=0$. We
also used the linear growth factor $g(z)$ to translate the $P(k)$ to
the initial redshift $z_{\rm in}$. 
It is convenient to rescale the initial density field $\delta_{\rm in}$ to
redshift $z$ using the linear growth factor, 
\be
\delta_L(\k,z)  = \frac{g(z)}{g(z_{\rm in})}\delta_{\rm in}(\k)\,.
\ee
We will refer to $\delta_L$ as the linear density perturbation in what
follows and will omit the redshift-dependence.

\begin{figure}
	\centering
    \includegraphics[width=0.49\linewidth]{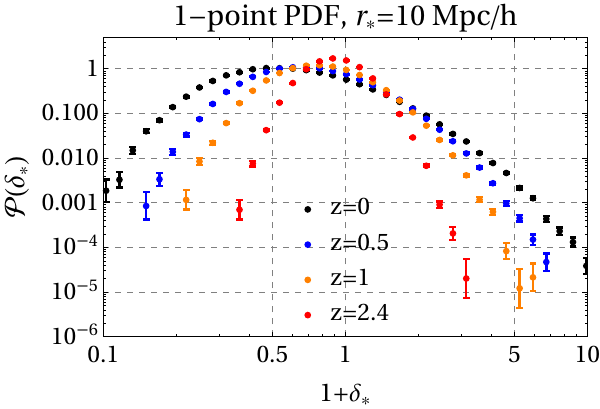}~~
	\includegraphics[width=0.49\linewidth]{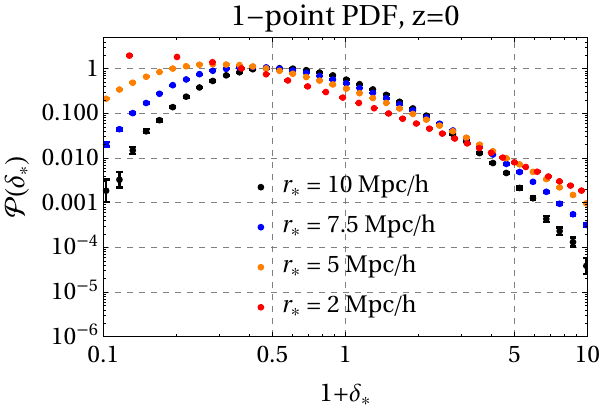}\\
	\caption{The one-point PDF $\mathcal{P}(\delta_*)$ measured from the N-body
          data. {\it Left:} $\mathcal{P}(\delta_*)$ for several redshifts at fixed
          comoving cell radius $r_*=10\,{\rm Mpc}/h$. {\it Right:}
          $\mathcal{P}(\delta_*)$ for several cell radii at $z=0$. 
The errorbars show the statistical uncertainties.}
	\label{fig:scal0}
\end{figure}

The one-point PDF $\mathcal{P}(\delta_*;z,r_*)$ is defined as the probability to observe
$\bar\delta_W$ within a given infinitesimal interval,
\be
\label{Pdef}
\mathcal{P}(\delta_*;z,r_*)d\delta_*=\text{Probability}
\{\delta_*<\bar\delta_W<\delta_*+d\delta_*\}\;.
\ee
It depends on the redshift $z$ and the cell radius $r_*$ as parameters; in what follows we will often suppress these arguments to keep notations concise. Fig.~\ref{fig:scal0} shows PDF measured from data of  
high-resolution N-body simulation \cite{HACC:2021sgt}.\footnote{See Sec.~\ref{sec:result} for the details of this measurement.}
The PDF varies by several orders of magnitude in the density range $1+\delta_*\in [0.1,10]$ and exhibits strong dependence on the redshift and cell radii. This exponential sensitivity is a hallmark of non-perturbative behavior similar to tunneling in quantum theory or large fluctuations in statistical physics and suggests using the saddle-point method which successfully describes the latter phenomena. 

The PDF can be evaluated as the functional 
integral with Gaussian weight over all linear density configurations that
produce desired non-linear density contrast at the final time. The
latter condition is conveniently imposed by the delta-function
inserted into the integration measure,
\be
\label{deltaD}
\delta_{\rm D}\Big(\delta_*-\bar\delta_W[\delta_{L}]\Big)
=\int_{-i\infty}^{i\infty}\frac{d\lambda}{2\pi i g^2}
\exp\bigg\{\frac{\lambda}{g^2}
\Big(\delta_*-\bar\delta_W[\delta_{L}]\Big)\bigg\}\;,
\ee
where $\bar\delta_W[\delta_L]$ describes the functional dependence of
$\bar\delta_W$ on the linear field. This leads to the following
expression \cite{Valageas:2001zr,Ivanov:2018lcg}, 
\be 
\label{eq:pdfLaplace2}
\mathcal{P}(\delta_*)=
\int_{-i\infty}^{i\infty}\frac{d\lambda}{2\pi i g^2}
 \exp\bigg\{\frac{\lambda\delta_*}{g^2}+w(\lambda)\bigg\}\,.
\ee 
where the function $w(\lambda)$ reads
\be
\label{w0}
w(\lambda)=\ln\left[{\cal N}^{-1}\int \mathscr{ D}\delta_L\exp\bigg\{-\frac{1}{g^2}
\int_\k\frac{|\delta_L(\k)|^2}{2P(k)}
-\frac{\lambda\bar \delta_W[\delta_L]}{g^2}
\bigg\}\right]\;,
\ee
and ${\cal N}$ is a normalization factor. Let us first consider high redshifts when the growth factor $g(z)$ is small. Then the integrals (\ref{eq:pdfLaplace2}), (\ref{w0}) can be evaluated in the saddle-point approximation. The results reads schematically: 
\be
\label{scaling}
\mathcal{P}(\delta_*)=
\exp\left\{-\frac{1}{g^2}\left(\alpha_0+\alpha_1 g^2 + \alpha_2
    g^4+...\right)\right\} \,. 
\ee
The leading term in this expansion $\alpha_0$ corresponds to the
exponential part given by the leading saddle-point configuration. The
first correction $\alpha_1g^2$ comes from the Gaussian integration over small
fluctuations around the saddle-point solution which is equivalent to
an one-loop calculation.\footnote{For the sake of the argument, we suppress the logarithmic dependence of $\alpha_1$ on $g$ which will be restored later, cf. Eq.~(\ref{Psp2}).}
This part produces the prefactor in front of the leading exponent.
The term $\alpha_2 g^4$ represents the two-loop correction to the
prefactor, and so on.
In this paper we focus on evaluation of the first two terms. 

Of course, the growth factor $g(z)$ is not small at low redshift --- in fact, we normalize it to 1 at $z=0$. 
Still, it is convenient to keep it as a formal expansion parameter. As we are going to see shortly, the true expansion goes in the smoothed linear density variance $\sigma_{r_*}^2(z)$ at the scale of the cell. This is proportional to $g^2(z)$ which thus provides a useful book-keeping tool to track the powers of $\sigma_{r_*}^2$ arising in the expansion.\footnote{Cf. Refs.~\cite{Blas:2015qsi,Blas:2016sfa} where a similar trick was used to keep track of the powers of the linear power spectrum in perturbation theory.} 
Note that the higher-loop contributions corresponding to higher powers of $g^2$ in (\ref{scaling}) 
are more important at lower redshifts.

The functional $\bar\delta_W[\delta_L]$ entering in Eq.~(\ref{w0})
encodes the relation between the initial linear perturbation and the
final non-linear density contrast into which it evolves. It therefore
depends on the non-linear dynamics of matter inhomogeneities and can
be complicated for a general window function $\tilde W$. However,
the situation is dramatically simplified for the top-hat $\tilde
W$. In this case the symmetry of the problem implies that the
saddle-point configuration $\hat\delta_L(\k)$ saturating the integral
(\ref{w0}) is spherically symmetric. Then its evolution is described
by equations of spherical collapse which, as long as matter obeys the
equivalence principle and before shell-crossing, establish one-to-one
mapping between the linear and non-linear density contrasts, 
\be
\label{eq:scmap}
\bar\delta_W=f(\bar\delta_{L,R(\bar\delta_W)})~~~~
\Longleftrightarrow~~~\bar\delta_{L,R(\bar\delta_W)}=F(\bar\delta_W)\;.
\ee
Here $\bar\delta_{L,R(\bar\delta_W)}$ is the linear density perturbation averaged
with a top-hat filter at
the Lagrangian radius corresponding to the Eulerian radius
$r_*$ and density contrast $\bar\delta_W$, 
\be
\label{Rstar0}
R(\bar\delta_W)=r_*(1+\bar\delta_W)^{1/3}\;.
\ee 
In $\Lambda$CDM cosmology the mapping (\ref{eq:scmap}) depends,
in principle, on the redshift $z$. However, this dependence happens
to be very weak, and the functions $f$ and $F$ are essentially the same
as in the Einstein--de Sitter (EdS) approximation
\cite{Ivanov:2018lcg}.\footnote{In particular, the relative difference
of the function $F(\delta_*)$ in the $\Lambda$CDM and EdS cosmologies
is at most $3\cdot 10^{-3}$ for extreme density contrasts and redshift
zero.} We will adopt the latter approximation in what follows since it
somewhat simplifies the analysis. The value of the cosmological
constant then affects the PDF only through the redshift dependence of
the growth factor $g(z)$. The analytic expressions for the functions $f$, $F$
are given in appendix~\ref{app:path}.

The mapping (\ref{eq:scmap}) allows one to find the saddle-point
solution and perform the integration over spherically symmetric fluctuations around
it analytically (see appendix~\ref{app:path}). 
This gives a part of the PDF, to which we refer as ``spherical PDF'':
\be
\label{Psp2}
{\cal P}_{\rm sp}(\delta_*;z,r_*)\equiv\frac{1}{\sqrt{2\pi
    g^2(z)}}\frac{\d\nu(\delta_*;r_*)}{\d\delta_*}
\e^{-\frac{\nu^2(\delta_*;r_*)}{2g^2(z)}}\;,
\ee
where
\be
\label{eq:nu}
\nu(\delta_*;r_*)\equiv \frac{F(\delta_*)}{\sigma_{R_*}}\;,
\ee
and 
\be
\label{sigmaRstar}
\sigma^2_{R_*} = \int_{\k}\,P(k)\,|W_{\rm th}(kR_*)|^2\,
\ee
is the linear density variance at the Lagrangian radius 
\be
\label{Rstar}
R_*=r_*(1+\delta_*)^{1/3}\;.
\ee 
Note that $\sigma_{R_*}$ depends on $\delta_*$ through the filtering
radius (\ref{Rstar}). 
Note also that the exponent is inversely proportional to
the density variance, which justifies our claim that the latter plays
the role of the saddle-point expansion parameter. 
The spherical PDF is nothing but a map of the Gaussian PDF for the linear density threshold $\nu_L=\delta_L/\sigma_L$ to non-linear densities using spherical collapse and mass conservation. It naturally appears in the excursion-set approach to the PDF \cite{Lam:2007qw,Paranjape:2011wa,Musso:2012qk} but by itself does not reproduces the correct mean $\langle\delta_*\rangle=0$. This problem can be remedied by correcting ${\cal P}_{\rm sp}(\delta_*)$ with an additional factor $1/(1+\delta_*)$ which then, however, spoils the PDF normalization.

\begin{figure}
	\centering
	\includegraphics[width=0.49\linewidth]{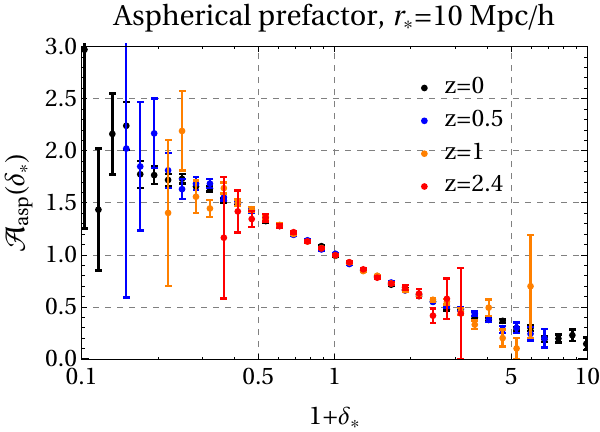}~~~
	\includegraphics[width=0.49\linewidth]{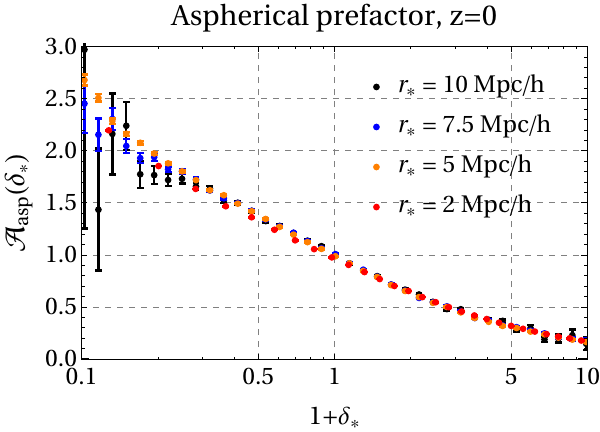}
	\caption{Aspherical prefactor measured from the N-body
          data. {\it Left:} $\Apr(\delta_*)$ for several redshifts at fixed
          comoving cell radius $r_*=10\,{\rm Mpc}/h$. {\it Right:}
          $\Apr(\delta_*)$ for several cell radii at $z=0$. 
The errorbars show the statistical uncertainties of the
measured PDF.}
	\label{fig:scal}
\end{figure}

As shown in \cite{Ivanov:2018lcg}, inclusion of the Gaussian integral over aspherical fluctuations in Eq.~(\ref{w0}) automatically produces the correct mean and normalization. It leads to an ``aspherical prefactor'' ${\cal A}_{\rm asp}$, so that the full PDF takes  
a factorized form
\be
\label{Ptot}
{\cal P}(\delta_*;z,r_*)=\Apr(\delta_*;z,r_*)\cdot {\cal P}_{\rm sp}(\delta_*;z,r_*)\;.
\ee 
To get a sense of the properties of $\Apr$, we can compare the formula
(\ref{Ptot}) to the PDF measured from data of high-resolution N-body
simulation \cite{HACC:2021sgt}. We take the measured PDF and
divide out the spherical part.
This gives us an estimate of
$\Apr(\delta_*)$ shown in Fig.~\ref{fig:scal}.
We observe a remarkable universality: The aspherical prefactor has
almost no dependence on redshift $z$ or the cell radius $r_*$. The
universality persists down to very short scales $r_*\simeq 2\,{\rm
  Mpc}/h$ which are strongly non-linear at zero redshift. As we
explain below, both properties are consistent with the saddle-point
expansion suggesting that it can work even at the non-linear
scales.

Given rather featureless shape of $\Apr(\delta_*)$ one can try to fit it by some simple function with a few free
parameters. A possibility of this kind is discussed in
appendix~\ref{sec:log} where we consider a fitting 
function with just two free parameters
(for each cell radius) which is shown to capture $\Apr$ with
reasonable accuracy. Such fitting, however, does not tell us anything
about the physical origin of $\Apr$ and our ability to describe it in
more general cosmologies than minimal $\Lambda$CDM. In what follows we
develop an accurate
theoretical model for $\Apr$ from first principles.

\subsection{Aspherical prefactor}
\label{sec:asp}

\begin{figure}[tb]
	\centering
	\includegraphics[width=0.3\linewidth]{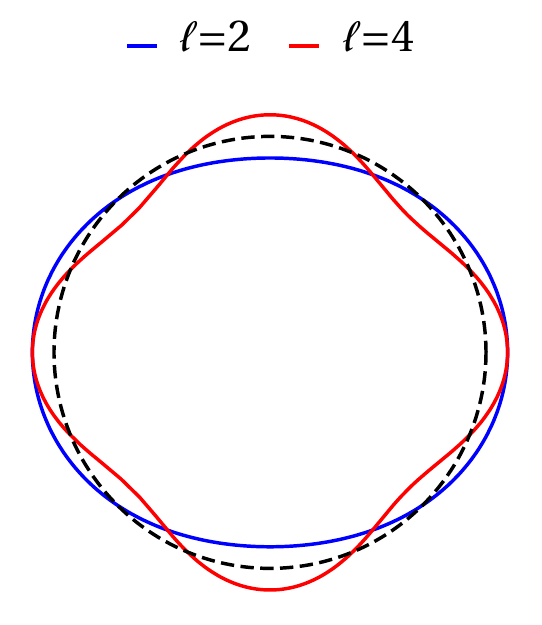}~~~\hspace{2cm}
	\includegraphics[width=0.31\linewidth]{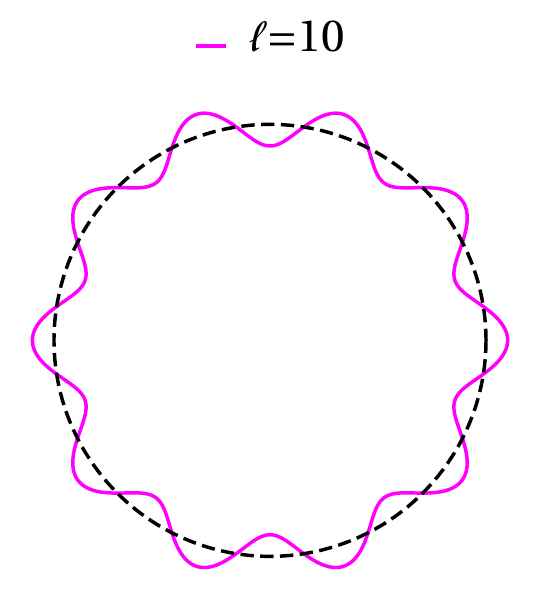}
	\caption{Pictorial representation of aspherical fluctuations in different multipole sectors. Dashed line shows the unperturbed cell boundary. 
 {\it Left:} $\ell=2$ and $\ell=4$. {\it Right:}
         $\ell=10$.}
	\label{fig:multi}
\end{figure}

The aspherical prefactor accounts for the Gaussian integral over small direction-dependent fluctuations of the linear density $\delta_L$ around the saddle-point solution in Eq.~(\ref{w0}). These aspherical fluctuations perturb the non-linear density $\bar\delta_W$ in the cell at the second order of perturbation theory. One can think of them as the perturbations of the comoving cell boundary --- the trajectories of particles ending up at the radius $r_*$ by the final time when the PDF is measured, see Fig.~\ref{fig:multi}. Clearly, the perturbations with different multipole numbers $\ell$ decouple and can be evaluated separately. We now briefly 
review the algorithm for calculation of
the aspherical prefactor developed in \cite{Ivanov:2018lcg}. 
We describe matter as the pressureless fluid and
work in the EdS approximation.\footnote{We have checked that in $\Lambda$CDM the result is the same.}
The details are given in appendix~\ref{app:Apr}. 

As a first step we
consider the evolution of linearized aspherical fluctuations
in the background of saddle point solution.
Expanding all quantities into 
background and first order perturbations, $\delta=\hat\delta+\delta^{(1)}$, etc.,
we write the Euler--Poisson equations to leading order in perturbations.
The linearized equations (\ref{EPppp})
contain, besides the density perturbation $\delta^{(1)}$, the
potentials $\Psi^{(1)}$ and $\Theta^{(1)}$ related to the
perturbations of velocity $u_i^{(1)}$ as follows,\footnote{We neglect
vorticity which is not generated until shell crossing.}
\be
\label{PsiThet}
u_i^{(1)}=-{\cal H}\d_i \Psi^{(1)}~,~~~~
\d_i u_i^{(1)}=-{\cal H}\Theta^{(1)}\;.
\ee
where $\H=\d_ta/a=H_0\e^{-\eta/2}$ is the conformal Hubble
parameter, $t$ is conformal time, and $a(t)$ is the scale factor normalized to $1$ at the present
time. 
In the last expression we introduced a new time variable related to the linear growth rate
\be
\label{eta}
\eta \equiv \ln g(t) =\ln a(t)\,.
\ee
Working in a particular multipole sector with $\ell\geq 1$, 
we find the first order perturbations
$(\delta_\ell^{(1)},\Theta_\ell^{(1)},\Psi_\ell^{(1)})$ 
for a set of initial
wavenumbers $\{k_I\}$, $I=1,\ldots,N$ representing discretized radial momentum space. We take the momentum lattice to be equidistant with the spacing $\Delta k=(k_N-k_1)/N$.

As a next step, we consider the evolution of the second-order monopole perturbation $\delta^{(2)}_0$
sourced by two aspherical fluctuations with a given $\ell$ and momenta $k_I$, $k_J$.
The problem reduces to two ordinary differential equations (\ref{mu2r2})
on the time-dependent cell radius $r_{IJ}(\eta)$ and the mass perturbation
inside it $\mu^{(2)}_{IJ}(\eta)$ with the sources (\ref{Ups}) quadratic in first-order aspherical perturbations with $k_I$ and $k_J$.
The sources are evaluated at the unperturbed comoving cell boundary $\hat r(\eta)$ corresponding to the trajectories of particles in the saddle-point spherical collapse solution which end up at $r_*$ at the final time $\eta=0$. 
The second-order perturbation of the mass gives us an element of the response matrix,
\be
\label{Qdiag}
Q_\ell(k_I,k_J)=\frac{3}{r_*^3}\mu^{(2)}_{IJ}(\eta=0)\;, 
\ee
which describes the second-order response of the non-linear density contrast
to the changes in the initial linear field.   

We repeat the above procedure for all pairs $I,J$ and construct the full response matrix for a fixed multipole $\ell$.
Then, we evaluate the fluctuation determinant in the $\ell$th sector \cite{Ivanov:2018lcg},
\be
\label{Dl}
\D_{\ell} = \det\left(\delta_{IJ}+2\hat\lambda (2\pi)^{-3} \Delta k\, k_I k_J
  \sqrt{P(k_I)}Q_{\ell}(k_I,k_J)
\sqrt{P(k_J)}\right)\,,
\ee
where 
$\hat\lambda$ is the
saddle-point value of the Lagrange multiplier given in Eq.~(\ref{eq:lambda}).
The final answer for the aspherical prefactor is given by the product of contributions from different multipole
sectors,
\be
\label{prefl>0}
\Apr= \prod_{\ell\geq 1}\D_\ell^{-(\ell+1/2)} \,.
\ee
Note that the power
$(\ell+1/2)$ comes from the number $(2\ell+1)$ of
independent spherical harmonics in the $\ell$th sector.

Let us
discuss several general properties of the aspherical prefactor that
follow from Eqs.~(\ref{Dl}), (\ref{prefl>0}). We observe that these
expressions do not explicitly depend on the growth factor $g$. Since
the latter controls the redshift dependence of the PDF, this implies
that $\Apr$ formally does not depend on $z$. Further, for a power-law
power spectrum, $P(k)\propto k^n$, the LSS properties are invariant
under the simultaneous rescaling of momenta and the power spectrum
amplitude, $P(k)\mapsto \alpha^{-n-3}P(\alpha k)$
\cite{Pajer:2013jj}. 
Applied to the PDF, this symmetry maps cells of different radii onto
each other, $r_*\mapsto \alpha^{-1}r_*$, simultaneously changing the
value of $g^2\mapsto \alpha^{-n-3}g^2$. The independence of $\Apr$ of
$g$ then implies its independence of the cell radius.

We have already noticed in Fig.~\ref{fig:scal} that $\Apr$ extracted
from N-body data is essentially independent of $z$ and $r_*$. This
supports the semiclassical expressions (\ref{Dl}), (\ref{prefl>0})  and
also suggests that $\Apr$ is saturated by modes in a relatively narrow
range of momenta where the power spectrum can be approximated as a
power-law. We are going to see below that the $z$ and $r_*$
independence of $\Apr$ is weakly broken by the multiplicative
counterterm 
which renormalizes the UV contribution of modes with high $\ell$ and
$k$. Another source of breaking is the two-loop correction,
  see Sec.~\ref{sec:result}.

\subsection{Contribution of UV modes}
\label{sec:WKB}

As discussed above, the computation of the fluctuation determinant for a general
$\ell$
requires solving the system of partial differential equations for the
set of functions
$(\delta_\ell^{(1)},\Theta_\ell^{(1)},\Psi_\ell^{(1)})$. But in the
limit of large orbital numbers $\ell\gg1$ one can use the
Wentzel--Kramers--Brillouin (WKB) approximation to reduce 
the equations on aspherical perturbations
to the ordinary differential equations in time which can be easily
solved. 
The value of this analysis is twofold. First, it allows us to
efficiently evaluate the contributions of high multipoles to the
aspherical prefactor within the fluid picture. Second, it elucidates
the structure of this contribution and provides an insight for its
renormalization.  

The details of this approach are worked out in appendix~\ref{app:WKB}
leading to the following key results. 
First, the WKB equations for linear perturbations have ultralocal form, i.e. they do not contain $r$-derivatives and become ordinary differential equations for time evolution along the flow lines of the background solution. 
Second, the elements of the response matrix \eqref{Qdiag} happen to be small, $Q_\ell(k,k')\propto (\ell+1/2)^{-2}$.
This allows us to approximate the
fluctuation determinant (\ref{Dl}) by the exponent of trace,
\be
\label{detdiag}
{\cal D}_\ell\approx \exp\left(2\hat{\lambda}\,\text{Tr}PQ_\ell\right)
=\exp\left(2\hat\lambda\int [dk]P(k)Q_\ell(k,k)\right)\,,
\ee 
where the integration measure $[dk]$ is defined in (\ref{kintradial}). 
We see that only the diagonal elements of the response matrix
contribute implying that it is sufficient to solve the second-order equations 
Eqs.~(\ref{mu2r2}) only for diagonal sources with $k_I=k_J=k$.

Third, the diagonal elements of the response matrix are sums of two parts. One of them
quickly oscillates with $k$. It gives rise to a quickly oscillating
contribution in $Q_\ell(k,k)$ which averages to zero in the integral
(\ref{detdiag}). The second part is slowly varying
and has the form,
\be
\label{QR}
Q_\ell(k,k)\simeq 
\frac{q(\vk)}{k^{2}(\ell+1/2)^{2}}\;,
\ee 
where the symbol $\simeq$ reminds that we have
discarded the oscillating part.
The function $q$ depends only on the ratio,
\be
\label{vk}
\varkappa\equiv\frac{k}{\ell +1/2}\;,
\ee
Using above scaling one can efficiently evaluate
the contribution of high-$\ell$ multipoles to the aspherical prefactor, for detail see Ref.~\cite{Ivanov:2018lcg}.

Let us now isolate the contribution of the UV modes into the aspherical prefactor. We sum over all
multipoles with $\ell\gg1$ and all
modes with $k\gg R_*^{-1}$, where $R_*$ is the Lagrangian radius of the cell defined in Eq.~(\ref{Rstar}).
The corresponding part of the aspherical prefactor can be written as
\begin{align}
\label{AprUV}
\Apr^{\rm UV}&\simeq\exp\bigg\{
\int\limits_{k\gg R_*^{-1}} \frac{dk\,P(k)}{(2\pi)^3}\sum_\ell\frac{1}{\ell+1/2}q\l\frac{k}{\ell+1/2}\r\bigg\}\notag\\
&\approx\exp\bigg\{
\int\limits_{k\gg R_*^{-1}} \frac{dk\,P(k)}{(2\pi)^3}\cdot
\dashint_{R_*^{-1}}^\infty \frac{d\vk}{\vk}\,q(\vk)\bigg\}\;,
\end{align}
where in the second line we have
approximated the sum over $\ell$ by an integral and used 
$d\ell/(\ell+1/2)=d\vk/\vk$. 
We observe that the expression inside the exponent factorizes into
an integral over $k$ of the power spectrum and a $\vk$-integral of the function $q(\vk)$ encoding the background dependence. The latter integral 
quickly
converges at the upper limit. It has a spurious
singularity at the lower end stemming from the breakdown of the WKB
approximation at $\vk\to R_*^{-1}$ but is rendered finite by a proper subtraction of the boundary term (see appendix \ref{app:WKB2}). 
We use the dash-integral sign to remind of this subtraction when it is needed. 

In deriving (\ref{AprUV}) we used the Euler--Poisson equations for ideal pressureless fluid (see appendix \ref{app:Apr}). This is a good description for modes with moderate $\ell$ and $k$, like those shown on the left of Fig.~\ref{fig:multi}. However, it breaks down for modes with high $\ell$ and $k$. These vary on such short scales that the perturbed trajectories of particles cross violating the single-stream approximation --- the modes experience shell-crossing. It is worth stressing that this shell-crossing should not be confused with that happening in the center of the cell on the leading spherical collapse solution for very large overdensities $\delta_*\gtrsim 7$ (see appendix \ref{app:path}). The latter redistributes matter in the central part of the cell but does not affect its total mass.\footnote{At least as long as it does not lead to an outflow through the cell boundary which does not happen for $\delta_*<9$ and $r\gtrsim 5\,{\rm Mpc}/h$.} 
On the other hand, the shell-crossing of short modes occur for all $\delta_*$ and perturbs the cell boundary (see Fig.~\ref{fig:multi}, right), thereby affecting the mass inside the cell. This effect cannot be discarded in the calculation of the PDF.

\begin{figure}[t]
	\centering
	\includegraphics[width=0.48\linewidth]{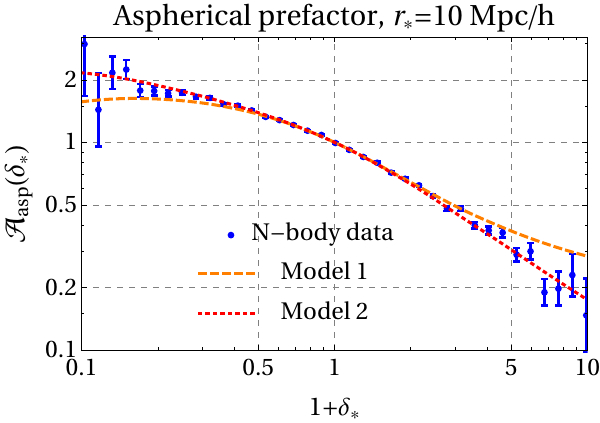}~~~
	\includegraphics[width=0.48\linewidth]{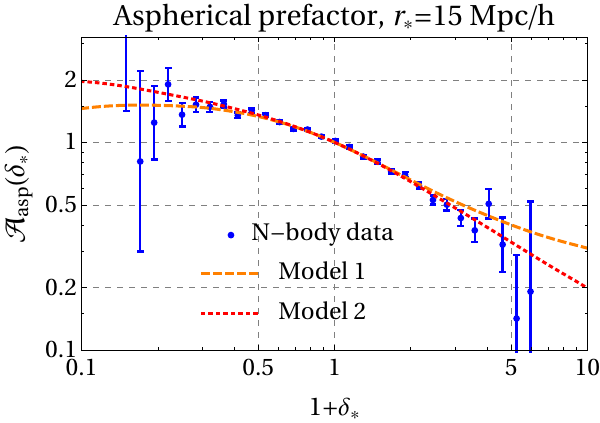}
	\caption{Comparison of Models 1 and 2 from Ref.~\cite{Ivanov:2018lcg} for renormalized aspherical prefactor with the N-body
          data at $z=0$. The cell radii are $r_*=10\,{\rm Mpc}/h$ (left) and $r_*=15\,{\rm Mpc}/h$ (right). 
          The theoretical models use 
          $\gamma_0=1.95\,({\rm Mpc}/h)^2$ 
          for the effective speed of sound at $z=0$ taken from the fit of the power spectrum (see Sec.~\ref{sec:result}). The dependence of the sound speed on the growth factor is approximated by a power-law with the index
          $m=8/3$.
           }
          \label{fig:models}
\end{figure}

Ref.~\cite{Ivanov:2018lcg} suggested to take this effect into account by renormalizing the $k$-integral in Eq.~(\ref{AprUV}). The key observation is that the same integral arises in the one-loop calculation of the power spectrum where it is corrected by a {\em counterterm} $\gamma$ also sometime referred to as the ``sound speed''. Physically, $\gamma$ is inversely proportional to the square of the cutoff momentum $k_{\rm sc}$ corresponding to the modes experiencing shell-crossing, $\gamma\propto k_{\rm sc}^{-2}$. 
Model 1 proposed in \cite{Ivanov:2018lcg} simply applies the renormalization with fixed $\gamma$ taken from the perturbation theory to the integral in (\ref{AprUV}). This provides a good description of the aspherical prefactor for moderate $\delta_*\in[-0.5,1]$ but deviated from the N-body data by $\sim 30\%$ at the tails, see Fig.~\ref{fig:models}. The second model of \cite{Ivanov:2018lcg} improves the description by estimating the $\delta_*$-dependence of the counterterm within the separate universe approximation. In this simplified approach the spherical collapse solution is approximated by a uniform density contrast reaching $\delta_*$ at the final time. The model happens to provide an accurate description of the data within the statistical errorbars, as shown in Fig.~\ref{fig:models} (more detailed comparison of Model 2 with the N-body data can be found in appendix~\ref{app:comp}). 

Despite the phenomenological success of Model 2, it is admittedly {\it ad hoc}. The true saddle-point profile is not uniform (see Fig.~\ref{fig:profiles} from appendix~\ref{app:path}) and changes significantly at the cell boundary where aspherical perturbations are most important. For this reason it was pointed out in \cite{Ivanov:2018lcg} that Model 2 should be taken with a grain of salt and, in the absence of a better understanding of the prefactor renormalization, its difference from Model 1 should be considered as the theoretical uncertainty. The goal of this work is to reduce the uncertainty by constructing a well-grounded counterterm model based on renormalization of the stress tensor due to short perturbations in the effective equations of motion.

\section{Counterterm prefactor and its relation to correlators}
\label{sec:rel}

Before proceeding to the effective stress tensor, it is instructive to discuss the connection between the renormalization of PDF and the correlation functions. There is a well-known relation between tree-level spherically-averaged cumulants and the physics of spherical collapse providing the leading PDF exponent \cite{Bernardeau:1992zw,Bernardeau:1994zd,Fosalba:1997tn}. In this section we scrutinize the relation at the one-loop level, focusing, in particular, on the counterterms on both sides of the relation.  

The EFT approach to LSS suggests to focus on the evolution and
statistics of long-wavelength perturbations in the quasi-linear
regime, while ``integrating out'' short non-linear modes. The effect of
the latter is then encapsulated into an effective stress tensor that
must be added to the r.h.s. of the Euler equation (\ref{EP2}) for long
modes. Application of this strategy to the PDF means restricting the
integral in Eq.~(\ref{w0}) to modes with $k<k_{\rm cut}$, where
$k_{\rm cut}$ is some cutoff scale, simultaneously modifying the map
$\bar\delta_W[\delta_L]$. In practice, it is conventional to remove
the cutoff by sending $k_{\rm cut}\to \infty$. Still, the effective
stress tensor and hence the modification of the map with respect to
the pure fluid picture persists even in this limit and results in a
{\it counterterm} contribution. By assumption, the latter is an
one-loop effect and in the perturbative power counting is proportional
to the coupling $g^2$. Thus, we write,\footnote{We are going to see that $\bar\delta_W^\ctr$ can have an additional weak dependence on $g$.}
\be
\label{deltactr}
\bar\delta_W[\delta_L]=\bar\delta_W^{\rm fluid}[\delta_L]
+g^2\bar\delta_W^{\rm ctr}[\delta_L]\;,
\ee 
where the first term is the pure fluid contribution given by solving
the sourceless Euler--Poisson system (\ref{EP}), and the second term
is the counterterm due to the stress tensor of short modes. 

\begin{figure}[tb]
	\centering
	\includegraphics[width=0.55\linewidth]{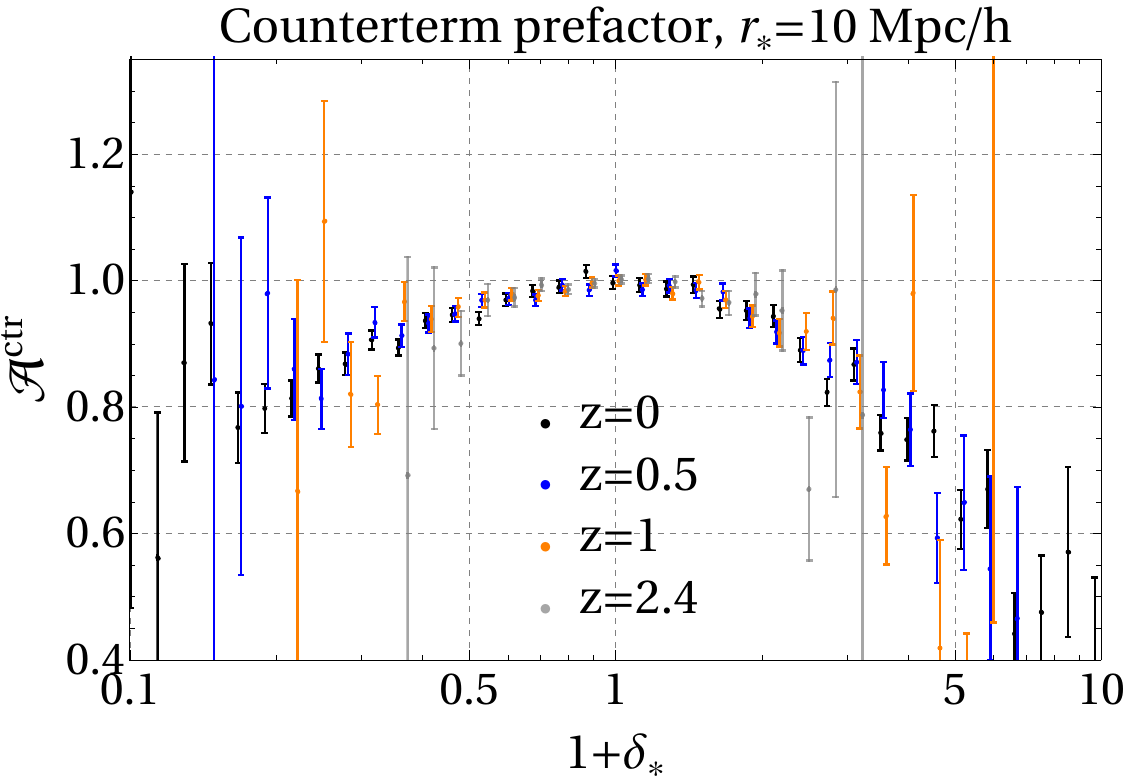}~~~
	\caption{Counterterm prefactor extracted from N-body
          simulations for cell radius ${r_*=10~{\rm Mpc}/h}$ and several
          values of the redshift. Error bars represent the
          statistical uncertainty of the simulations. }
	\label{fig:Actr_sim}
\end{figure}

Since the counterterm is $\mathcal{O}(g^2)$, it does not affect the saddle-point
solution. At one-loop level its contribution is simply given by
evaluating $\bar\delta_W^{\rm ctr}$ on the saddle point and leads to
a multiplicative renormalization of the aspherical prefactor,
\be
\label{Arenorm}
\Apr={\cal A}^{\rm ctr}\cdot \Apr^{\rm fluid}\;,
\ee
where $\Apr^{\rm fluid}$ is given by the product of determinants
(\ref{prefl>0}) computed within the fluid approximation, and 
\be
\label{Actrdef}
\Actr=\exp[-\hat\lambda\bar\delta_W^\ctr[\hat\delta_L]]
\ee
is the {\it counterterm prefactor}. To get an idea of its magnitude, we
divide the aspherical prefactor extracted from N-body simulations by
the computed value of $\Apr^{\rm fluid}$. The result is shown in
Fig.~\ref{fig:Actr_sim}. We observe that 
$\Actr$ varies from $1$ at $\delta_*=0$ to $\sim 0.8$ ($\sim
0.5$) at extreme under- (over-) densities. In other words, it
represents a moderate, but significant correction to the ideal fluid
approximation.
We want to develop an algorithm for computing 
$\Actr$ as function of the density contrast
$\delta_*$, the cell radius $r_*$ and redshift $z$. 
This task for arbitrary $\delta_*$ will be addressed in
the next section. Presently, we focus on the vicinity of the point
$\delta_*=0$ and establish the relation between the Taylor expansion
of $\ln[\Actr(\delta_*)]$ and the counterterms in the correlation
functions. 
 
We notice that Eq.~(\ref{w0}) can be viewed as the generating function
for connected correlators with the source $-\lambda W/g^2$,
\bseq
\label{wfunc0}
\be
\label{wfun}
w(\lambda)=\ln\left\langle\exp\left(-\frac{\lambda}{g^2}
\int_\k W(kr_*)\delta(\k)\right)\right\rangle
=
\sum_{n=2}^\infty\frac{1}{n!}\left(-\frac{\lambda}{g^2}\right)^n c_{W,n}
\ee
where 
\be
\label{c0}
c_{W,n}=
\int_{\k_1}...\int_{\k_n} W(k_1r_*)...W(k_nr_*)\,
\langle\delta(\k_1)\delta(\k_2)...\delta(\k_n)\rangle_\text{connected}
\ee
\eseq
is the connected $n$-point correlator filtered with the window
function in each argument. Within the perturbation theory every such
correlator can be split into a tree-level and loop parts that scale differently with $g$,
\be
\label{cWsplit}
c_{W,n}=g^{2(n-1)} c_{W,n}^\tree+g^{2n}c_{W,n}^\text{1-loop}
+\mathcal{O}(g^{2(n+1)})\;.
\ee
Grouping terms of the same order in the generating function, we obtain
the decomposition,
\be
\label{w}
w(\lambda)=\frac{w^\tree(\lambda)}{g^2}+w^\oneloop(\lambda)+\mathcal{O}(g^2)\;. 
\ee
Substituting this onto Eq.~(\ref{eq:pdfLaplace2}) and evaluating the
integral by the saddle point yields the PDF in the form,
\be 
\label{eq:pdfLaplace3}
\mathcal{P}(\delta_*)=\frac{1}{\sqrt{2\pi g^2}} 
\bigg(\frac{d^2w^\tree}{d\lambda^2}\Big|_{\hat\lambda}\bigg)^{-1/2}
\e^{w^\oneloop(\hat\lambda)}
\cdot\e^{\frac{1}{g^2}(\hat\lambda\delta_*+w^\tree(\hat\lambda))}\;,
\ee
where $\hat\lambda$ and $\delta_*$ are related by the saddle-point
condition, 
\be
\label{saddle}
\delta_*+\frac{d w^\tree}{d\lambda}\Big|_{\hat\lambda}=0\;.
\ee

Of course, the PDF (\ref{eq:pdfLaplace3}) must be the same as
(\ref{Ptot}) which gives us the expression 
\be
\label{ASPnu}
\Apr=\sqrt{1+\nu\frac{d^2\nu}{d\delta_*^2}\left(\frac{d\delta_*}{d\nu}\right)^2}
\exp\left[ w^\oneloop(\hat\lambda)\right]\;,
\ee
where we have used the relations (\ref{saddle}) and (\ref{eq:lambda})
to express the first factor (recall also the definition of $\nu$
(\ref{eq:nu})). This formula establishes a connection between the
aspherical prefactor and the one-loop contribution into the cumulant generating
function. Note that while its derivation assumes the validity of
the saddle-point approximation,
it does not rely on any model for dynamics. 

We now observe that the first factor in Eq.~(\ref{ASPnu}) is
determined exclusively by the spherical collapse saddle-point solution
and therefore is not UV sensitive. All UV counterterms are contained
in the second factor, so we have
\be
\label{Actrwctr}
\ln[\Actr(\delta_*)]=w^\ctr(\hat \lambda)\;,
\ee
where $w^\ctr(\lambda)$ is composed of one-loop counterterms for all
correlators using  Eqs.~(\ref{wfunc0}). Let us work out this relation
up to the cubic terms. From the equations of appendix~\ref{app:path} one
can obtain,\footnote{Alternatively, one can use Eq.~(\ref{saddle}) and
obtain
\[
\hat\lambda=-\frac{1}{c_{W,2}^\tree}\bigg(\delta_*
-\frac{c_{W,3}^\tree}{2(c_{W,2}^\tree)^2}\delta_*^2+\mathcal{O}(\delta_*^3)\bigg)\;,
\]
which implies the identities,
$c_{W,2}^\tree=\sigma_{r_*}^2$, 
$c_{W,3}^\tree=-6\sigma_{r_*}^4
\left(\frac{4}{21}-\frac{\xi_{r_*}}{\sigma_{r_*}^2}\right)
$.
}
\be
\hat\lambda(\delta_*)=-\frac{1}{\sigma_{r_*}^2}\bigg(
\delta_*+3\bigg(\frac{4}{21}-\frac{\xi_{r_*}}{\sigma_{r_*}^2}\bigg)\delta_*^2
+\mathcal{O}(\delta_*^3)\bigg)\;,
\ee
where the tree-level variance $\sigma_{r_*}^2$ and smeared correlation
function $\xi_{r_*}$ are defined by Eqs.~(\ref{sigmaRstar}),
(\ref{xiRstar}) with $R_*$ replaced by $r_*$.
This give us, 
\be
\label{Actrd3}
\ln[\Actr(\delta_*)]=
\frac{\delta_*^2}{2\sigma_{r_*}^4} c_{W,2}^\ctr
+\frac{\delta_*^3}{6\sigma_{r_*}^6}\left( c_{W,3}^\ctr+
18\left(\frac{4}{21}-\frac{\xi_{r_*}}{\sigma_{r_*}^2}
\right)
\sigma_{r_*}^2c_{W,2}^\ctr
\right)
+\mathcal{O}(\delta_*^4)
\,,
\ee
where 
\be
\label{cnctr}
c_{W,n}^\ctr=g^{-2n}\int_{\k_1}...\int_{\k_n}
W(k_1r_*)...W(k_n r_*)\,\langle
\delta(\k_1)...\delta(\k_n)\rangle^\ctr\;.
\ee
Let us discuss this result. First, we note the absence of a linear
term. This is consistent with the fact that the first derivative of
the aspherical prefactor at $\delta_*=0$ is entirely determined by the
spherical collapse dynamics and is insensitive to UV
\cite{Ivanov:2018lcg}. Second, the coefficients of the quadratic and
cubic terms are directly related to the counterterms in the power
spectrum and bispectrum that have been widely studied in the
literature
\cite{Carrasco:2012cv,Pajer:2013jj,Carrasco:2013mua,
Baldauf:2014qfa,Angulo:2014tfa,Foreman:2015uva,
Baldauf:2015aha,Lazeyras:2019dcx,Steele:2020tak}. 

The power-spectrum counterterm has the form, 
\be
\label{PSctr}
\langle\delta(\k)\delta(\k')\rangle^\ctr=-2\gamma g^2 k^2
P(k)\,(2\pi)^3\delta_{\rm D}(\k+\k')\;,
\ee
where $\gamma$ is a redshift-dependent parameter of dimension
$(\text{length})^2$, referred to as the
``effective sound speed''. By comparing the EFT predictions with the
results of N-body simulations it has been shown to approximately 
follow a power-law
dependence on the growth factor, 
\bseq
\label{gammaz}
\begin{gather}
\label{gammazfit}
\gamma(z)\simeq \gamma_0\, [g(z)]^m\;,\\
\label{gammazval}
\gamma_0\sim 1.5\, (\Mpch)^2~,~~~~~~m\sim 8/3\;.
\end{gather}
\eseq
Both $\gamma_0$ and $m$ depend on 
cosmology and the details of the
analysis (inclusion or not of two-loop corrections, the precise
$k$-range used for measurements, etc.). 
The scaling (\ref{gammazfit}) is compatible with
the approximation of a power-law power spectrum; we will elaborate
more on this in the next section. Substituting (\ref{PSctr}) into
Eq.~(\ref{cnctr}) we obtain,
\be
\label{cW2ctr}
c_{W,2}^\ctr(z)=-\frac{2\gamma(z)}{g^2(z)}\Sigma_{r_*}^2\;,
\ee
where
\be
\label{sigmactr}
\Sigma^2_{r_*}=\int_\k \,k^2 P(k)|W_{\rm th}(kr_*)|^2 \,.
\ee
The relation between $\gamma$ and the Taylor expansion of $\Actr$ will
be used below as a prior on the counterterm prefactor model.

The bispectrum counterterm is more complicated. It involves three more
parameters whose measurement from N-body data is challenging. Such
measurement has been performed in \cite{Steele:2020tak}, but for different
cosmological parameters than in the N-body simulations used in this
paper. Since the dependence of the bispectrum counterterm on cosmology
is yet to be explored, we presently do not include it in the
calibration of $\Actr$, leaving this task for future.  

Note that, in contrast to the fluid part of the prefactor, the
counterterm depends on redshift. This dependence is,
however, rather mild. Using (\ref{gammaz}), (\ref{cW2ctr}) we obtain, 
\be
\label{Actrz}
\ln[\Actr]\propto [g(z)]^{m-2}\;.
\ee
Since $(m-2)>0$ this implies that the counterterm
contribution to the PDF decreases with increasing redshift, which is expected
because the LSS is more linear at early epochs.

\section{Counterterm model}
\label{sec:ctr}

In this section we develop a procedure for calculation of the
counterterm prefactor. We first review the EFT approach to LSS following
Refs.~\cite{Baumann:2010tm,Carrasco:2012cv} and discuss the complications that arise in applying it to 
the PDF. We then derive an estimate for the shell crossing scale in
the background of saddle-point solution. Finally, we propose a model
for the
counterterm stress tensor renormalizing the contribution of short modes and
discuss the qualitative features of the resulting~$\Actr$.

\subsection{Stress tensor in EFT of LSS}
\label{sec:tau}

At the fundamental level,
matter composed of non-relativistic collisionless particles of mass
$M$ is
described by the phase-space density   
$f(\x,\textbf{p},t)$ where $\x$ is the comoving spatial coordinate of
a particle, $\textbf{p}=aM\frac{d\x}{dt}$ is 
its canonical momentum, and $t$ is the conformal
time. The first two moments of the distribution function over momentum
give the local matter density and average flow velocity,
\be
\label{rhouf}
\rho(\x,t)=\frac{M}{a^3}\int f(\x,{\bf p},t)d^3p~,~~~~~
{\bf u}(\x,t)=\frac{\int {\bf p} f(\x,{\bf p},t)d^3p}{\int f(\x,{\bf
    p},t)d^3p}\;. 
\ee
Phase-space density conservation leads to the Vlasov equation,  
\be
\label{vlasov}
\frac{\d f}{\d t}+\frac{\textbf{p}}{aM}\cdot\nabla f
-aM\nabla\Phi\cdot\frac{\d f}{\d \textbf{p}}=0\;,
\ee
where $\Phi$ is the Newtonian potential.
Taking moments of Eq.~(\ref{vlasov}) over ${\bf p}$ one obtains the
Boltzmann hierarchy of equations relating various moments of the
distribution to each
other. The fluid equations (\ref{EP}) correspond to truncating this
hierarchy down to first two moments --- matter density and flow
velocity. Such truncation is not justified for short-wavelength
perturbations which undergo shell-crossing breaking the
single-stream approximation and generating velocity dispersion
\cite{Pueblas:2008uv}. 

The EFT of LSS overcomes this problem by restricting to
long modes. Technically, this is implemented by smoothing
the distribution function and Newtonian potential on length scales
$\Delta x\sim k_{\rm cut}^{-1}$,
\bseq
\label{longdef}
\begin{align}
\label{longdef1}
&f^l(\x,{\bf p},t)=\int d^3 x'\,k_{\rm cut}^3 {\cal S}\big(k_{\rm
  cut}(\x-\x')\big)f(\x',{\bf p},t)\;,\\
\label{longdef2}
&\Phi^l(\x,t)=\int d^3 x'\,k_{\rm cut}^3 {\cal S}\big(k_{\rm
  cut}(\x-\x')\big)\Phi(\x',t)\;,
\end{align}
\eseq
where the smoothing function ${\cal S}({\bf y})$ has unit width and is
normalized in such a way that its integral is equal to 1. Its precise
form is unimportant; for example, it can be chosen as a Gaussian. The
wavenumber cutoff is assumed to lie below the shell-crossing scale
$k_{\rm cut}\lesssim k_{\rm sc}$. 

Let us stress that the smoothing function ${\cal S}$ should not be confused with the window function $\tilde W$ used to construct the PDF. The latter is associated with the physical length scale $r_*$ which can be varied in the measurements and serves as a parameter of the PDF. In our study we assume it to lie in the mildly non-linear range, so that the corresponding momentum scale is well below the shell-crossing, $r_*^{-1}\ll\ksc$. Besides, while in this work we focus on the top-hat window in coordinate space, other choices for $\tilde W$ are in general possible and the PDF is expected to depend on them \cite{Bernardeau:2015khs,Ivanov:2018lcg}. On the other hand, the smoothing function ${\cal S}$ is just a theoretical tool to isolate the contribution of short non-linear modes with wavelengths hierarchically smaller than the cell size, $k_{\rm cut}^{-1}\ll r_*$. It appears only in the intermediate steps of the derivation and does not enter in the final answer.  

Applying smoothing to the Vlasov equation (\ref{vlasov}) and then
taking its moments, we find that the continuity equation (\ref{EP1})
is not modified with the long-wavelength density 
$\delta^l$ and velocity ${\bf u}^l$ defined using the
smoothed distribution (\ref{longdef1}). The Poisson equation
(\ref{EP3}) is also left intact by the smoothing procedure thanks to
its linearity. On the other hand, the Euler equation (\ref{EP2})
receives on the r.h.s. a source of the form,
\be
\label{EPnew2}
-\frac{1}{1+\delta^l}\d_j\tau_{ij}\;,
\ee
with an effective stress tensor
\be
\label{stress}
\tau_{ij}=(1+\delta^l)\sigma_{ij}^l
+\frac{2}{3\mathcal{H}^2}\bigg([\d_i\Phi^s\d_j\Phi^s]^l
-\frac{1}{2}\delta_{ij}[\d_k\Phi^s\d_k\Phi^s]^l\bigg)\;.
\ee
It consists of two physically distinct parts to which we will refer
as {\it kinetic} and {\it potential}. The first is related to
the velocity dispersion smoothed on long scales,
\be
\label{disp0}
\sigma_{ij}^l=\frac{\int (v_i-u_i^l)(v_j-u_j^l) f^l\,d^3 p}{\int f^l\,d^3
  p}~,~~~~~~
v_i\equiv\frac{p_i}{aM}\;.
\ee
Note that the adiabatic initial conditions for LSS correspond to vanishing
$\sigma_{ij}$. Its non-zero value is generated in the course of
evolution due to shell crossing which occurs in the short non-linear
modes. The potential contribution
is associated to the fluctuations of the
gravitational potential at short scales
defined as the difference between the full potential and its
smoothed part,
\be
\label{Phishort}
\Phi^s\equiv\Phi-\Phi^l\;.
\ee
Note that while smoothing of any linear function of $\Phi^s$ gives
zero by definition, this is not true for quadratic expressions present in
(\ref{stress}). The potential part is crucial for the decoupling of the
virialized structures from the large-scale evolution: it 
ensures that the effective stress-energy tensor vanishes for
fully virialized systems \cite{Baumann:2010tm}. 
The stress tensor $\tau_{ij}$ encapsulates the effect of
short modes beyond the cutoff on the long-distance dynamics. 

In principle, $\tau_{ij}$ depends both on the states of short and long
modes. We get rid of the former dependence by averaging over random
initial conditions for 
short modes. Using angular brackets with
subscript $s$ to denote the result, we conclude that 
$\langle \tau_{ij}\rangle_s$ is a
functional of the long-wavelength perturbation only.\footnote{Apart from
  this deterministic component, $\tau_{ij}$ contains also a stochastic
contribution. However, its effect on LSS statistics is subdominant
\cite{Carrasco:2012cv,Pajer:2013jj} and
we neglect it in this paper.} In general, this functional is
complicated and depends on the full non-linear configuration of the
long mode and its history. The situation is simplified in the case
when the long mode can be treated perturbatively. Then $\langle
\tau_{ij}\rangle_s$ can be systematically expanded in powers of the
density contrast $\delta^l$, the tidal tensor
$\d_i\d_j\Phi^l-(1/3)\delta_{ij}\Delta\Phi^l$ and their spatial
derivatives. The expansion is still non-local in time, but this issue
can be treated consistently at each order of the perturbation theory
in $\delta^l$. 

Presence of $\langle\tau_{ij}\rangle_s$ leads to additional terms in
the cosmological correlation functions. For concreteness, let us focus
on the
one-loop power spectrum. At this level, the only
relevant term in the stress tensor is 
\be
\label{taufirst}
\langle\tau_{ij}(\x,\eta)\rangle_s=\tilde d(\eta)\delta_{ij}\delta^l(\x,\eta)\;,
\ee
where $\tilde d(\eta)$ is a time-dependent coefficient of dimension
$(\text{length})^2$. On dimensional grounds, it must be proportional
to the inverse cutoff squared, $\tilde d\sim k_{\rm cut}^{-2}$. Substituting (\ref{taufirst}) into perturbed Euler equation gives correction to
the density contrast
\be
\label{deltacorr}
\delta^l(\k,\eta)\ni
-\tilde\gamma(\eta)k^2\delta_L(\k,\eta)\;,
\ee
where $\tilde\gamma(\eta)$
is a time integral of 
$\tilde d$, convolved 
with the fluid equations' Green function~\cite{Ivanov:2022mrd}. 
This correction is small as long as $k$ is well below
the cutoff, ${k^2\tilde \gamma\!\sim\! k^2/k_{\rm cut}^2\!\ll\! 1}$. 
It leads to the counterterm in the power spectrum,
\be
\label{Plctr}
\langle\delta^l(\k,\eta)\delta^l(\k',\eta)\rangle^{\rm ctr}
=-2k^2\tilde \gamma(\eta)g^2(\eta)P(k)\,(2\pi)^3\delta_{\rm D}(\k+\k')\;,
\ee   
which adds up to the standard one-loop contribution from the modes
below the cutoff computed using the ideal fluid approximation. A
detailed analysis shows that the counterterm enters into the final
result in the combination with the most UV sensitive part of the
one-loop contribution controlled by the integral of the power
spectrum from $k$ to $k_{\rm cut}$. This combination reads,
\be
\label{comb1}
\tilde\gamma(\eta)+g^2(\eta)\frac{122\pi}{315}\int_k^{k_{\rm cut}}
\frac{dq\,P(q)}{(2\pi)^3}\;.
\ee

Up to now, we have been working with finite cutoff $k_{\rm
cut}$. However, it is convenient to extend the momentum integrals to infinity, absorbing the change into the
redefinition of the counterterm coefficient,
\be
\label{gammaredef}
\tilde\gamma\mapsto \gamma=\tilde\gamma
-g^2\frac{122\pi}{315}\int_{k_{\rm cut}}^\infty
\frac{dq\,P(q)}{(2\pi)^3}\;.
\ee
This corresponds to formally extending the fluid approximation to
arbitrarily short modes which, however, does not lead to
inconsistencies since spurious contribution coming from the incorrect
treatment of the short modes gets renormalized by the proper
adjustment of the counterterm. Recalling the relation between $\gamma$
and the effective stress tensor, we see that we can perform the
renormalization already at the level of the equations of motion where
we can treat all modes on equal footing with an appropriate choice of
the counterterm stress $\tau_{ij}^{\rm ctr}$. In what follows we
will not distinguish long and short modes and will omit the
corresponding superscripts.  

Let us discuss the time dependence of $\gamma$. On dimensional grounds
one expects $\gamma$ to be inversely proportional to the square of the
physical cutoff of the ideal fluid theory which is set by the
shell-crossing scale,
\be
\label{gammaksc}
\gamma(\eta)\propto \big(\ksc(\eta)\big)^{-2}\;.
\ee 
In the perturbation theory around homogeneous background the
shell-crossing scale coincides with the non-linear scale $k_{\rm NL}$
at which the density variance reaches order-one, 
\be
\label{kNL}
\langle\delta^2\rangle_{\kNL}=g^2(\eta)\cdot 4\pi\int_0^{\kNL(\eta)}
[dk] P(k)\sim 1\;.
\ee
We can move one step further by approximating the power spectrum in
the relevant range of wavenumbers by a power-law, 
\be
\label{plPS}
P(k)\propto
k^n\;. 
\ee
Then Eq.~(\ref{kNL}) implies 
\be
\label{kscscal}
\ksc\simeq \kNL\propto \big(g(\eta)\big)^{-\frac{2}{n+3}}~,~~~~~~~
\gamma(\eta)\propto \big(g(\eta)\big)^{\frac{4}{n+3}}\;.
\ee
The latter scaling has been already mentioned in Eq.~(\ref{gammaz})
where we can now identify $m=4/(n+3)$.

Which part of the above reasoning applies to the
non-perturbative PDF? It is easy to see that the derivation of the
effective stress tensor (\ref{stress}) does not rely on any properties
of the perturbation theory and thus holds for the non-perturbative case as
well. The separation into long and short modes still holds as long as
the shell crossing scale $\ksc$ is much higher than the inverse cell
radius $r_*^{-1}$. The latter sets the scale of long modes in this
problem. We can further send the cutoff to infinity at the
expense of retaining the counterterm stress $\tau_{ij}^{\rm ctr}$. What is missing, however, is the possibility to
restrict only to a finite number of terms in the Taylor expansion of 
$\tau_{ij}^{\rm ctr}$ in the long-mode overdensity. Indeed, the
saddle-point solution $\hat\delta$ is in general not small and all
terms in the Taylor expansion contribute equally. Allowing for
arbitrary coefficients in front of them in the strict application of EFT rules would
lead to a complete loss of predictive power. Thus, we are compelled to
{\em model} the $\hat\delta$ dependence of $\tau_{ij}^{\rm ctr}$
invoking reasonable physical assumptions. This necessarily introduces
some degree of model dependence that must be treated as a theoretical
uncertainty. We now proceed to the description of our model.

\subsection{Shell-crossing in the background of spherical collapse}
\label{sec:ksc}

The key element of the model is the shell-crossing scale $\ksc$ in the vicinity of the Lagrangian radius $R=R_*$ which corresponds to the unperturbed boundary of the cell in the Lagrangian space. The spherical collapse dynamics maps the latter to the trajectory $\hat r(\eta)$ in the Eulerian space, such that $\hat r(-\infty)=R_*$, $\hat r(0)=r_*$. To estimate the dependence of $\ksc$ on $\eta$ and $r$ in the vicinity of this trajectory, we use the linear perturbation theory on top of the saddle-point spherical collapse solution. Of course, such estimate is not precise since $\ksc$ characterizes intrinsically non-linear dynamics of the short modes. Still, we expect it to reproduce the right scalings. The logic here is the same as in the derivation of Eqs.~(\ref{kNL}), (\ref{kscscal}) --- we use the linear theory to identify the critical momenta at which it breaks down.

Consider the trajectory of a matter particle belonging to the
saddle-point spherical collapse solution perturbed by a linear fluctuation. It has fixed Lagrangian coordinates $X_i$ and time-dependent Eulerian coordinates $x_i$. The latter obey the
equations, 
\be
\label{traject}
\frac{dx_i}{d\eta}=-\partial_i{\hat \Psi}(\eta,\x)-\partial_i\Psi^{(1)}(\eta,\x)\;,
\ee
where $\hat\Psi$ is the background part of the velocity potential and
$\Psi^{(1)}$ is the fluctuation. Linearizing this equation, we obtain,
\be
\label{traject1}
\frac{d x_i^{(1)}}{d\eta}=-\d_i\d_j\hat\Psi(\eta,\hat\x)\,x_j^{(1)}
-\d_i\Psi^{(1)}(\eta,\hat\x)\;, 
\ee
where both velocity potentials are now evaluated on the unperturbed
trajectory $\hat x_i(\eta)$ satisfying
\be
\label{traject2}
\frac{d\hat x_i}{d\eta}=-\d_i\hat\Psi(\eta,\hat\x)\;.
\ee 
In this way the perturbation $\x^{(1)}$ can be considered as function of time $\eta$ and unperturbed coordinates $\hat\x$. 

\begin{figure}[t]
	\centering
    \includegraphics[width=0.8\linewidth]{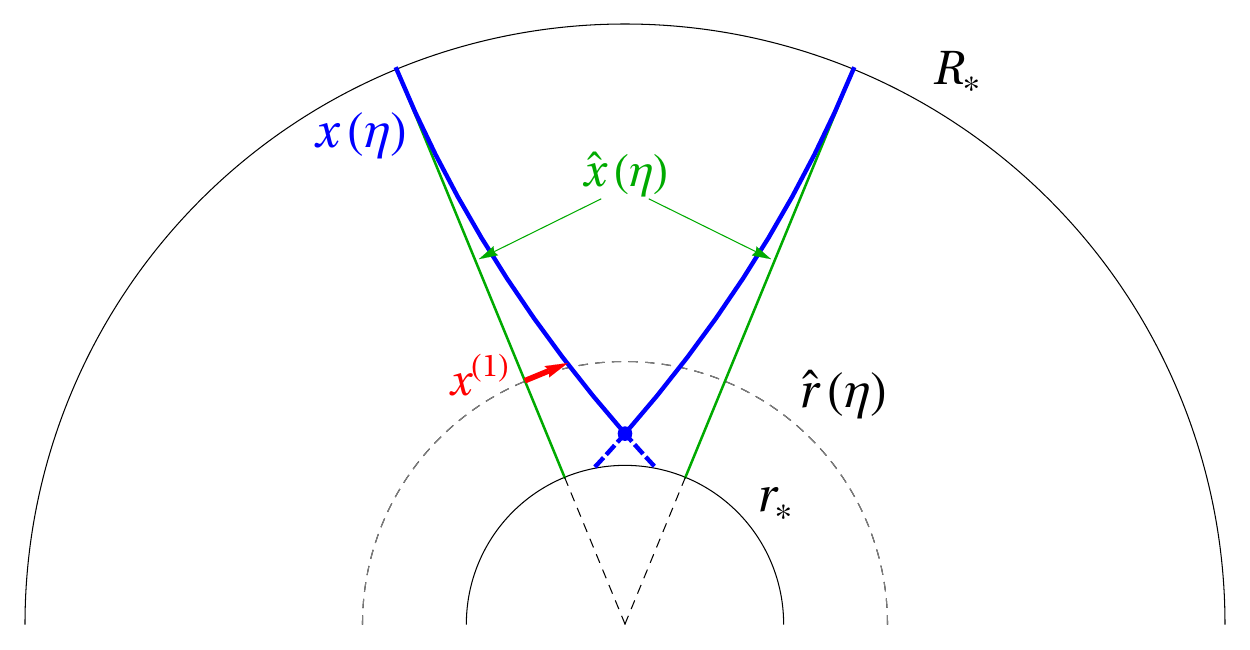}
	\caption{Trajectories of two matter particles (blue curves) 
 moving along the flow lines of the background saddle-point
    solution (straight green lines) perturbed by linear fluctuations~$\x^{(1)}$ (red arrow). 
    The particles evolve from the initial Lagrangian radius $R_*$
    to the final radius of the cell~$r_*$.
    The shell-crossing occurs when the variance of the perturbation
    becomes comparable to the unperturbed distance between the two particles (blue dot). 
    }
	\label{fig:shellcross}
\end{figure}

Let us for concreteness consider the case of overdensity $\delta_*>0$, so that $R_*>r_*$. The amplitude of the perturbation $\x^{(1)}$ grows as the particle moves towards smaller radii, see a schematic picture of the particle trajectory in Fig.~\ref{fig:shellcross}. Two particles starting from different points at $r=R_*$ may come closer to each other and their trajectories may cross. It is worth stressing that the shell-crossing responsible for the counterterm prefactor happens in the vicinity of the cell boundary, rather than in its center, see Fig.~\ref{fig:shellcross}. 
Shell-crossing corresponds to a singularity of
the Euler-space density contrast,
\begin{align}
1+\delta(\eta,\x)&=\left(\det\frac{\d x_i}{\d X_j}\right)^{-1}
=\left[\det\left(\frac{\d \hat x_i}{\d X_j}+\frac{\d x_i^{(1)}}{\d X_j}\right)\right]^{-1}\notag\\
&=\left(\det\frac{\d \hat x_k}{\d X_l}\right)^{-1} 
\left[\det\left(\delta_i^j+\frac{\d x_i^{(1)}}{\d \hat x_j}\right)\right]^{-1}\;.
\label{determ}
\end{align}
The first factor here is finite since the spherical collapse background solution is regular. Thus, the singularity must arise from the second factor controlled by the perturbations. For the determinant to vanish, the elements of the matrix 
\be
\label{dispmatr}
\foc^j_{i}\equiv-\frac{\d x_i^{(1)}}{\d \hat x_j}
\ee
must be of order one.

To estimate them, consider
the trace $\foc\equiv\foc_i^i$
which characterizes the focusing of near-by particle trajectories. Combining Eqs.~(\ref{traject1}) and (\ref{traject2}) yields for its evolution along the unperturbed trajectory (see derivation in appendix~\ref{app:ksc}):
\be
\label{displ}
\frac{d\foc}{d\eta}=\Delta\Psi^{(1)}=\Theta^{(1)}\;.
\ee
This result is easy to understand intuitively: the increased focusing of particles is given by the convergence of the velocity perturbation. Thus, to find $\foc$ at a given moment of time $\eta$ and position $\hat\x$, we trace back the trajectory of particle from this point along the unperturbed spherical collapse solution to its original Lagrangian position ${\bf X}$. Let us denote this trajectory as $\hat\x(\eta',{\bf X})$. In this way $\foc$ becomes a function of time and the Lagrangian coordinate:
\be
\foc(\eta,{\bf X})=\int_{-\infty}^\eta d\eta'\,\Theta^{(1)}\big(\eta',\hat \x(\eta',{\bf X})\big).
\ee

We can now calculate the variance of $\foc$. It is the same for all points with the same Eulerian radius $\hat r$ corresponding to Lagrangian radius $R$.  
Expanding
$\Theta^{(1)}$ in spherical harmonics and using the basis
$\Theta_\ell^{(1)}(\eta,r;k_I)$ introduced in Sec.~\ref{sec:asp}, we
obtain the contribution of modes with $k<k_{\rm max}$, 
\begin{align}
\label{shell_cross}
    \left\langle\foc^2\right\rangle_{k_{\rm max}}
&=\sum_{\ell}\frac{2\ell+1}{4\pi}
\int^{k_{\rm max}} [dk]P(k)
\left|\int_{-\infty}^{\eta} d\eta'\,
\Theta_\ell^{(1)}\big(\eta',\hat r(\eta',R);k\big)\right|^2\notag\\
	&\approx 4\pi\int^{k_{\rm max}} 
[dk]P(k)\cdot \int_{1/R}^\infty
\frac{d\varkappa}{(2\pi)^2\varkappa}
\left|\int_{-\infty}^{\eta} d\eta'\,\upvartheta_{\ell}(\eta',R;\vk)\right|^2\;.
\end{align}
In the last equality we used the WKB approximation
for short modes (see appendix~\ref{app:WKB1}) and replaced the sum over
$\ell$ with the integral over $\vk$. 
Note that the expression has factorized due to the universal behavior of the WKB
functions. The first factor --- the
integral of $k$ --- describes the statistical properties of the
initial conditions. Whereas the second factor
\be
\label{D}
D(\eta;R)=\l\int_{1/R}^\infty\frac{d\varkappa}{(2\pi)^2
\varkappa}\left|\int_{-\infty}^{\eta} 
d\eta'\,\upvartheta_{\ell}(\eta',R;\vk)\right|^2\r^{1/2}\;
\ee 
describes the growth of the short-mode velocity
divergence in the background of the saddle-point solution. Note that it depends on time and the Lagrangian radius of the matter shell comoving with the background flow. Through the latter it implicitly depends on the 
eventual
density contrast $\delta_*$ which determines the background
solution. 

\begin{figure}
	\centering
	\includegraphics[width=0.48\linewidth]{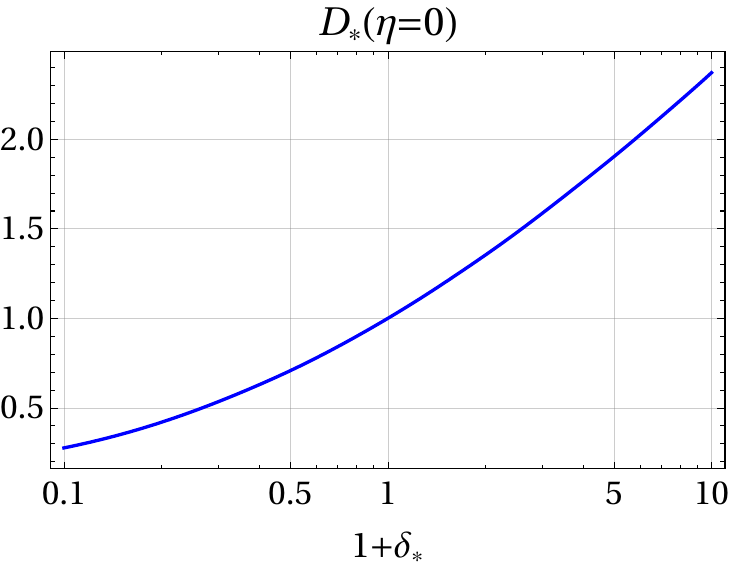}~~~
	\includegraphics[width=0.50\linewidth]{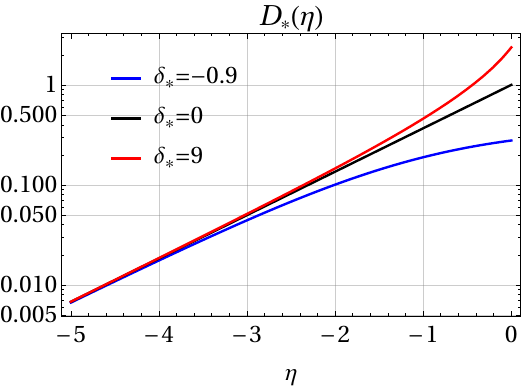}
	\caption{Cumulative growth factor of short modes 
in the background of the
          saddle-point solution evaluated at the cell boundary.
 {\it Left:} Dependence on the density contrast for $\eta=0$. {\it Right:}
         Time dependence for three different density contrasts.}
	\label{fig:growth}
\end{figure}

Using the formulas from appendix~\ref{app:WKB1} one can show that
for $\delta_*=0$ the growth factor (\ref{D}) becomes independent of $R$ and 
reduces to $\e^\eta\equiv g(\eta)$ --- the usual
linear growth rate. However, for $\delta_*\neq 0$ it sizeably
deviates from $g(\eta)$.
As we will see in the next subsection, we will only need its
value and the value of its first radial derivative 
at
the cell boundary $\hat r(\eta)$ (which corresponds to the fixed Lagrangian radius $R=R_*$). 
In more detail, we introduce 
\be
\label{Dstar}
D_{*}(\eta)\equiv D(\eta,R_*)~,~~~~~~~[\ln D_*(\eta)]'\equiv 
\left(\frac{\d\hat r(\eta,R)}{\d R}\right)^{-1}
\d_R\ln D(\eta,R)\bigg|_{R=R_*}\;.
\ee
In the left panel of Fig.~\ref{fig:growth}
we show the dependence of $D_{*}$ on $\delta_*$ for $\eta=0$. The
right panel of Fig.~\ref{fig:growth} shows the time dependence of $D_{*}$
for three representative density contrasts. 

We are now ready to estimate the shell crossing scale by equating 
\be
\label{crit}
\left\langle\foc^2\right\rangle_{\ksc}\sim 1\;.
\ee
Using the power-law power spectrum approximation, we get
(cf.~(\ref{kscscal})), 
\be
\label{ksc}
k_{\rm sc}(\eta,R)\propto \big(D(\eta,R)\big)^{-m/2}\;.
\ee
We will use this scaling in our counterterm model below.

Of course, the definition of $\ksc$ in non-linear background is not
unique. Instead of focusing on the trace of the matrix
(\ref{dispmatr}), one can consider the variance of its individual
elements, or some combination thereof. A few options are discussed in
appendix~\ref{app:ksc} and shown to give qualitatively similar
results. We use them to estimate the theoretical uncertainty of the
model.

\subsection{Counterterm stress for spherical collapse} 
\label{sec:couterterm}

The symmetry of the background implies that the only non-vanishing
components of the counterterm stress tensor are $\tau_{rr}^{\rm
  ctr}\equiv \tau_{\paral}^{\rm ctr}$ and 
$\tau_{\mu\nu}^{\rm ctr}=r^2g_{\mu\nu}\tau_\perp^{\rm ctr}$
where $\mu,\nu=\theta,\phi$ are the angular directions, and
$g_{\mu\nu}$ is the metric on the unit two-sphere. Both
$\tau_{\paral}^{\rm ctr}$ and $\tau_\perp^{\rm ctr}$ are independent
of the angles. The force appearing in Eq.~(\ref{EPnew2}) and entering
the Euler equation is purely radial. It perturbs the trajectory of the
cell boundary and thereby modifies the mass inside it. 
Repeating the derivation in appendices~\ref{app:Apr2} and~\ref{app:Apr4} 
we obtain
that the perturbations of the cell raius and mass satisfy the system of equations,
\bseq
\label{murctr}
\begin{align}
\label{muctr}
&\dot\mu^{\rm ctr}+A(\eta)\,\dot r^{\rm ctr}+\frac{dA}{d\eta}\, r^{\rm
  ctr}=0\;,\\
\label{rctr}
&\ddot r^{\rm ctr}+\frac{1}{2}\dot r^{\rm ctr}
+B(\eta)\, r^{\rm ctr}+\frac{3}{2\hat r^2}\mu^{\rm
  ctr}=-\Upsilon_\Theta^{\rm ctr}\;,
\end{align}
\eseq
where the functions $A(\eta)$, $B(\eta)$ are defined in \eqref{A}, \eqref{B}, and
the {\it counterterm source} reads
\be
\label{Upsctr}
\Upsilon_\Theta^{\rm ctr}(\eta)=\frac{1}{\H^2(1+\hat\delta)}\d_j
\tau_{ij}^{\rm ctr}\Big|_{i\to r}
=\frac{1}{\H^2(1+\hat\delta)}\bigg(\d_r\tau_{\paral}^{\rm ctr}
+\frac{2}{r}\tau_{\paral}^{\rm ctr}
-\frac{2}{r}\tau_\perp^{\rm ctr}\bigg)\bigg|_{\eta,\hat r(\eta)}\;.
\ee
Here $\hat\delta(\eta,r)$ is the density contrast of the unperturbed spherical collapse solution, and all quantities are evaluated at the unperturbed moving cell boundary $\hat r(\eta)$. 
We observe that these equations and the source term have the same structure
as 
the equations (\ref{mu2r2}) and the source (\ref{Ups2}) 
governing the second-order
perturbations of the density contrast in the cell.
As
in that case, Eqs.~(\ref{murctr}) are to be solved with the boundary
conditions: 
\be
\label{murctrbc}
r^{\rm ctr}\big|_{\eta=0}=0~;~~~~~
\e^{-\eta}\mu^{\rm ctr},~\e^{-\eta}\left(\dot r^{\rm ctr}+\frac{d\ln A}{d\eta}\,r^{\rm ctr}\right)
\to 0
\text{ at } \eta\to-\infty\;,
\ee
which uniquely fix the solution. The first condition enforces that the radius of the cell at the final time remains unperturbed and still equals $r_*$, whereas the second condition implies that the perturbation is truly second-order and does not contain any admixture of the first-order perturbations which behave as $\propto \e^\eta$ at early epoch.  Once Eqs.~(\ref{murctr}) are solved, 
the counterterm prefactor (\ref{Actrdef}) at $\eta=0$ is 
determined through 
\be
\label{deltamuctr}
\bar\delta_W^{\rm ctr}=\frac{3}{r_*^3}\mu^{\rm ctr}(\eta=0)\;.
\ee

The counterterm source must renormalize the UV-sensitive contributions
coming from the perturbative calculation.
In Sec.~\ref{sec:WKB} we have discussed the contributions of the
short modes with $k\gg R_*^{-1}$ to the prefactor. 
However, the same result will be obtained if we
first integrate these modes out and construct their effective stress
tensor.
Indeed, consider a slab of
 modes with momenta $k$, such that 
\be
\label{slab}
1/R_*\ll k_1<k<k_2\ll k_{{\rm sc}*}\;,
\ee
so that these modes, while being short, are still within the validity of the
fluid description. The velocity dispersion of such modes with respect
to the background flow is 
\be
\label{disp}
\sigma_{ij}^l= \langle u_i^{(1)} u_j^{(1)}\rangle_s
=\H^2\langle \d_i\Psi^{(1)} \d_j\Psi^{(1)}\rangle_s\;,
\ee
where averaging is over the ensemble of random mode
realizations. 
We substitute this into the kinetic part of the stress
tensor (\ref{stress}) and expand $\Psi^{(1)}$ into spherical
harmonics.
The radial (parallel) and transverse (perpendicular) parts
of the stress tensor
have precisely the form of the first terms in \eqref{Ssigmas}
obtained in the perturbative analysis.
Performing the same manipulations with the potential part of 
the effective stress (\ref{stress}), 
we reproduce the remaining terms in \eqref{Ssigmas}.
The agreement of the short-wavelength contribution obtained 
through the effective stress tensor with the perturbative calculation  
constitutes a consistency check of our approach.
Integrating over short modes and using the WKB approximation, we obtain the kinetic and potential
contributions to the stress tensor as a function of the Lagrangian radius, 
\begin{align}
\label{UpsSigma}
\tau_\a^a=2\H^2\int_{k_1}^{k_2} 
\frac{dk\,P(k)}{(2\pi)^3}\cdot
\int_{R^{-1}}^{\infty}\frac{d\vk}{\vk}\,\chi^a_\alpha(\eta,R;\varkappa)\;,
\qquad a=\kin,\pot;~~\alpha=\paral,\perp\;,
\end{align}
The expressions for the WKB fields $\chi^a_{\alpha}$ are given in
appendix~\ref{app:WKB1}.

The extension of
$k$-integration in \eqref{UpsSigma} up to infinity would be illegitimate because of the breakdown of the fluid approximation, so these effective stress tensors must be
renormalized. 
Given the distinct physical origin of the kinetic and potential
stresses, it is natural to renormalize them separately. We now
construct the counterterm stresses $\tau_\alpha^{\kin,\ctr}$ and $\tau^{\pot,\ctr}_\alpha$ as
follows. We assume that the expressions \eqref{UpsSigma}
capture faithfully the effect of short modes up to $k\lesssim
k_{{\rm sc}}$. The modes with $k\gtrsim \ksc$, however, are not described
correctly. The counterterms must fix this discrepancy, so they should
scale as
\be
\tau_\alpha^{a,\ctr}\sim 2\H^2\int_{\ksc}^\infty 
\frac{dk\,P(k)}{(2\pi)^3}\cdot
\int\frac{d\vk}{\vk}\,\chi^a_\alpha\;.
\ee
For a power-law power spectrum (\ref{plPS}) with $-3<n<-1$ we have a
chain of scaling relations
\be
\label{intscals}
\int_{\ksc}^\infty dk\,P(k)\propto \ksc^{n+1}\propto
\frac{1}{\ksc^2}\cdot\ksc^{n+3}\propto 
\frac{1}{\ksc^2}\int^{\ksc}[dk] P(k)
\propto\frac{1}{\ksc^2 D^2}\propto D^{m-2}\;,
\ee
where in the last two equalities we used Eqs.~(\ref{shell_cross}),
(\ref{crit}) and (\ref{ksc}). This motivates us to make the following
Ansatz,
\be 
\label{eftpartau}
\tau^{a,\ctr}_\alpha(\eta,R)=\zeta^a\cdot 2\H^2 \big(D(\eta,R)\big)^{m-2}
\!\!\int_{R^{-1}}^\infty\frac{d\vk}{\vk}\;\chi^a_\alpha(\eta,R;\vk)\;,
\qquad a=\kin,\pot;~~\alpha=\paral,\perp\;,
\ee
where $\zeta^a$ are constant parameters of dimension
$(\text{length})^2$, independent of time, space or density contrast. We will refer to them as the kinetic and potential counterterm amplitudes.
In the spirit of EFT,
they must be treated as nuisance parameters in fitting the PDF data
with the model.

For the PDF calculation we need to go from the stress tensor to the
$\Upsilon$-sources. Substituting (\ref{eftpartau}) into
(\ref{Upsctr}) we obtain,\footnote{From now on we omit the ``ctr'' index on the counterterm sources, to avoid cluttered notations.} 
\be 
\label{eftpar}
\Upsilon^{a}(\eta)\!=2\zeta^a [D_*(\eta)\big]^{m-2}\bigg[
\dashint_{R_*^{-1}}^\infty\!\frac{d\vk}{\vk}\;\upsilon^a(\eta;\vk)
+\frac{(m\!-\!2)[\ln
  D_*(\eta)]'}{1+\hat\delta|_{\eta,\hat r(\eta)}}\!
\int_{R_*^{-1}}^\infty\!\frac{d\vk}{\vk}\,\chi^a_{\paral}(\eta,R_*;\vk)\bigg].
\ee
where we introduced the rescaled WKB sources,\footnote{Note that radial derivative in the expression for $\upsilon^a$ is taken with respect to the Eulerian radius at fixed time. Since $\chi_{\paral}^a$ depends on the Lagrangian radius, we have $\d_r\chi_{\paral}^a=(\d \hat r(\eta,R)/\d R)^{-1}\d_R\chi_{\paral}^a$. }
\be
\label{upschi}
\upsilon^a=\frac{1}{1+\hat\delta}\bigg[\d_r\chi^a_{\paral}
+\frac{2}{r}
\chi^a_{\paral}-\frac{2}{r}\chi^a_\perp\bigg]\bigg|_{\eta,\hat
  r(\eta)}\;. 
\ee
Eq. \eqref{eftpar} is our final expression for the counterterm sources used in the
numerical procedures. It contains three free parameters: the counterterm scaling index $m$ and the kinetic and potential counterterm amplitudes $\zeta^\kin$, $\zeta^\pot$. The rest of it is determined by the dynamics of the spherical collapse and small perturbations on top of it.

\subsection{Synthesis}
\label{sec:synth}

Let us summarize the procedure for calculating the counterterm
prefactor in the PDF:
\begin{enumerate}
\item Fix the values of $m$, $\zeta^\kin$,
  $\zeta^\pot$.
\item Find the saddle-point
  solution for the density contrast $\delta_*$ .
\item
Solve the WKB equations (\ref{WKBNLO}), (\ref{PsiPhiLO})
along the flow line corresponding to the cell boundary $\hat
r(\eta)$. 
\item Compute the cumulative growth factor $D_*$, Eq.~(\ref{D}). 
\item Evaluate $\upsilon^\kin$, $\upsilon^\pot$ using
  Eqs.~(\ref{UpsF}) and compute their
  $\vk$-integrals. 
\item Construct the sources (\ref{eftpar}) and solve
  Eqs.~(\ref{murctr}) separately for 
$\Upsilon^{\kin}$ and
  $\Upsilon^{\pot}$. This gives the counterterm densities
  $\bar\delta^{\kin}_{W}$ and $\bar\delta^{\pot}_{W}$.
\item Substitute them into Eq.~(\ref{Actrdef}) to obtain
\be
\label{Actra}
{\cal A}^{a}=\exp[-\hat\lambda \bar\delta^{a}_{W}]~,\qquad
a=\kin,\pot\;.
\ee
The full counterterm prefactor is the product
$\Actr={\cal A}^{\kin}\cdot{\cal A}^{\pot}$. 
\item Repeat the steps 2 to 7 for different $\delta_*$.
\end{enumerate}

\begin{figure}
	\centering
	\includegraphics[width=0.48\linewidth]{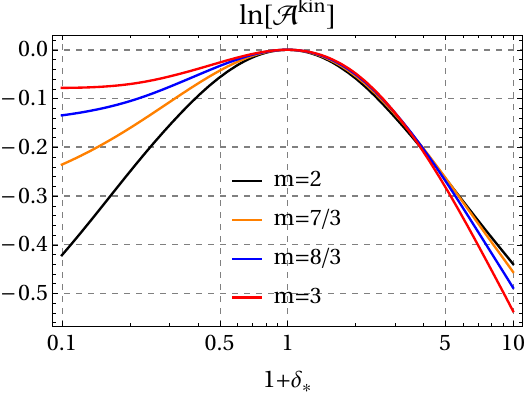}~~
	\includegraphics[width=0.48\linewidth]{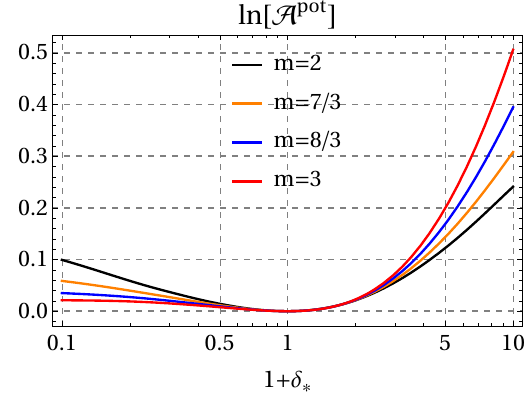}
	\caption{Contributions to the counterterm prefactor at zero
          redshift for $r_*=10\,\Mpch$. The parameters $\zeta^\kin$,
          $\zeta^\pot$ are fixed to $(1\,\Mpch)^2$. 
	Both contributions are evaluated for the cosmological
        parameters of the Farpoint simulation~\cite{HACC:2021sgt}. 
}
	\label{fig:Actrms}
\end{figure}

We have focused so far on the counterterm prefactor at the present
epoch. One might think that for computing the PDF at non-zero
redshift, $z_1>0$, one has to repeat the whole procedure from the
start. Fortunately, this is not the case thanks to the invariance of
EdS with respect to time-translations in $\eta$. Consider the PDF at
$\eta_1<0$. The saddle-point solution corresponding to $\delta_*$ in
this case is obtained simply by translating the solution corresponding
to $\delta_*$ at the present epoch back in time by~$\eta_1$,
\be
\label{saddle1}
\hat\delta_1(\eta,r)=\hat\delta_0(\eta-\eta_1,r)\;.
\ee
The linear perturbations translated backward by the same amount are
also solutions of the relevant equations. However, they do not obey
the asymptotic initial conditions (\ref{linit1}), differing in
normalization. To make them properly normalized, we have to multiply
them by a constant factor $\e^{\eta_1}$. This implies that the new
growth factor and fields $\chi^a_\alpha$ are related to the old ones by 
\bseq
\begin{align}
\label{Dshift}
&D_{1}(\eta,R)=\e^{\eta_1}D_{0}(\eta-\eta_1,R)=g(\eta_1)
D_{0}(\eta-\eta_1,R)\;,\\
\label{upsshift}
&\chi^a_{1\alpha}(\eta,R;\vk)=\e^{2\eta_1}\chi^a_{0\alpha
  }(\eta-\eta_1,R;\vk)
=g^2(\eta_1)
\chi^a_{0\alpha}(\eta-\eta_1,R;\vk)\;.
\end{align}
\eseq
Hence, the new sources are
\be
\label{Upsshift}
\Upsilon^{a}_{1}(\eta)=\big[g(\eta_1)\big]^m\,
\Upsilon^{a}_{0}(\eta-\eta_1)\;.
\ee 
By linearity of Eqs.~(\ref{murctr}), we have for the resulting
counterterm prefactors,
\be
\label{Actrz1}
\ln{\cal A}^{a}_{1}=\big[g(\eta_1)\big]^{m-2}\,\ln{\cal A}^{a}_{0}\;.
\ee
Note that two powers of $g(\eta_1)$ have canceled since the
definition (\ref{deltactr}) of the counterterm overdensity contains an
explicit factor $g^2$. Thus, once we know the counterterm prefactor at
the present epoch, we can easily find it at any other redshift by
multiplying $\ln\Actr$ with $\big[g(z)\big]^{m-2}$. This reproduces
the scaling (\ref{Actrz}) already encountered in Sec.~\ref{sec:rel}.

\begin{figure}[tb]
	\centering
	\includegraphics[width=0.48\linewidth]{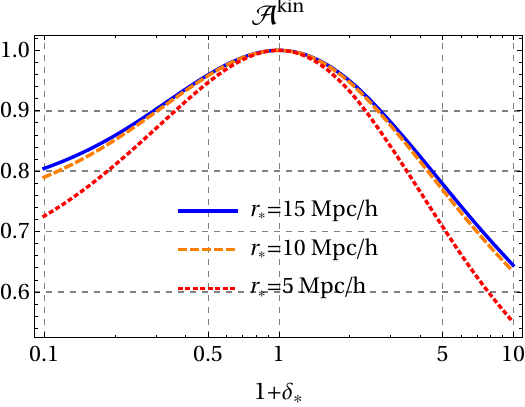}~~~
	\includegraphics[width=0.48\linewidth]{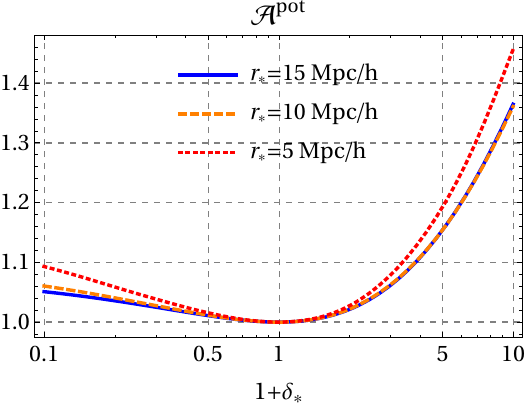}
	\caption{Contributions to the counterterm prefactor at redshift
          zero for several cell radii. Cosmological parameters are
          the same as in the Farpoint simulation
          \cite{HACC:2021sgt}. The model parameters are 
          $\zeta^\kin=\zeta^\pot=(1\,\Mpch)^2$, $m=2.33$.}
	\label{fig:Actr}
\end{figure}

In Fig.~\ref{fig:Actrms} we illustrate the typical behavior of
${\cal A}^{\kin}$, ${\cal A}^{\pot}$ by plotting their logarithms at the
present epoch for $\zeta^\kin=\zeta^\pot=(1\,\Mpch)^2$ and several
values of the scaling index $m$. The cell radius is $r_*=10\,\Mpch$ and the cosmological
parameters are the same as in the Farpoint N-body simulation
\cite{HACC:2021sgt}  
(see Sec.~\ref{sec:result}). One makes several observations. First,
$\ln{\cal A}^{\kin}$ and $\ln{\cal A}^{\pot}$ have opposite signs. Second,
they both have zero first derivatives at $\delta_*=0$, in agreement
with the model-independent considerations in
Sec.~\ref{sec:rel}. Third, the counterterms are more important for
overdensities $\delta_*>0$. This appears natural since the overdense
regions are expected to be more non-linear that the underdense ones,
so the effects of the UV renormalization are larger there. We also
plot the counterterms for several cell radii in  
Fig. \ref{fig:Actr}. As expected, the counterterms become stronger for
smaller cells, though the dependence on the radius is quite weak.

The eventual values of the three model parameters $m$, $\zeta^\kin$, $\zeta^\pot$ are to be determined from the best fit of the data. In principle, they must be the same for all cell radii $r_*$.  
In practice, however, it is more convenient to determine them
independently for different cells, and check a posteriori the
consistency of the measured 
values.
On the other hand, they are taken to be common for all redshift bins. 
Note that in the
actual fits to the PDF, one can impose priors following from
the connection between $\Actr$ and the counterterms of the $n$-point
correlators discussed in Sec.~\ref{sec:rel}. This strategy is
illustrated in the next section.

\section{Comparison with N-body data}
\label{sec:result}

In this section we compare our theoretical PDF model with the N-body
data. For this purpose we use the results of cosmological Farpoint
simulation \cite{HACC:2021sgt}. The latter achieves a very high mass
resolution, 
 $M_{\rm particle}=4.6\cdot 10^7h^{-1}M_\odot$, by evolving $12288^3$
 particles in a $1\,({\rm Gpc}/h)^3$ volume. The force softening was
 chosen $\sim0.8\,{\rm kpc}/h$ and the initial conditions were set at 
$z_i=200$ using the Zeldovich approximation. The simulation was run
for $\Lambda$CDM cosmology with the Planck-2018 cosmological
parameters: 
\be
\label{Planckparam}
\Omega_m=0.310~,~~~~
\Omega_b=0.049~,~~~~ h=0.677~,~~~~ n_s=0.967~,~~~~\sigma_8=0.810\;. 
\ee
To accelerate the numerical computations, we use randomly downsampled simulation snapshots containing 
$1\%$ of the total number of particles. 

We first analyze the power spectrum in order to determine the
``effective sound speed'' $\gamma$ and its dependence on redshift. We
generate the linear power spectrum $P(k)$ with the cosmological
parameters of the simulation using the Boltzmann code 
 \texttt{CLASS} \cite{Blas:2011rf} and compute one-loop correction to
 it with \texttt{CLASS-PT} \cite{Chudaykin:2020aoj}. We then extract
 the nonlinear power spectrum $P^{\rm sim}(k)$ from the simulation
 snapshots at redshifts $z=\{0,\,0.2,\,0.5,\,0.8,\,1.0,\,1.5,\,2.4\}$ 
using the Yamamoto estimator~\cite{Yamamoto:2005dz}
 implemented in \texttt{nbodykit} \cite{Hand:2017pqn},
 and
 fit it with the theoretical template.
We restrict the fitting range of momenta by $k_{\rm
 max}=0.3\,h/\Mpc$ and include the theoretical error following
Ref.~\cite{Baldauf:2016sjb} with the correlation length $\Delta k_{\rm
corr}=0.1\,h/\Mpc$. Inclusion of the theoretical error makes our
results insensitive to the precise choice of $k_{\rm max}$. 

\begin{figure}
	\centering
	\includegraphics[width=0.5\linewidth]{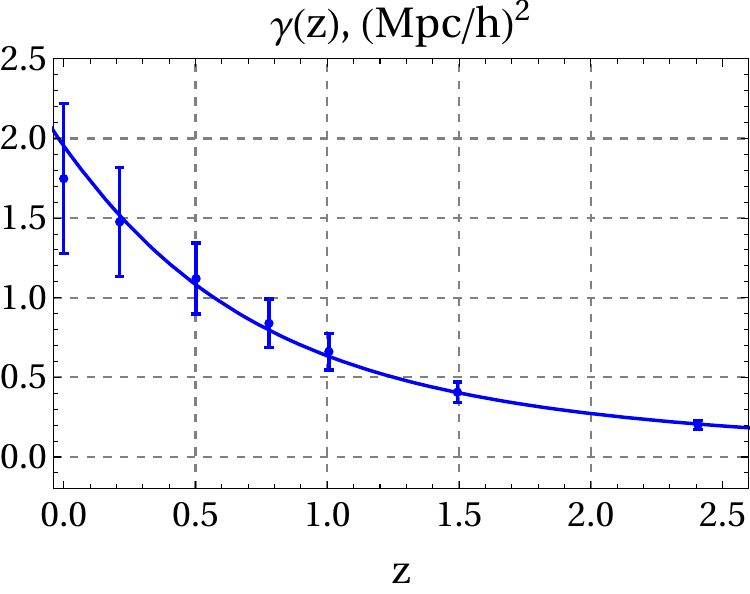}
	\caption{Effective sound speed measured from simulations at
          different redshifts. The power law fit (\ref{gammazfit})
          with parameters (\ref{mdpl}) is also shown.}
	\label{fig:gammaz}
\end{figure}

Figure~\ref{fig:gammaz} shows the measured values of $\gamma$ at
different redshifts together with the best power-law fit of the form
(\ref{gammazfit}). The parameters of the fit and their correlation are
\be
\label{mdpl}
\gamma_0=(1.95\pm 0.26)(\Mpch)^2\;,\qquad m=2.26\pm 0.21\;,\qquad
{\rm corr}(\gamma_0,m)=0.85\;.
\ee
We observe that these values somewhat differ from the values
(\ref{gammazval}) previously reported in the literature
\cite{Carrasco:2013mua,Angulo:2014tfa,Foreman:2015uva,Baldauf:2015aha,
Ivanov:2018lcg,Lazeyras:2019dcx,Steele:2020tak}. In 
this regard, it is worth pointing out that the previous works used
WMAP-based cosmology with higher $h$ and lower $\Omega_m$. As we discuss in
appendix~\ref{app:gamma}, this difference in the cosmological
parameters is likely responsible for the shift in $\gamma_0$ and $m$.

We now turn to the PDF. We cover the simulation volume with spherical
cells of radius $r_*$ placing their centers at the vertices of
regular cubic grid with side $2r_*$. In other words, we allow the
neighboring cells to touch each other, but not overlap. This is done to suppress correlations between the cells. 
The counts of particles in the cells are binned logarithmically in
$1+\delta_*$ with the bin width $\Delta\ln(1+\delta_*)=0.127$. 
The PDF ${\cal P}^{\rm sim}(\delta_{*,n})$ in the $n$th bin
  is then estimated as the ratio of the number of cells ${\cal N}_n$
  falling into this bin over the total number of cells. We assume the
  errors to follow the Poisson statistics, 
$\Delta \P^{\rm sim}(\delta_{*,n})/ \P^{\rm sim}(\delta_{*,n})\simeq
{\cal N}_n^{-1/2}$, and be uncorrelated between the bins. These
simplifying assumptions appear reasonable for cells of sizes considered in this
work. A more careful treatment would include covariance which can be either extracted from large suits of N-body simulations, such as Quijote \cite{Villaescusa-Navarro:2019bje}, or estimated theoretically \cite{Repp:2020etr,Bernardeau:2022sva,Uhlemann:2022znd}. We leave inclusion of covariance for future.

The Zeldovich initial conditions are known to affect the PDF measured in
N-body simulations through long-lived transients
\cite{Scoccimarro:1997gr,Crocce:2006ve,Uhlemann:2019gni}. In appendix~\ref{app:ZA} we
estimate the systematic error in ${\cal P}^{\rm sim}$ induced by this
effect and impose a quality cut: we use only the PDF data for which
the systematic error does not exceed the statistical uncertainty. We
have found this cut to be relevant for PDF of underdensities,
$\delta_*<0$, for the smallest cell radius analyzed in this work
$r_*=5\,\Mpch$. In all other cases the systematic error happens to be
negligible. 

The measured PDF is used to estimate the aspherical prefactor. We
divide ${\cal P}^{\rm sim}$ by the spherical PDF, Eq.~(\ref{Psp2}),
which we calculate using the linear power spectrum. To account for the
particle shot noise and the finite width of the bin, we multiply
$\P_{\rm sp}(\delta_*)$ with the Poisson probability ${\cal P}({\cal
  N}|\delta_*)$ of having ${\cal N}$ particles inside the cell with a
given density contrast and average over the bin. 
In this way we
obtain $\langle {\cal P}_{\rm sp}\rangle_n$ --- the spherical PDF in
the $n$th bin. The measured aspherical prefactor is then defined as
\be
\label{AprData}
\Apr^{\rm sim}(\delta_{*,n})=
\frac{\P^{\rm sim}(\delta_{*,n})}{\langle\P_{\rm sp}\rangle_{n}}\;.
\ee

The data on the aspherical prefactor are fitted using the theoretical
model developed in the previous sections. The ``fluid'' part 
$\Apr^{\rm fluid}$ is evaluated using the method described
in Sec.~\ref{sec:theory}. We use the open-source code \texttt{AsPy}
\cite{AsPy} to compute the contributions of multipoles with
$1\leq\ell\leq 9$, whereas the cumulative contribution of high
multipoles with $\ell\geq 10$ is computed with the WKB approximation.
This is then multiplied by the counterterm prefactor
$\Actr$ evaluated following the algorithm of Sec.~\ref{sec:ctr}, and
the parameters $\zeta^\kin$, $\zeta^\pot$, $m$ are fitted to
reproduce $\Apr^{\rm sim}$. In the fitting procedure we impose two
Gaussian correlated priors following from the fit of the power spectrum
(\ref{mdpl}). The prior on the scaling index $m$ is applied directly, whereas the second
prior constrains the linear combination
of the counterterm amplitudes $\zeta^\kin$ and $\zeta^\pot$ 
which appears in the second derivative of $\Actr(\delta_*)$ at
$\delta_*=0$ and is proportional to $\gamma_0$ (see
Eqs.~(\ref{Actrd3}), (\ref{cW2ctr})). The product of the aspherical
prefactor and the spherical PDF gives the model PDF $\P^{\rm mod}$. 

\begin{table}[h!]
\begin{center}
\begin{tabular}{|l|c|c|c|c|}
\hline
 & $r_*=15$ Mpc$/h$ & $r_*=10$ Mpc$/h$ & $r_*=7.5$ Mpc$/h$& $r_*=5$ Mpc$/h$\\
\hline
$z=0$ & 0.255 & 0.478 & 0.717 & 1.205\\ \hline
$z=0.5$ & 0.151 & 0.283 & 0.425 & 0.714\\ \hline
$z=1$ & 0.094 & 0.176 & 0.264 & 0.444\\ \hline
$z=2.4$ & 0.035 & 0.066 & 0.098 & 0.165\\ \hline
\end{tabular}
\caption{
Filtered density variance $g^2\sigma^2_{r_*}$ 
for the cell radii and redshifts used in the analysis.
}
\label{tab:sigmas}
\end{center}
\end{table} 

This analysis is performed for four cell radii
$r_*=\{15,\,10,\,7.5,\,5\} \Mpch$ and
four redshift slices 
$z=\{0,\,0.5,\,1.0,\,2.4\}$. 
The linear density variance $g^2\sigma_{r_*}^2$ for various cell radii and redshifts is shown in Tab.~\ref{tab:sigmas}.
We consider cells of different radii separately. This is done for two reasons. First, we want to avoid cross-correlation between the corresponding PDFs. This aspect can, in principle, be addressed in future studies by taking into account the PDF covariance which is, however, outside the scope of this work. Second, our theoretical model uses the saddle-point expansion which is expected to hold in the mildly non-linear regime. Thus, we do not expect the model to work equally well for larger and smaller radii. Our goal is to assess these differences. 
We now describe the results.

\subsection{Large cells}
\label{sec:result1}

\begin{figure}
	\centering
	\includegraphics[width=0.48\linewidth]{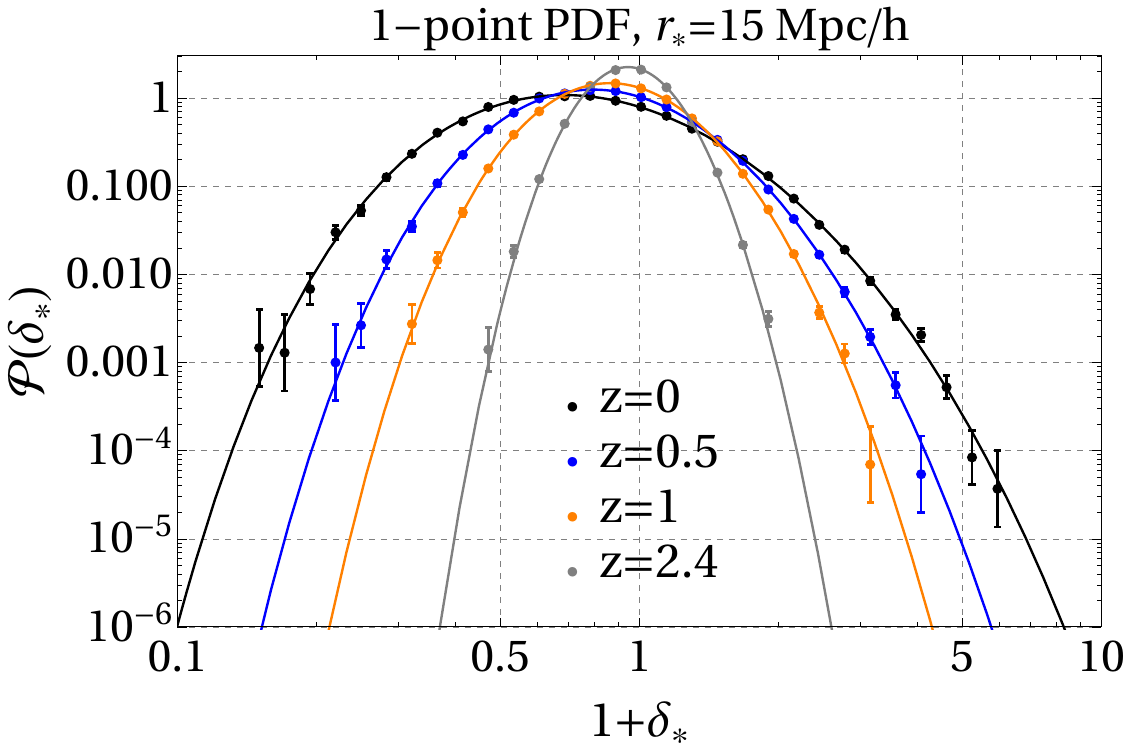}~~
	\includegraphics[width=0.48\linewidth]{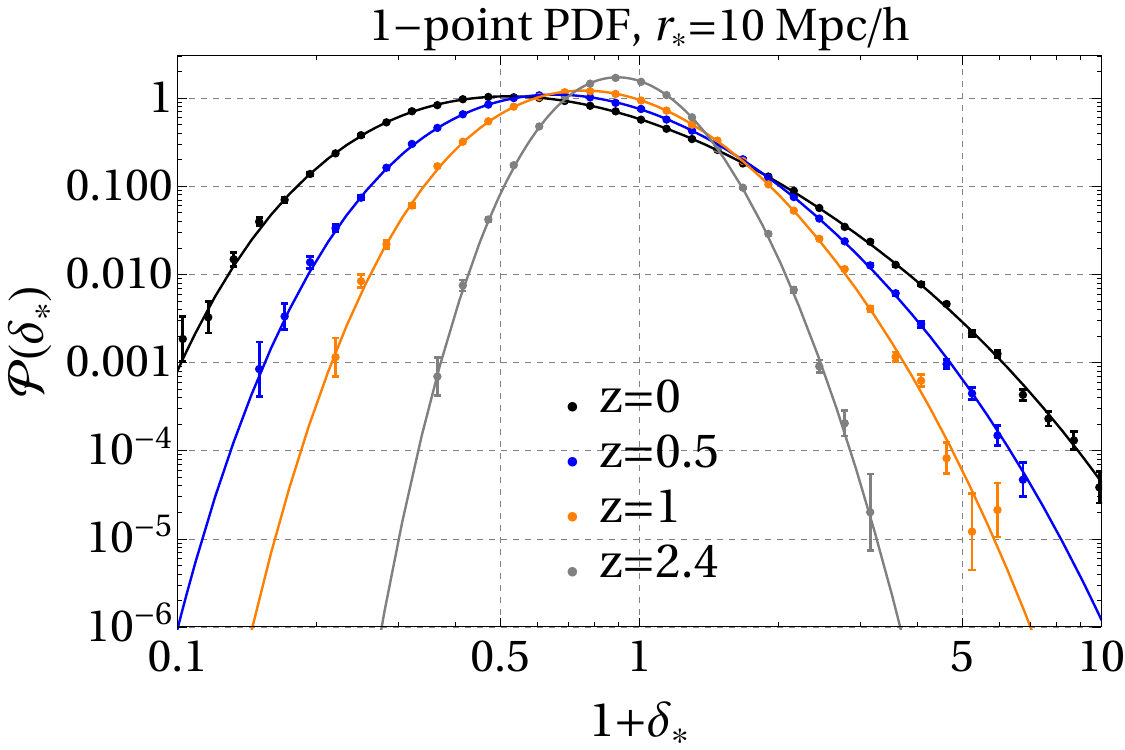}\\
~\\
	\includegraphics[width=0.48\linewidth]{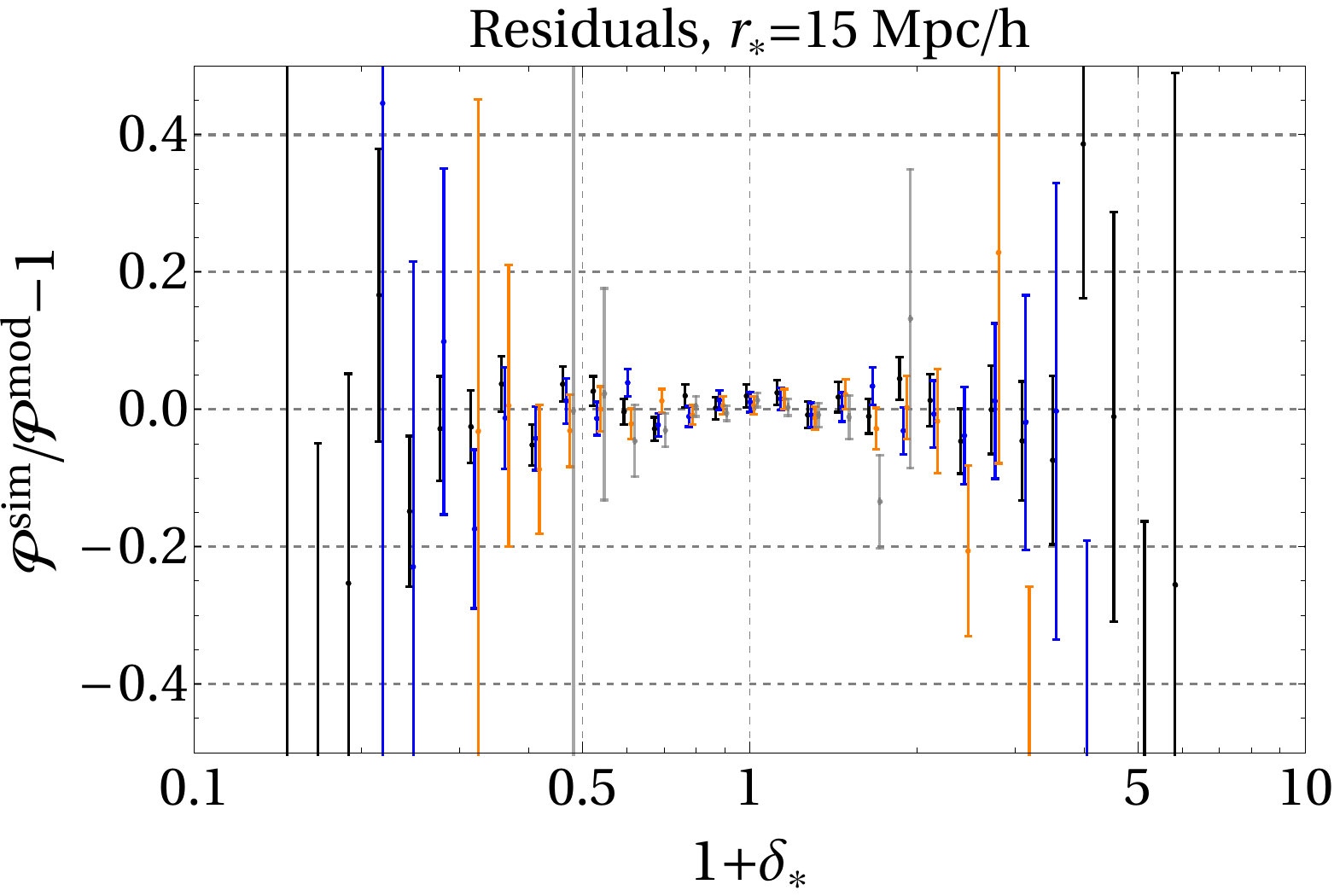}~~
	\includegraphics[width=0.48\linewidth]{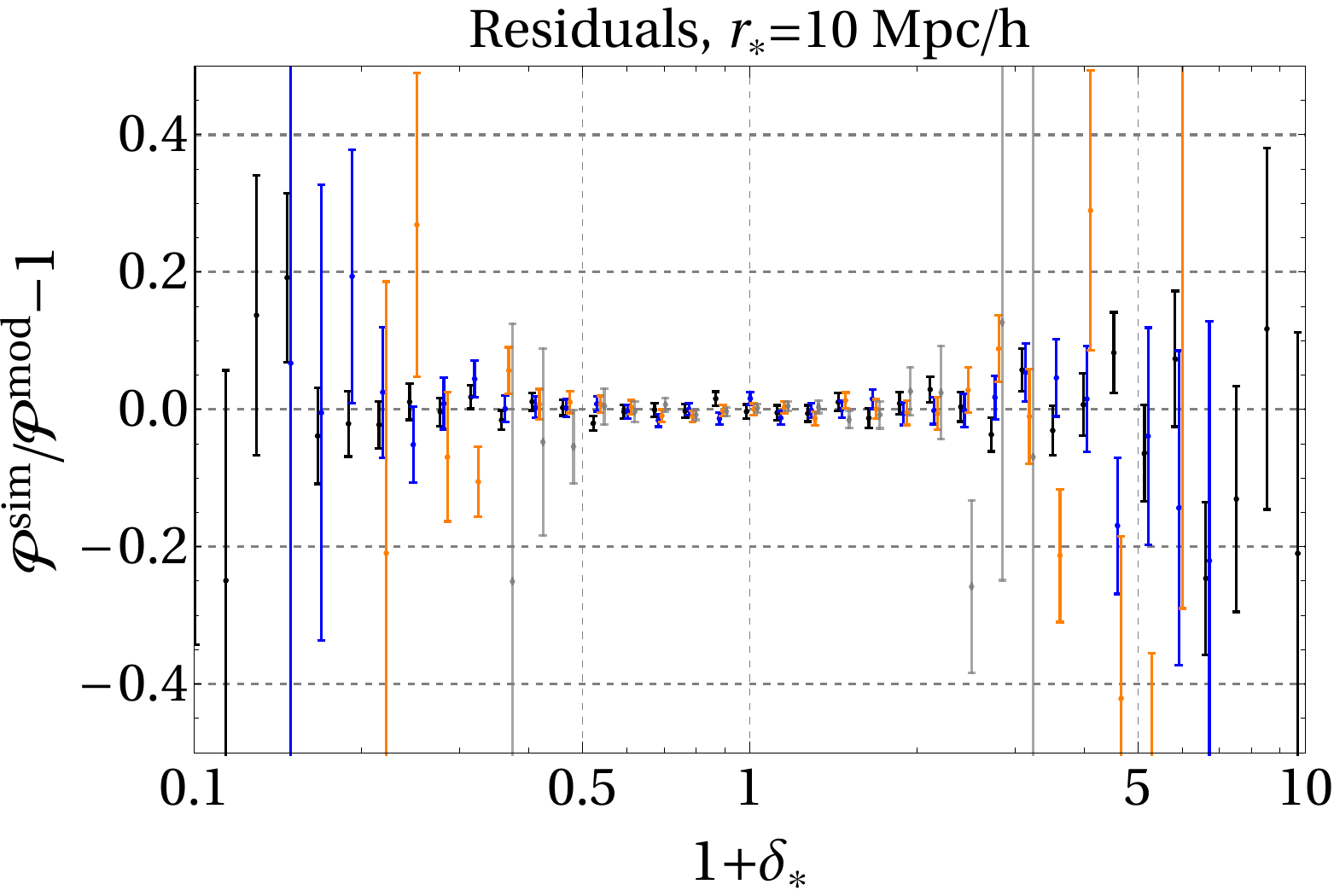}
	\caption{Comparison between the PDF measured in the simulation (points)
          and the best-fit theoretical model (lines) for cell radii
          $r_*=15\,\Mpch$ ({\it left}) and $r_*=10\,\Mpch$ ({\it right}) and for
          four redshift values $z=\{0,\,0.5,\,1.0,\,2.4\}$. Lower
          panels show residuals. For presentation purposes the residuals
          corresponding to different redshifts
          are slightly shifted in the horizontal direction.}
	\label{fig:far_1}
\end{figure}

Figure \ref{fig:far_1} shows the comparison between the theoretical
model and the PDF obtained from the $N$-body simulation for cell radii
$r_*=15\Mpch$ and $10\Mpch$. The agreement is very good with the residuals
inside the statistical uncertainty of the simulations for the range of
densities $0.1\leq 1+\delta_*\leq 10$ and all redshifts. At moderate density
contrasts $0.5\lesssim 1+\delta_*\lesssim 2$ the model reproduces the
PDF with per cent precision. 

\begin{table}[ht]
\begin{center}
\begin{tabular}{|c|c|c|c|c|c|}
\hline 
$r_*,\,\Mpch$ & $m$ & $\zeta^\kin$ & $\zeta^\pot$ & $\chi^2/N_{\rm
  dof}$ & $\gamma_0$\\
\hline 
$15$ &  $2.23\pm 0.15$ & $0.53\pm0.54$ & $-2.6\pm 1.4$ & $72/82$ &
$2.04\pm 0.16$\\
\hline
$10$ &  $2.14\pm 0.10$ & $0.48\pm0.29$ & $-2.03\pm0.69$  & $108/110$ &
$1.73\pm0.08$\\ 
\hline
\end{tabular}
\end{center}
\caption{Best fit parameters of the counterterm model, $\chi^2$ per number of degrees of freedom, and the derived effective speed of sound $\gamma_0$ for 
 large cell
  radii.
\label{tab:large}
}
\end{table}

The good quality of the fit is confirmed
by the values of $\chi^2$ listed in
Table~\ref{tab:large}. This table also summarizes the fitted
parameters $m$, $\zeta^\kin$, $\zeta^\pot$. We observe that they
coincide for the two radii within the error bars. This provides a
consistency check of our model since, as mentioned in
Sec.~\ref{sec:couterterm}, they are expected to be $r_*$-independent.  
Last column contains the derived parameter $\gamma_0$ calculated from
the second derivative of the counterterm prefactor at $\delta_*=0$. We
see that for $r_*=10\,\Mpch$ 
the PDF prefers slightly lower best-fit values of $\gamma_0$
and $m$ than those derived from the power spectrum (see
Eq.~(\ref{mdpl})), though still within one standard deviation of the
latter. It is also worth noting that the error bars listed in
Table~\ref{tab:large} suggest that PDF may provide a better precision
in determination of these parameters. However, a proper treatment of
uncertainties should include the theoretical error of the model
associated with two-loop contributions into the prefactor (see below)
whose estimation is outside the scope of this paper.

As another consistency check, we compute the norm, mean and variance from the theoretically constructed PDF. These are listed in Table~\ref{tab:nmv}. We see that the deviations of the norm from unity and the mean from zero never exceed $\text{a few}\times 10^{-3}$. The non-linear density variance extracted from the PDF matches with per cent accuracy the result of the 1-loop EFT calculation (last column in the table). This is consistent with the discussion in Sec.~\ref{sec:rel} and provides a non-trivial confirmation of validity of the saddle-point expansion. Note that in our method the non-linear variance is not used as an input, but instead comes as the result of the theoretical modeling. Only its counterterm part is fitted from the data.

\begin{table}[h!]
\begin{center}
\begin{tabular}{|c|c|c|c|c|c|c|}
\hline
 \multicolumn{2}{|c|}{} &
 ${\rm norm}-1$ &
 $\langle\delta_*\rangle$ &
 $\langle\delta_*^2\rangle$ &
 $\sigma^2_{\rm EFT}$ &
 $\langle\delta_*^3\rangle/\langle\delta_*^2\rangle^2$\\
\hline
\multirow{4}{*}{$r_*=15\Mpch$} &
$z=0$ &
$-6.2\cdot10^{-3}$ &
$-7.1\cdot10^{-3}$ &
$0.260$ &
$0.262$ &
$3.35$\\ 
& $z=0.5$ & 
$-2.9\cdot10^{-3}$ &
$-3.1\cdot10^{-3}$ &
$0.153$ &
$0.154$ &
$3.30$\\ 
& $z=1$ &
$-1.3\cdot10^{-3}$ &
$-1.3\cdot10^{-3}$ &
$0.095$ &
$0.095$ &
$3.27$\\ 
& $z=2.4$ & 
$-3.3\cdot10^{-5}$ &
$-5.6\cdot10^{-5}$ &
$0.035$ &
$0.035$ &
$3.23$\\ \hline
\multirow{4}{*}{$r_*=10\Mpch$} &
$z=0$ &
$-3.7\cdot10^{-4}$ &
$-2.1\cdot10^{-3}$ &
$0.533$ &
$0.532$ &
$3.63$\\ 
& $z=0.5$ & 
$1.2\cdot10^{-4}$ &
$-6.7\cdot10^{-4}$ &
$0.306$ &
$0.304$ &
$3.65$\\ 
& $z=1$ &
$2.8\cdot10^{-4}$ &
$-2.9\cdot10^{-4}$ &
$0.185$ &
$0.185$ &
$3.56$\\ 
& $z=2.4$ & 
$3.3\cdot10^{-4}$ &
$6.4\cdot10^{-5}$ &
$0.067$ &
$0.067$ &
$3.45$\\ \hline
\end{tabular}
\caption{
The norm, mean, variance and skewness from the best-fit theoretical PDF 
as well as the filtered variance computed using the 1-loop in the EFT power-spectrum. 
}
\label{tab:nmv}
\end{center}
\end{table} 

In the last column of Table~\ref{tab:nmv} we also present the skewness $S_3\equiv\langle\delta_*^3\rangle/\langle\delta_*^2\rangle^2$ computed from the PDF.
This can be contrasted with the tree-level result: $S_3^{\rm tree}=3.21$ for $r_*=15\Mpch$ and $3.39$ for $r_*=10\Mpch$. We see that the full skewness is larger than the tree-level contribution and has mild but significant dependence on the redshift. It would be interesting to compare these results with the perturbative EFT calculation.

Our counterterm model involves several choices which are not unique.
One of them is the choice
of the shell-crossing momentum $\ksc$. We have re-fitted the PDF
using alternative definitions of $\ksc$ which we implemented by
replacing the growth factor $D_*$ in the counterterm sources
(\ref{eftpar}) with $D_{\paral}$ or $D_\perp$ defined in
appendix~\ref{app:ksc}. We found almost identical results: the
best-fit PDFs differ from our baseline model by less than $1\%$ even
at the tails. This robustness with respect to the choice of $\ksc$ is
partially due to the small best-fit value of $(m-2)$ in the Planck
cosmology which suppresses the sensitivity of the PDF to the
position-dependent growth factor.  
   
One more choice is the split of the counterterm stress into kinetic and
potential parts with independent amplitudes $\zeta^\kin$ and
$\zeta^\pot$. 
One may wonder if this freedom can be reduced. We study this possibility in appendix~\ref{app:comp} where we find that setting one of $\zeta^\kin$ or $\zeta^\pot$ to zero can actually provide a good quality of the fit. Moreover, the counterterm prefactor for $\zeta^\pot=0$ happens to be very close to the Model 2 of \cite{Ivanov:2018lcg} which explains the phenomenological success of this model. On the other hand, a reduce model with $\zeta^\kin=\zeta^\pot$ cannot reproduce the data with required precision.
All in all, we believe that keeping $\zeta^\kin$ and $\zeta^\pot$ as independent free parameters is physically motivated and more precise data may require non-zero values for both of them.  
On the
other hand, introducing more free parameters\footnote{Recall that the most
general expression for the counterterm stress allowed by EFT can contain
arbitrary functions of the density and the tidal tensor.} is unlikely to add
anything to the quality of the fit.

\begin{figure}[t]
	\centering
	\includegraphics[width=0.5\linewidth]{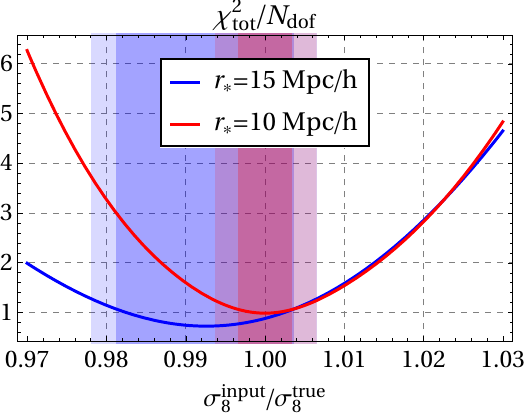}
	\caption{Quality of the fit of the measured PDF with the
          theoretical model computed using the
          rescaled value of $\sigma_8$. Shaded bands show the
          $1\sigma$ (dark) and $2\sigma$ (light) intervals around the best-fit
          values. Note that the upper boundaries of the blue and
          magenta bands almost coincide.}
	\label{fig:sigma8}
\end{figure}

One might be worried that having three free parameters precludes the
model from being predictive. This is not the case as demonstrated by
the following exercise. We artificially change the amplitude of the
linear power spectrum, traditionally parameterized by $\sigma_8$, and
use the rescaled power spectrum to compute the theoretical PDF. As
before, we adjust $m$, $\zeta^\kin$, $\zeta^\pot$ to best fit the
data. Despite this freedom, we find that the quality of the fit
quickly degrades when $\sigma_8$ deviates from its true value, see
Fig.~\ref{fig:sigma8}. The figure suggests that our PDF model is highly
sensitive to the value of $\sigma_8$, perhaps at sub-per cent
level. This statement, however, should be taken with a grain of salt
since the proper assessment of the sensitivity must include an estimate of the
theoretical error which we leave for future.

\subsection{Small cells}
\label{sec:result2}

We now repeat the fitting procedure for cells of smaller radii
$r_*=7.5\,\Mpch$ and $5\,\Mpch$. In the latter case we impose a lower
cut on the density contrast, ${(1+\delta_*)_{\rm
  min}=\{0.28,\,0.36,\,0.44,\,0.6\}}$ at $z=\{0,\,0.5,\,1.0,\,2.4\}$ to
suppress the systematics in the N-body data due to the use of
Zeldovich initial conditions (see appendix~\ref{app:ZA}).

\begin{figure}[t]
	\centering
	\includegraphics[width=0.48\linewidth]{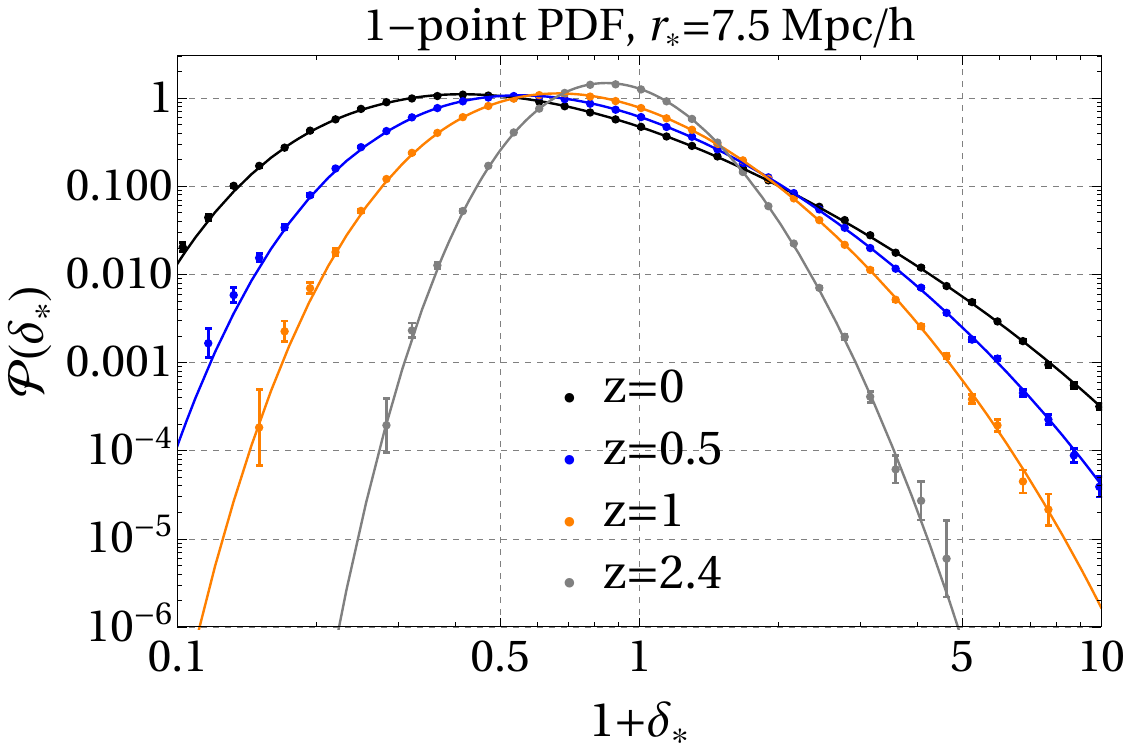}~~
	\includegraphics[width=0.48\linewidth]{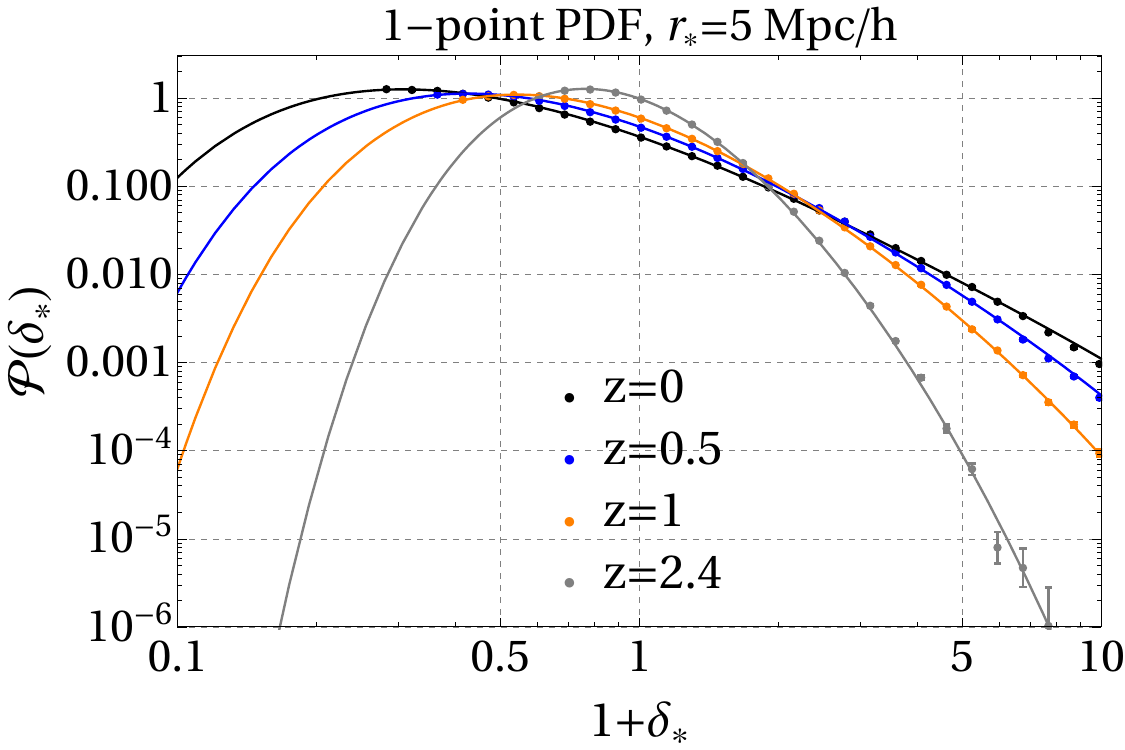}\\
~\\
	\includegraphics[width=0.48\linewidth]{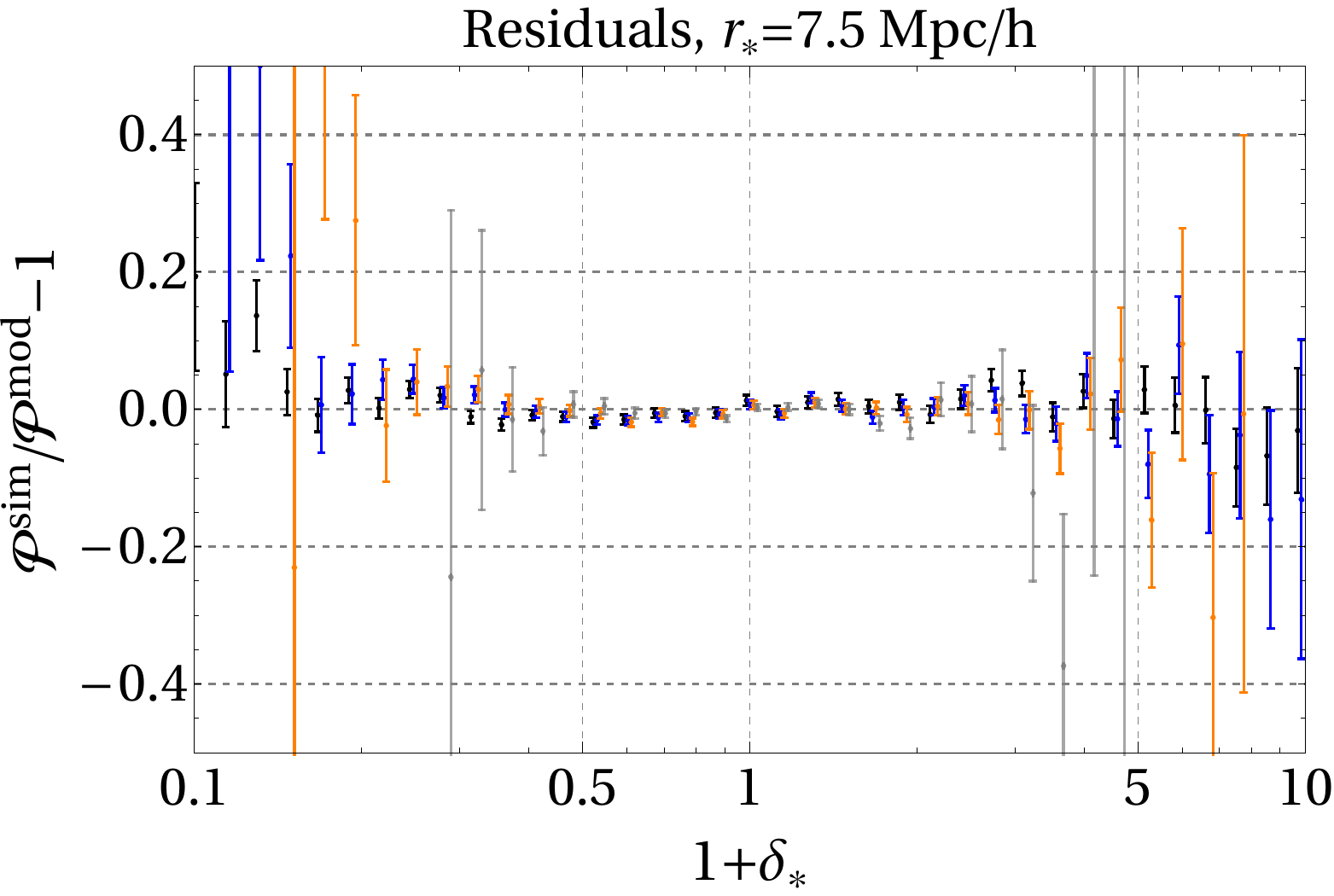}~~
	\includegraphics[width=0.48\linewidth]{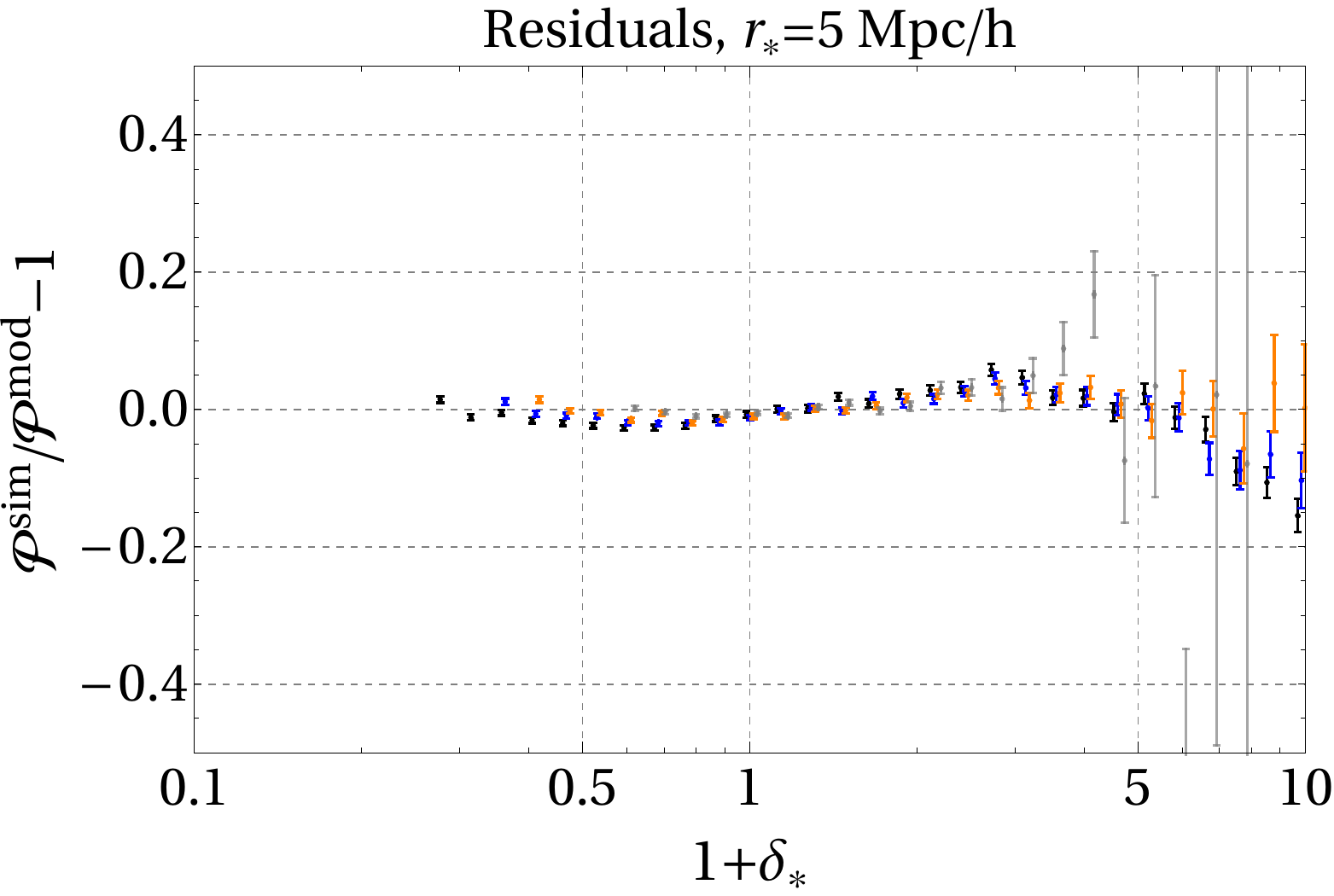}
	\caption{Comparison between the PDF measured in the simulation (points)
          and the best-fit theoretical model (lines) for cell radii
          $r_*=7.5\,\Mpch$ ({\it left}) and $r_*=5\,\Mpch$ ({\it right}) and for
          four redshift values $z=\{0,\,0.5,\,1.0,\,2.4\}$. Lower
          panels show residuals. For presentation purposes the residuals
          corresponding to different redshifts
          are slightly shifted in the horizontal direction. }
	\label{fig:far_2}
\end{figure}

Upper panels in Fig.~\ref{fig:far_2} suggest that the model still
captures the behavior of the PDF quite well. However, when looking at
the residuals, we start seeing tensions. The difference between the
best-fit model and the measured PDF reaches $\sim 5\%$ which is well
beyond the statistical uncertainties. This discrepancy is reflected in
the large $\chi^2$ listed in Table~\ref{tab:small}. There are also
clear tensions between the values of $m$ and $\gamma_0$ for different
cell radii implying that the model fails to reproduce the data with
required precision.

\begin{table}[ht]
\begin{center}
\begin{tabular}{|c|c|c|c|c|c|}
\hline 
$r_*,\,\Mpch$ & $m$ & $\zeta^\kin$ & $\zeta^\pot$ & $\chi^2/N_{\rm
  dof}$ & $\gamma_0$\\
\hline 
$7.5$ &  $2.37\pm 0.06$ & $1.06\pm0.18$ & $-0.49\pm 0.36$ & $225/125$ &
$1.71\pm 0.05$\\
\hline
$5$ &  $2.12\pm 0.02$ & $0.95\pm 0.06$ & $-0.13\pm0.12$  & $965/100$ &
$1.57\pm0.02$\\ 
\hline
\end{tabular}
\end{center}
\caption{Best fit parameters of the counterterm model, $\chi^2$ per number of degrees of freedom, and the derived effective speed of sound $\gamma_0$ for 
 small cell
  radii.
\label{tab:small}
}
\end{table}

\begin{figure}[t]
	\centering
	\includegraphics[width=0.48\linewidth]{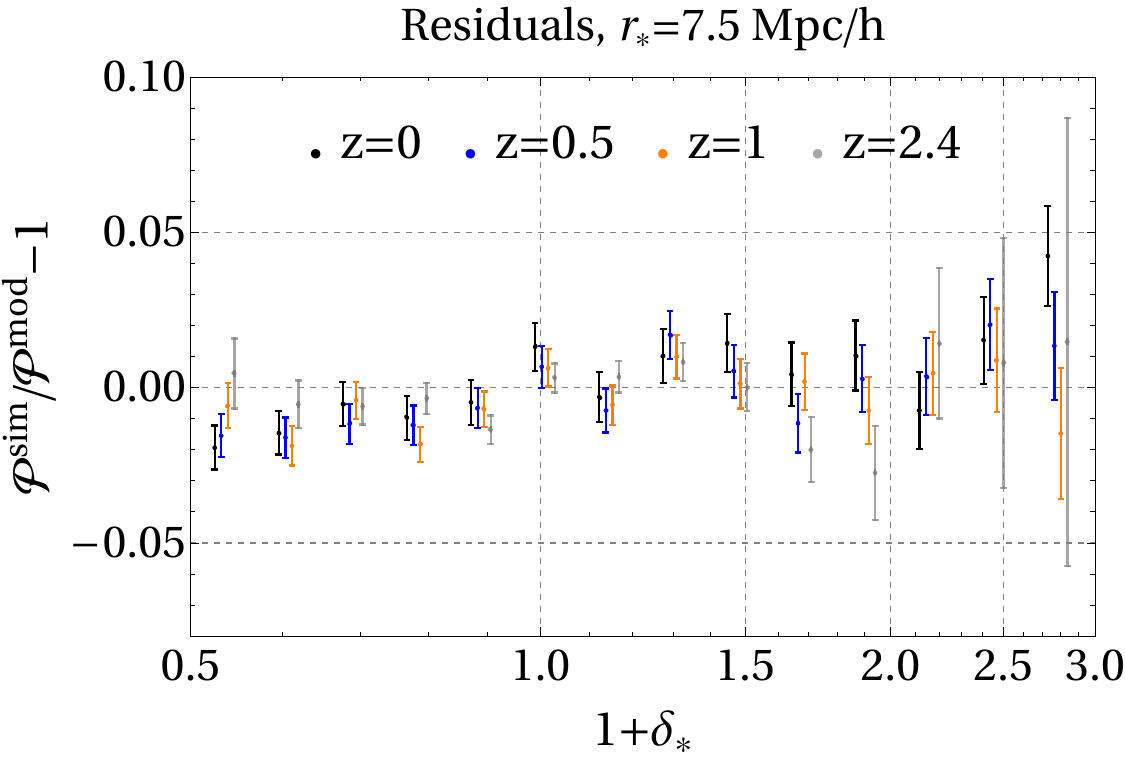}~~
	\includegraphics[width=0.48\linewidth]{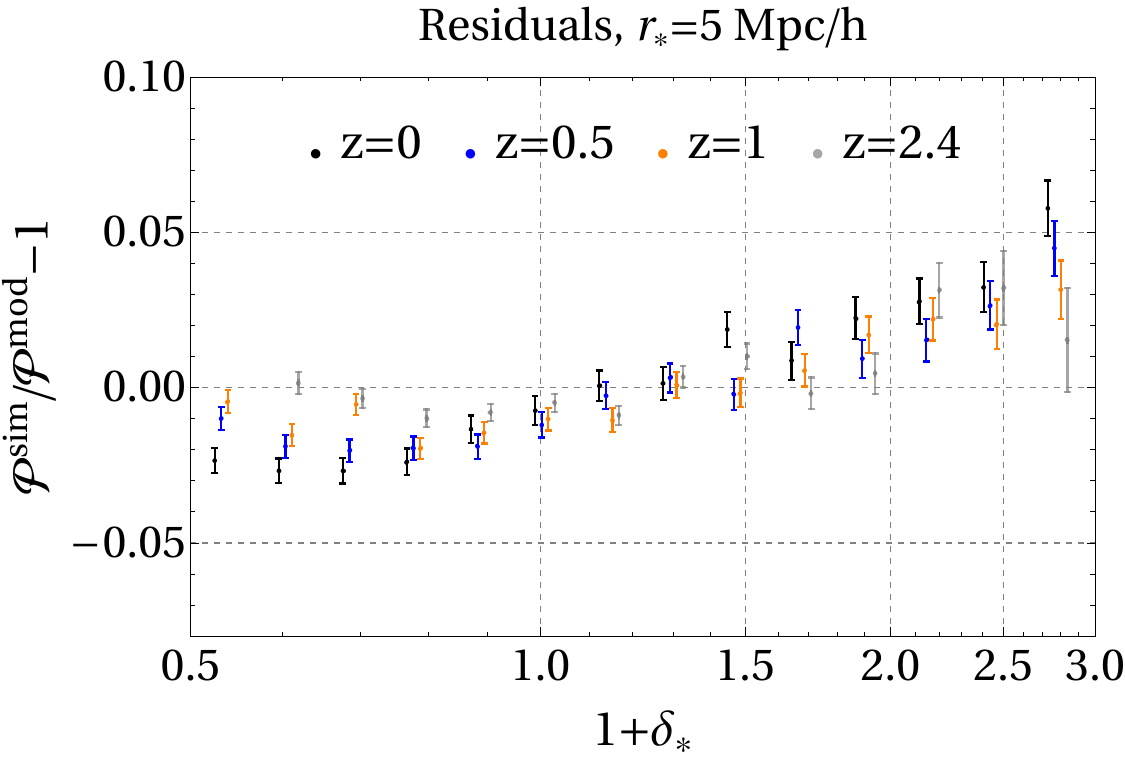}
	\caption{Residuals in the central region of densities between the PDF 
extracted from the N-body simulation and the theoretical
model proposed in this paper. The cell radii are
$r_*=7.5\Mpch$ ({\it left}) and $r_*=5\Mpch$ ({\it right}). The
redshift values are given in the legend.} 
	\label{fig:far_22}
\end{figure}

To understand the cause of this failure, we zoom on the central part of the density range in
Fig.~\ref{fig:far_22}. We
observe that the residuals exhibit a linear trend at $\delta_*=0$. For
$r_*=5\,\Mpch$ we also see that $\P^{\rm sim}/\P^{\rm
  mod}|_{\delta_*=0}$ significantly deviates from $1$. Both these
discrepancies somewhat weaken at larger redshifts. Such behavior
cannot be attributed to a deficiency in the counterterm
model. Indeed, in Sec.~\ref{sec:rel} we derived that, independently of
its modeling, the one-loop counterterm can affect only the second and
higher derivatives of the PDF at $\delta_*=0$.  

Thus, we interpret the discrepancy as the first detection of the
two-loop corrections to the PDF. It is encouraging to find that they
are relatively suppressed even down to cell radii as small as
$5\,\Mpch$. However, they clearly must be taken into account in future
analysis of the PDF at these scales. A precise calculation of the two-loop
corrections appears challenging. A more feasible
approach will be to estimate them approximately and include in the PDF
model as the theoretical uncertainty in the spirit of \cite{Baldauf:2016sjb}. We
leave this interesting task for future.

\section{Conclusions}
\label{sec:con}

In this work we developed an analytic description of the
non-perturbative one-point matter PDF, supplementing the approach of
\cite{Ivanov:2018lcg} with a physically motivated model for the one-loop counterterm
prefactor. It is based on modeling the
counterterm stress tensor which renormalizes the effect of
short-wavelength modes on the saddle-point spherical collapse solution. We have validated the model using the data
from high-resolution Farpoint simulation \cite{HACC:2021sgt} 
and found a per cent
level agreement down to cell radii $r_*=10\,\Mpch$ and redshift
$z=0$. 

The model contains three redshift-independent free parameters
that in the spirit of EFT of LSS
must be fitted from the data. Two combinations of these parameters
already appear in the EFT template for the matter power
spectrum. Thus, if the power spectrum is measured with high precision,
our PDF model adds only a single parameter on top, which can be
further connected to the bispectrum counterterms. In practice,
however, as our comparison with N-body data indicates, the PDF can
provide comparable or better constraints on the counterterm
coefficients than the correlation functions. This suggests doing a
joint analysis or using information from the power spectrum and
bispectrum as priors
when fitting the PDF. 

For cells of radius $5\,\Mpch<r_*<10\,\Mpch$ we detected a few per
cent deviations between the model and the N-body data which we
identified as the two-loop correction. The smallness of this
corrections is encouraging. Still, it is significant and may become visible even at larger
radii with real or
N-body data with larger statistics. 
Rather than trying to calculate it precisely, a more
promising path would be to include it as the theoretical uncertainty in the
one-loop PDF template. We plan to explore this interesting direction in future.

The matter PDF by itself is not directly observable. More work is needed to connect it to the CiC statistics of galaxies. This will require inclusion of 
redshift-space distortions and galaxy bias. Our analysis of
renormalization naturally reveals a picture of nonlinear short-scale
modes evolving along the flow lines of the background saddle-point
solution. This picture appears similar to that used in the Lagrangian
approach to the PDF biasing proposed in \cite{Friedrich:2021xff}. It
will be very interesting to explore this connection in detail. 

Another
important task is inclusion of the PDF covariance 
which is crucial for its application to constraining
the cosmological parameters or theories beyond
$\Lambda$CDM. In our work we restricted to non-intersecting cells which are essentially uncorrelated. With the proper treatment of covariance the statistical power of PDF can be increased by allowing for overlapping cells.  
Interesting progress in theoretical modeling of PDF has been made recently
in \cite{Repp:2020etr,Bernardeau:2022sva,Uhlemann:2022znd}. Alternatively, the covariance can be extracted from large suits of N-body simulations, such as Quijote \cite{Villaescusa-Navarro:2019bje}. 

Our approach can be applied to other non-perturbative statistics, such as
generalizations of the PDF using a non-top-hat window function \cite{Bernardeau:2015khs} 
and
multiple cells
\cite{Uhlemann:2016wug}, or correlations between the non-linear matter density and long-wavelength modes~\cite{Jamieson:2020wxf}. 
Yet one more promising route is modeling of the lensing shear and convergence PDF \cite{Boyle:2020bqn,Boyle:2022msq}. 
Finally, counts-in-cells are closely related to the $k$-nearest-neighbor \cite{Banerjee:2020umh,Banerjee:2021cmi} and 
the void 
size \cite{Contarini:2022mtu} distribution functions.
We believe our work sets firm grounds for analytical understanding of these interesting observables.

\paragraph{Acknowledgments}
We are indebted to Nicholas
Frontiere and Thomas Uram for assistance in the use of the Farpoint
simulation data. 
We thank  Niayesh
Afshordi, Neal Dalal, Alexander
Kaurov, Nickolas Kokron and Mehrdad Mirbabayi for fruitful discussions.
The work of A.C. is supported by the RFBR grant 20-02-00982.  
The work of S.S. is supported by
the Natural Sciences and Engineering Research Council (NSERC) of Canada.
Research at Perimeter Institute is supported in part by the Government
of Canada through the Department of Innovation, Science and Economic
Development Canada and by the Province of Ontario through the Ministry
of Colleges and Universities. 
This research was enabled in part by support provided by Compute
Ontario 
(www.computeontario.ca) and Digital Research Alliance of Canada (alliancecan.ca).
The numerical analysis of this work was partially performed on the
Helios cluster at the Institute for Advanced Study.

\appendix

\section{Conventions}
\label{app:notations}

We use the same conventions as in \cite{Ivanov:2018lcg}. For
completeness, we summarize them here.

We define the Fourier transform as, 
\be
\delta(\x)=\int_\k \delta(\k)
\e^{i\k\cdot\x}\,, 
\ee
where the integration measure in momentum space is
\be
\label{measurek}
\int_{\bf k} = \int \frac{d^3k}{(2\pi)^3}\;.
\ee
We also use the concise notation for the radial integral in momentum
space, 
\be
\label{kintradial}
\int [dk]=\int_0^\infty \frac{k^2 dk}{(2\pi)^3}\;,
\ee 
The power spectrum is defined as,
\be
\langle \delta(\k)\delta(\k')\rangle = (2\pi)^3\delta_{\rm D}(\k+\k')P(k)\,,
\ee
where $\delta_{\rm D}(\k)$ is the Dirac delta-function.

We use the following definition for the spherical harmonics:
\bseq
\label{sphharm}
\begin{align}
\label{sphharm0}
&Y_0(\theta,\phi)=1\;,\\
&Y_{\ell m}(\theta,\phi)=\frac{(-1)^{\ell+m}}{2^\ell\ell !}
\bigg[\frac{2\ell+1}{4\pi}\frac{(\ell-|m|)!}{(\ell+|m|)!}\bigg]^{1/2}
\e^{im\phi}
(\sin\theta)^{|m|}\bigg(\frac{d}{d\cos\theta}\bigg)^{\ell+|m|}(\sin\theta)^{2\ell},
\notag\\
&\qquad\qquad\qquad\ell>0\;,~~~~-\ell<m<\ell\;.
\label{sphharmlm}
\end{align}
\eseq
They obey the relations, 
\be
\label{Yprop}
\Delta_{\Omega}Y_{\ell m}=-\ell (\ell+1)Y_{\ell m}~,~~~~ 
Y_{\ell m}(-{\bf n})=(-1)^\ell Y_{\ell m}({\bf n})~,~~~~
Y_{\ell m}^*({\bf n})=Y_{\ell,-m}({\bf n})\;,
\ee
where $\Delta_{\Omega}$ is the the Laplace–Beltrami operator on the 2-dimensional
sphere.
All harmonics are orthogonal and normalized to 1 when integrated over
a 2d sphere, except the monopole that has the norm $4\pi$,
\be
\label{harmnorm}
\int d\Omega\, Y_{\ell
  m}\,Y^*_{\ell'm'}=(4\pi)^{\delta_{0\ell}}\delta_{\ell
  \ell'}\delta_{mm'}\,, 
\ee
where $\delta_{ij}$ is the Kronecker delta symbol. 

We expand the fields over spherical harmonics in position and Fourier
space as,
\bseq
\begin{align} 
&\delta(\x)=\delta_0(r)+\sum_{\ell >0}\sum_{m=-\ell}^\ell \delta_{\ell
  m}(r)\, Y_{\ell m}(\x/r)\;,\\
&\delta(\k)=\delta_0(k)+\sum_{\ell >0}\sum_{m=-\ell}^\ell (-i)^\ell \,
\delta_{\ell m}(k)\, Y_{\ell m}(\k/k)\;.
\label{eq:Ylk}
\end{align}
\eseq
Due to the relations (\ref{Yprop}) we have,
\be
\label{deltalmprop}
\big(\delta_{\ell m}(r)\big)^*=\big(\delta_{\ell,-m}(r)\big)~,~~~~~~
\big(\delta_{\ell m}(k)\big)^*=\big(\delta_{\ell,-m}(k)\big)\;.
\ee
The coefficient functions in the above expansions are related by,
\be
\label{eq:dlfourier}
 \delta_{\ell m}(r)=4\pi \int [dk] \; 
j_\ell (kr)\,\delta_{\ell m}(k)\,,
\ee
where $j_\ell(x)$ is the spherical Bessel function of order $\ell$. It
is related to the  
Bessel function of the first kind via
\be 
\label{jJ}
j_\ell(x)=\sqrt{\frac{\pi}{2x}}J_{\ell+1/2}(x)\,.
\ee
The first few functions are,
\be
\label{j012}
j_0(x)=\frac{\sin x}{x}~,~~~~
j_1(x)=\frac{\sin x}{x^2}-\frac{\cos x}{x}~,~~~~
j_2(x)=\bigg(-\frac{1}{x}+\frac{3}{x^3}\bigg)\sin x-\frac{3}{x^2}\cos x\;.
\ee
Spherical Bessel functions $j_\ell(kr)$ with different arguments $k$
form an orthogonal basis on the half-line with the normalization 
\be
\label{Besselnorm}
\int_0^{\infty} dr \,r^2j_\ell(k'r) j_\ell(kr)
=\frac{\pi}{2k^2}\delta_{\rm D}(k-k')\,.
\ee
They are eigenmodes of the radial part of the Laplace operator,
\be
\label{Besseleq}
\d_r^2 j_\ell(kr)+\frac{2}{r}\d_rj_\ell(kr)-\frac{\ell(\ell+1)}{r^2}
 j_\ell(kr)=-k^2 j_\ell(kr)\;.
\ee

\section{Spherical PDF}
\label{app:path}

\begin{figure}
	\centering
	\includegraphics[width=0.48\linewidth]{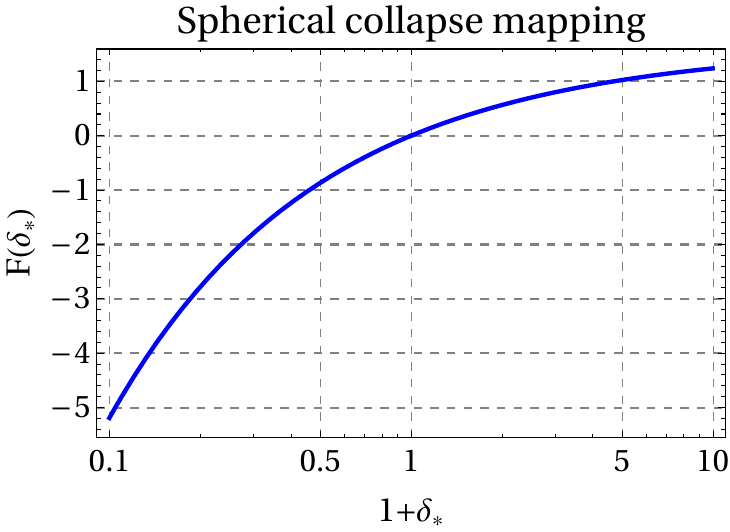}
~~~~\includegraphics[width=0.465\linewidth]{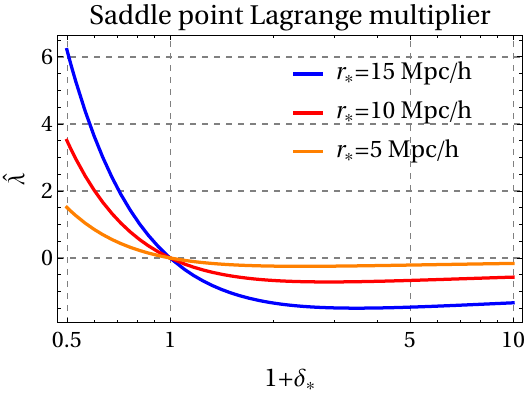}
\caption{{\it Left:} The spherical collapse map $F(\delta_*)$ for EdS
  cosmology. {\it Right:} The saddle-point value of the Lagrange
  multiplier $\hat\lambda$ as function of density contrast 
for several cell radii.}
	\label{fig:FEdS}
\end{figure}

In this appendix we derive the ``spherical'' part of the PDF
(\ref{Ptot}) accounting for the contribution of the saddle
configuration and the monopole fluctuations around it. We start from
the saddle-point equations for the integrals (\ref{eq:pdfLaplace2}),
(\ref{w0}) in the limit $g\to 0$. Taking derivatives of the
expressions in the exponent with respect to $\delta_L$ and $\lambda$
we obtain,  
\begin{subequations}
\label{eq:sp}
\begin{align}
\label{sp1}
& \frac{\delta_L(\k)}{P(k)}+\lambda \frac{\d
  \bar\delta_W}{\d \delta_L(\k)}=0\,,\\ 
\label{sp2}
& \bar\delta_W[\delta_L]=\delta_* \,.
\end{align} 
\end{subequations}
Since the window function and the power spectrum are invariant under
rotations and assuming that the solution to (\ref{eq:sp}) is unique, we
conclude that the saddle-point configuration $\hat\delta_L(k)$ is
spherically symmetric. This 
implies that $\hat\delta_L(k)$ and $\bar\delta_W$ are related by the
equations of spherical collapse which are exactly
solvable.\footnote{We assume that the saddle-point solution does not
  experience shell crossing. Strictly speaking, this assumption is
  violated at $z=0$ 
 for extreme overdensities $\delta_*\gtrsim 7$
  considered in this paper. 
Still, the shell crossing occurs 
in the center of the cell and does not invalidate the spherical
collapse mapping between average densities discussed below.
\label{foot:map}}
The solution is conveniently written in terms of relations between the
linear and nonlinear density contrasts taken at the same time slice
and averaged over the Lagrangian and the Eulerian radii, respectively,
\be
\label{sphmap}
\bar\delta(r)=f\big(\bar\delta_L(R)\big)~,~~~~
\bar\delta_L(R)=F\big(\bar\delta(r)\big)~,~~~~
R=r\,\big(1+\bar\delta(r)\big)^{1/3}\;,
\ee
where
\be
\label{bardeltadef}
\bar\delta(r)=\frac{3}{r^3}\int_0^r r'^2dr'\,\delta(r')~,~~~~~~
\bar\delta_L(R)=\frac{3}{R^3}\int_0^R R'^2dR'\,\delta_L(R')\;.
\ee
For the EdS cosmology the mapping (\ref{sphmap}) does not depend on
time and the functions $f$, $F$ are defined as follows. Introduce four
functions,
\bseq
\label{FGplusminus}
\begin{align}
\label{FGplus}
&{\cal
  F}_+(\theta)=\frac{9(\theta-\sin\theta)^2}{2(1-\cos\theta)^3}-1\;,
&&{\cal G}_+(\theta)=\frac{3}{20}[6(\theta-\sin\theta)]^{2/3}\;,\\
\label{FGminus}
&{\cal
  F}_-(\theta)=\frac{9(\sh\theta-\theta)^2}{2(\ch\theta-1)^3}-1
&&{\cal G}_-(\theta)=-\frac{3}{20}[6(\sh\theta-\theta)]^{2/3}\;.
\end{align}
\eseq
Then 
\be
\label{EdSfF}
f(x)=\begin{cases}
{\cal F}_+\big({\cal G}_+^{-1}(x)\big)\,,&x>0\\
{\cal F}_-\big({\cal G}_-^{-1}(x)\big)\,,&x<0
\end{cases}~~,\qquad~~
F(x)=\begin{cases}
{\cal G}_+\big({\cal F}_+^{-1}(x)\big)\,,&x>0\\
{\cal G}_-\big({\cal F}_-^{-1}(x)\big)\,,&x<0
\end{cases}~~,
\ee
where ${\cal G}_+^{-1}$ stands for the inverse function to ${\cal
  G}_+$, etc. The derivation of these expressions can be found, e.g.,
in \cite{Ivanov:2018lcg} (see also appendix~\ref{app:ZA}). We show the plot of
the function $F$ in the left panel of Fig.~\ref{fig:FEdS}. For small
density contrasts we have,
\be
\label{Fsmalldelta}
F(\delta_*)=\delta_*-\frac{17}{21}\delta_*^2
+\frac{2815}{3969}\delta^3+O(\delta_*^4)\;. 
\ee 

\begin{figure}
	\centering
	\includegraphics[width=0.48\linewidth]{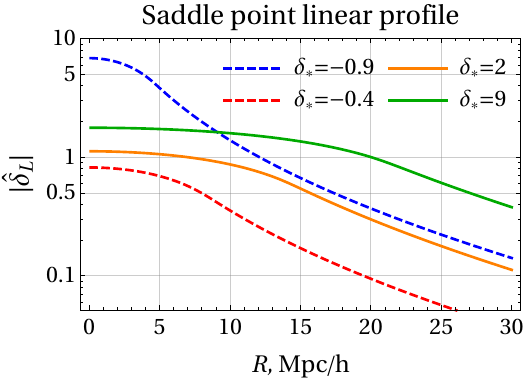}
~~~\includegraphics[width=0.48\linewidth]{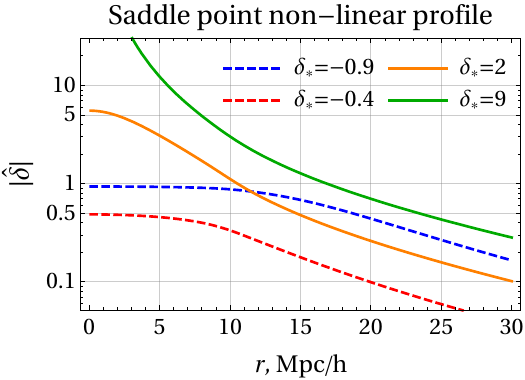}
\caption{Saddle-point linear profile in Lagrangian position space (left) and non-linear profile at $z=0$ as a function of an Eulerian radius (right) for several values of $\delta_*$. 
For $\delta_*<0$, the absolute value of the negative density contrast is shown (dashed lines).
The cell radius is $r_*=10\Mpch$. 
}
	\label{fig:profiles}
\end{figure}

The mapping (\ref{sphmap}) allows us to find the saddle point and
perform the integral over spherical fluctuations exactly. The 
shortest way to do it is to notice that for spherically symmetric
configurations we can replace the $\delta$-function (\ref{deltaD})
with a constraint on the linear density contrast, 
\be
\label{deltas}
\delta_{\rm D}\Big(\delta_*-\bar\delta_W[\delta_L]\Big)
=
C[\delta_*,\delta_L]\cdot
\delta_{\rm D}\Big(F(\delta_*)-\bar\delta_L(R_*)\Big)\;,
\ee
where $R_*$ is defined in (\ref{Rstar}) and the coefficient
\be
\label{Cfirst}
C[\delta_*,\delta_L]=F'(\delta_*)-\frac{R_*\d_R\bar\delta_L(R_*)}{3(1+\delta_*)}
\ee
has been inserted to ensure that the integral of the r.h.s. over
$\delta_*$ gives unity. This yields for the spherical part of the
PDF,
\begin{align}
{\cal P}_{\rm sp}(\delta_*)=&{\cal N}_0^{-1} \int\limits_{\rm spherical} 
{\cal D}\delta_L\,
\exp\bigg\{-\frac{4\pi}{g^2}\int[dk]\frac{|\delta_L(k)|^2}{2P(k)}\bigg\}
\,
\delta_{\rm
  D}\Big(\delta_*-\bar\delta_W[\delta_L]\Big)\notag\\
=&{\cal N}_0^{-1} \int_{-i\infty}^{i\infty}\frac{d\tilde\lambda}{2\pi ig^2}
\exp\bigg\{\frac{\tilde\lambda F(\delta_*)}{g^2}\bigg\}\notag\\
&\times\int\limits_{\rm spherical} 
{\cal D}\delta_L\,C[\delta_*,\delta_L]
\exp\bigg\{-\frac{4\pi}{g^2}
\int[dk]\bigg[\frac{|\delta_L(k)|^2}{2P(k)}+\tilde\lambda
 W_{\rm th}(kR_*)\delta_L(k)\bigg]
\bigg\},
\label{Pspint}
\end{align}
where in the second line we used Eq.~(\ref{deltas}) and
Fourier-transformed the $\delta$-function. This is a Gaussian
integral with a prefactor $C[\delta_*,\delta_L]$ which is a linear
function of $\delta_L$. Such integrals are exactly evaluated by the
saddle point method. It is straightforward to find the saddle
configuration, 
\bseq
\label{sphersaddle}
\begin{align}
\label{eq:deltahatlin}
&\hat{\delta}_L(k)=\frac{F(\delta_*)}{\sigma^2_{R_*}} P(k)\,W_{\rm
  th}(kR_*)\,,\\
\label{tildelambda}
&\hat{\tilde\lambda}=-\frac{F(\delta_*)}{\sigma_{R_*}^2}\;,
\end{align}
\eseq
where the density variance $\sigma_{R_*}^2$ is given by
Eq.~(\ref{sigmaRstar}). 
The resulting linear density profiles in the Lagrangian space $\hat \delta_L(R)$ for $\Lambda$CDM power spectrum and $r_*=10\,{\rm Mpc}/h$ are shown on the left of Fig.~\ref{fig:profiles} for several values of $\delta_*$. On the right we also show the corresponding non-linear profiles $\hat\delta(r)$ in Eulerian space at $z=0$ obtained by applying the map (\ref{sphmap}), (\ref{bardeltadef}). 
For $\delta_*\gtrsim 7$ the linear density profile in the center exceeds the EdS collapse threshold 
$1.686$ and thus the innermost part of $\hat\delta(r)$ diverges. This, however, does not invalidate the map between the averaged densities, as noted in footnote~\ref{foot:map}. 
 
Putting the expressions (\ref{sphersaddle}) back into
(\ref{Pspint}) we obtain,
\be
\label{Psp1}
{\cal P}_{\rm sp}(\delta_*)=\frac{\hat C(\delta_*)}{\sqrt{2\pi
    g^2\sigma_{R_*}^2}}
\,\e^{-\frac{F^2(\delta_*)}{2g^2\sigma_{R_*}^2}}\;. 
\ee
Here $\hat C(\delta_*)$ is 
the coefficient $C[\delta_*,\delta_L]$ evaluated at the saddle
point,
\bseq
\be
\hat C(\delta_*)=F'(\delta_*)+\frac{F(\delta_*)}{1+\delta_*}
\bigg(1-\frac{\xi_{R_*}}{\sigma_{R_*}^2}\bigg)
\ee
with
\be
\label{xiRstar}
\xi_{R_*}=4\pi\int[dk]\,\frac{\sin kR_*}{kR_*}\,W_{\rm th}(kR_*)P(k)\;.
\ee
\eseq
Finally, introducing the variable $\nu$ as in (\ref{eq:nu}) and
observing that $d\nu/d\delta_*=\hat C(\delta_*)/\sigma_{R_*}$ brings
the spherical PDF to the form (\ref{Psp2}).
Notice
that the spherical PDF is by itself normalized, $\int {\cal P}_{\rm
  sp}(\delta_*)\,d\delta_*=1$. On the other hand, it does not obey the
zero-average condition 
$\langle\delta_*\rangle_{\rm sp}\equiv\int\delta_* {\cal P}_{\rm
  sp}(\delta_*)\,d\delta_*\neq 0$. This is, of course, due to the fact
that ${\cal P}_{\rm sp}$ is not the full PDF since it does not take
into account the contribution of aspherical fluctuations. As discussed
in \cite{Ivanov:2018lcg}, the full PDF (\ref{Ptot}) does lead to zero
average density contrast, $\langle\delta_*\rangle\equiv\int\delta_* {\cal
  P}(\delta_*)\,d\delta_*= 0$. 

For calculation of the aspherical prefactor, we need the saddle-point
value of the Lagrange multiplier $\lambda$ entering into the integral
of the full PDF (\ref{eq:pdfLaplace2}). Comparing the Fourier
representations of the two $\delta$-functions in (\ref{deltas}), we
see that it is related to the ``spherical'' Lagrange multiplier
$\tilde\lambda$ by the factor $C$. So, we have, 
\be 
\label{eq:lambda}
\hat\lambda(\delta_*)=-\frac{F(\delta_*)}{\sigma^2_{R_*}}\hat
C(\delta_*)
=-\nu\frac{d\nu}{d\delta_*}\,,
\ee
Note that $\hat\lambda$ depends on the density contrast and on the
cell radius $r_*$, but is independent of the redshift. We plot it for
several cell radii in the right panel of Fig.~\ref{fig:FEdS}.

\section{A fit for the aspherical prefactor}
\label{sec:log}

The featureless behavior of the aspherical prefactor extracted
from N-body simulations in Fig.~\ref{fig:scal} suggests
to fit it in the considered range of density contrasts by a polynomial
in $\ln(1+\delta_*)$,
\be
\label{Aprfit}
\Apr^{\rm fit}=a_0+a_1\ln(1+\delta_*)+a_2\ln^2(1+\delta_*)
+a_3\ln^3(1+\delta_*)\;.
\ee
The coefficients $a_0$, $a_1$ here are not free and can be found from
the following consistency conditions. First, the normalization of the
PDF implies 
\be
\label{normid}
\int {\cal P}(\delta_*)\, d\delta_*\approx \Apr\big|_{\delta_*=0}=1\;,
\ee
where we have used Eq.~(\ref{Ptot}) and retained only the term of the
zeroth order in $g^2$. Second, due to the mass conservation and
translation invariance the full 1-point PDF must have zero mean,
\be
\label{meanid}
\langle\delta_*\rangle\equiv \int \delta_*\, {\cal
  P}(\delta_*)\,d\delta_*=0\;. 
\ee
Again using Eq.~(\ref{Ptot}) and retaining the leading contributions
in $g^2$ we obtain the identity
\be
\label{meanid1}
\frac{d\Apr}{d\delta_*}\bigg|_{\delta_*=0}=-\frac{1}{2}
\bigg(\frac{d\delta_*}{d\nu}\bigg)^{-2}
\frac{d^2\delta_*}{d\nu^2}\bigg|_{\nu=0}
=\frac{4}{21}-\frac{\xi_{r_*}}{\sigma_{r_*}^2}\;.
\ee
Thus, we conclude that the magnitude and the slope of the aspherical
prefactor at the origin are fixed, irrespective of the details of its
modeling.\footnote{This is true for the one-loop prefactor. Higher
  loops introduce corrections of order $g^2$, see
  Sec.~\ref{sec:result2}.}
In particular, for the fitting function (\ref{Aprfit}) it implies,
\be
\label{fitconstr}
a_0=1~,~~~~~a_1=\frac{4}{21}-\frac{\xi_{r_*}}{\sigma_{r_*}^2}\;.
\ee

\begin{figure}
	\centering
	\includegraphics[width=0.48\linewidth]{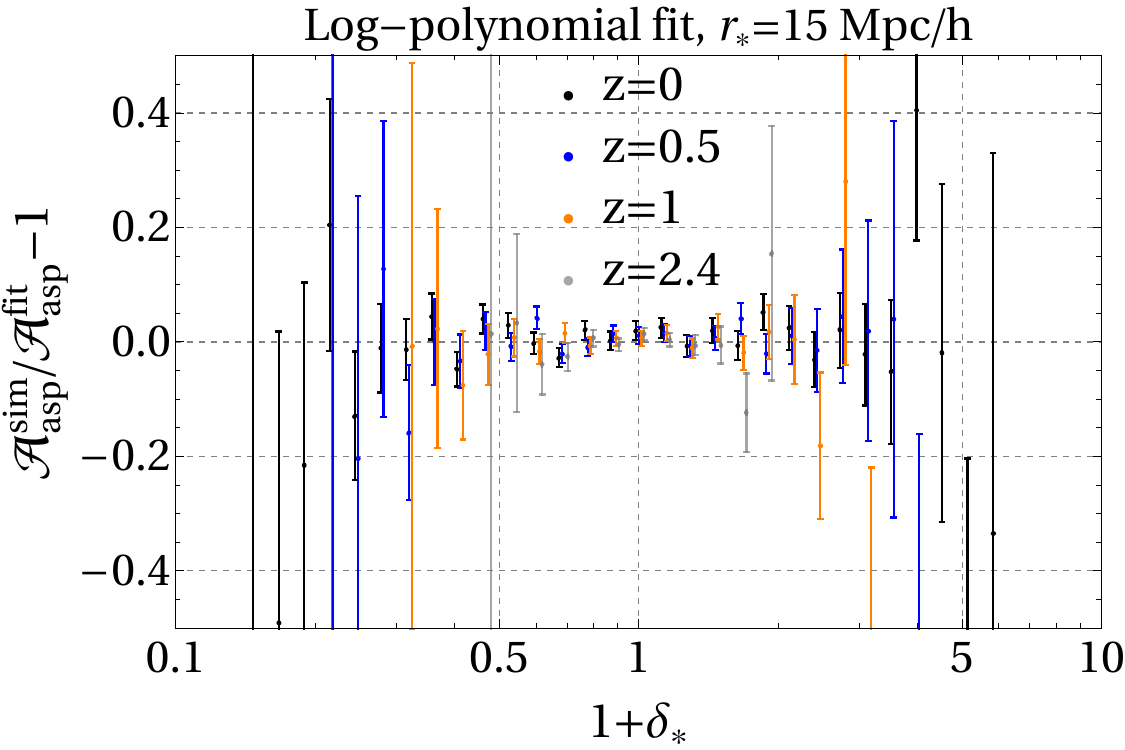}~~
	\includegraphics[width=0.48\linewidth]{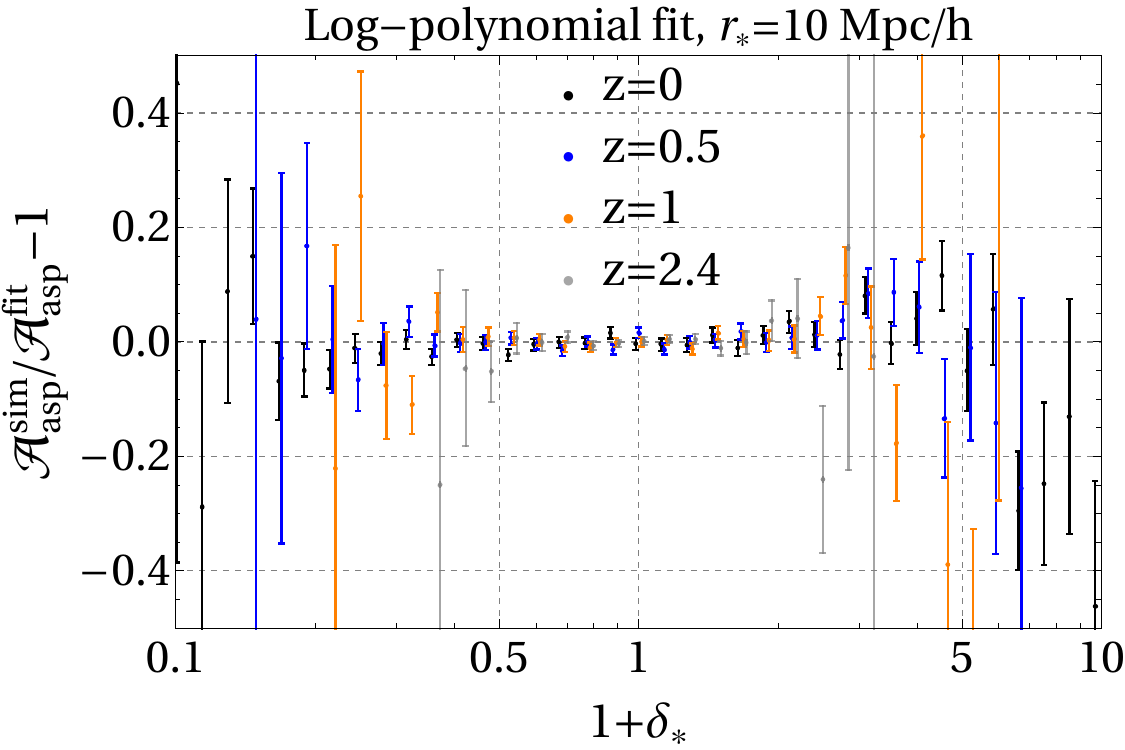}\\
~\\
	\includegraphics[width=0.48\linewidth]{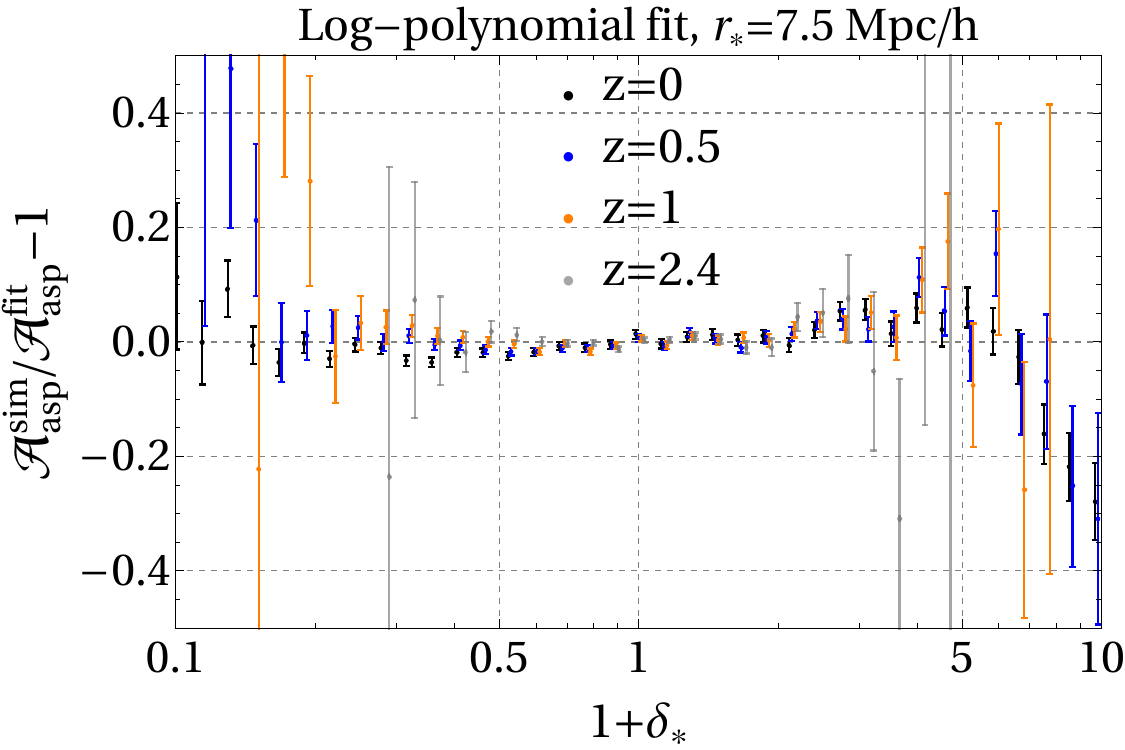}~~
	\includegraphics[width=0.48\linewidth]{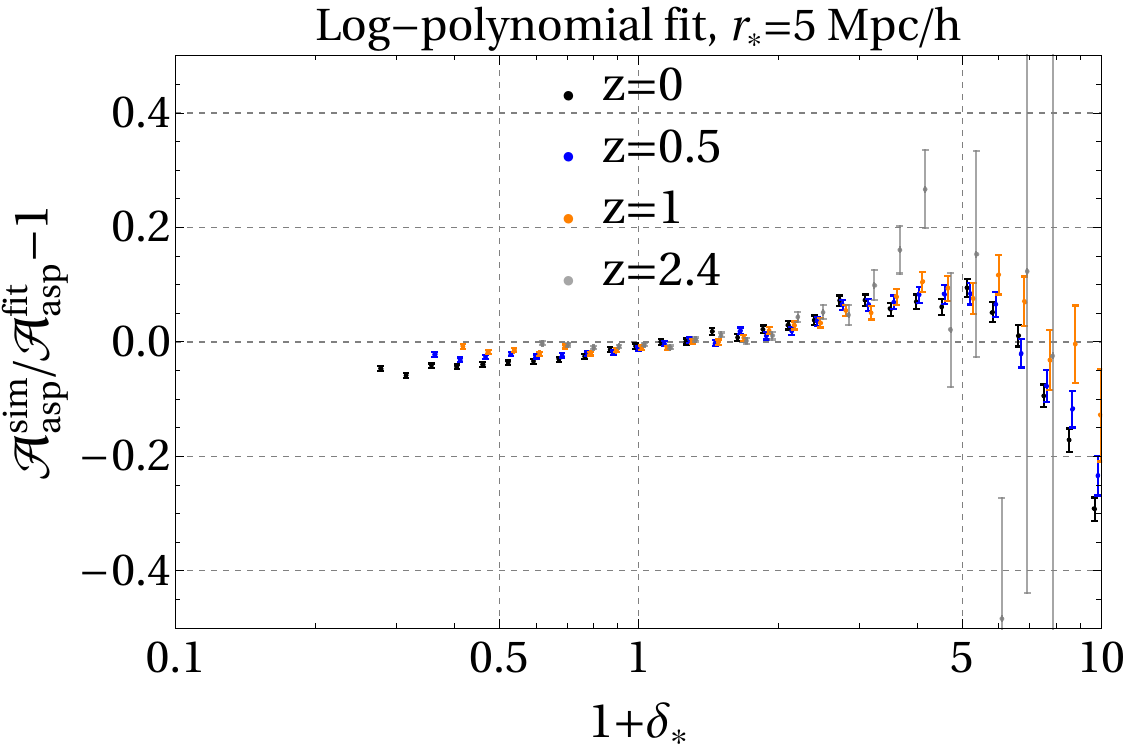}
	\caption{Residuals between the aspherical prefactor extracted
          from N-body data and the log-polynomial fit (\ref{Aprfit})
          for different redshifts and cell radii. The data for
          $r_*=5\,\Mpch$ are cut at small densities as in
          Sec.~\ref{sec:result2} 
          to suppress the systematic errors due to Zeldovich
          initial conditions.  }
	\label{fig:fitres}
\end{figure}

\begin{table}[ht]
	\begin{center}
		\begin{tabular}{|c|c|c|c|c|}
			\hline 
			$r_*,\,\Mpch$ & $a_1$ & $a_2$ & $a_3$ &
                        $\chi^2/N_{\rm dof}$\\
			\hline
			$15$ &  $-0.535$ & $-0.020\pm0.01$ & $0.054\pm
                        0.01$ & $68.5/83$\\
\hline
			$10$ &  $-0.565$ & $0.031\pm 0.003$ &
                        $0.034\pm0.002$  & $136/111$\\ 
\hline
			$7.5$ &  $-0.584$ & $0.044\pm 0.002$ &
                        $0.030\pm 0.001$  & $344/126$\\
\hline
			$5$  & $-0.608$ & $0.057\pm 0.001$ &
                        $0.0269\pm 0.0006$  & $2769/101$\\
			\hline
		\end{tabular}
	\end{center}
	\caption{Parameters of the log-polynomial fit \eqref{Aprfit} to the
          one-loop aspherical prefactor for different cell
          radii. $a_0=1$ for all cases, whereas $a_1$ is 
computed using \eqref{fitconstr}. Only $a_2$ and $a_3$ are 
fitted from the N-body data. The last column gives the $\chi^2$
per number of degrees of freedom.}
		\label{tab:Apref}
\end{table}

The remaining parameters $a_2$, $a_3$ are free and we fit them to the
aspherical prefactor extracted from simulations using the procedure
described in Sec.~\ref{sec:result}. We choose these parameters to be
redshift-independent, but allow them to vary with the cell radius
$r_*$.  
The residuals of the fit are shown in 
Fig.~\ref{fig:fitres}. The fit is quite good for the two larger radii 
$r_*=15\,\Mpch$, and $10\,\Mpch$. It correctly captures the behavior of the
prefactor near the origin, whereas there are some slight discrepancies
in the tail. In particular, for $r_*=10\,\Mpch$ the fit is in
$1.9\sigma$ tension with the data 
as follows from the $\chi^2$-value listed in the
Table~\ref{tab:Apref}.

\begin{figure}
	\centering
	\includegraphics[width=0.48\linewidth]{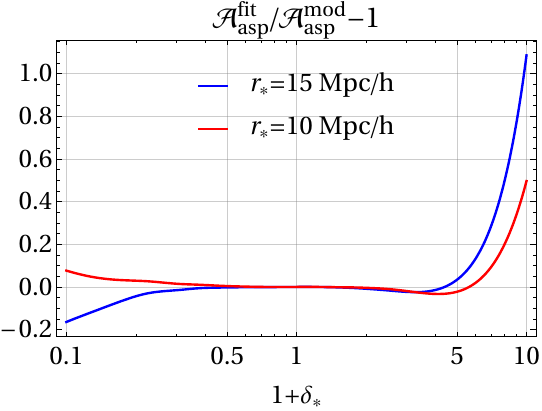}s
	\caption{The difference between the log-polynomial fit
          (\ref{Aprfit}) 
and the full theoretical model for the aspherical prefactor 
          for cell radii $r_*=15\,\Mpch$ and $10\,\Mpch$ at
          $z=0$. Note that for $r_*=15\,\Mpch$ the N-body data 
          have large uncertainties at $1+\delta_*\lesssim 0.2$ and
 $1+\delta_*\gtrsim 5$, which makes the fit at
 these densities poorly constrained.} 
	\label{fig:Apref2}
\end{figure}

We note that the performance of the log-polynomial fit is somewhat worse than
of the theoretical PDF model developed in the main part of the
paper. In this regard, it is instructive to compare the model and the
fit directly. We do it in Fig~\ref{fig:Apref2} where we plot the
difference between the fit (\ref{Aprfit}) to the aspherical prefactor
and the theoretically modeled value. To compute the latter we use
the counterterm parameters listed in
Table~\ref{tab:large}. We see that the fit is close to the
theoretical model in
the underdense region and at moderate overdensities. However, at large
overdensities $\delta_*\gtrsim 5$ the deviation becomes significant and
reaches $40\%$ for $r_*=10\,\Mpch$. It is even larger for
$r_*=15\,\Mpch$, but here it can be partially attributed to the fact
that the N-body data used in the fit are not constraining the tails
of the distribution.

For smaller radii $r_*=7.5\,\Mpch$ and $r_*=5\,\Mpch$ the
log-polynomial fit
deteriorates. It is unable to reproduce the slope of the prefactor at
the origin and the discrepancies at the overdense tail reach $\sim
30\%$. The values of $\chi^2$ are very poor, see
Table~\ref{tab:Apref}. In principle, the fit can be improved by
letting the parameters $a_0$, $a_1$ to be free and by allowing all parameters to
vary with the redshift. This, however, obviously reduces its
constraining power.

As
discussed in Sec.~\ref{sec:result2}, for small 
radii even the theoretical
model fails to precisely reproduce the data due to 2-loop corrections.
The crucial advantage of the theoretical model, however, is that using
it we can understand the cause of the failure and develope
well-defined strategies to systematically improve the accuracy.

\section{Dynamical equations}
\label{app:Apr}

We describe dark matter and baryons as a single self-gravitating
fluid. In this appendix we summarize the equations for the system
assuming the fluid to be pressureless. The effective stress tensor
arising from short modes which experience shell crossing and are
not captured by the fluid picture is considered in
Sec.~\ref{sec:ctr}. 

The standard Euler--Poisson equations read,
\bseq
\label{EP}
\begin{align}
\label{EP1}
&\frac{\d\delta}{\d t}+\d_i\big((1+\delta)u_i\big)=0\;,\\
\label{EP2}
&\frac{\d u_i}{\d t}+{\cal H}u_i+(u_j\d_j) u_i+\d_i \Phi=0\;,\\
\label{EP3}
&\Delta \Phi=\frac{3{\cal H}^2}{2}\Omega_{m}(t)\,\delta\;,
\end{align}
\eseq
where $t$ is the conformal time, ${\cal H}$ is the conformal Hubble
parameter, and $\Omega_m(t)$ is the time-dependent matter
fraction in the energy budget of the universe.
We neglect vorticity and write the peculiar velocity in
terms of scalar potentials,
\be
\label{upot}
u_i=-\H \d_i\Psi~,~~~~~\d_iu_i=-\H\Theta~,~~~~~
\Delta\Psi=\Theta\;.
\ee
Taking divergence of the Euler Eq.~(\ref{EP2}) and eliminating
$\Delta\Phi$ with (\ref{EP3}) we get an evolution equation for
$\Theta$ containing $\Psi$ and $\delta$.  
Assuming EdS cosmology, we set $\Omega_{m}(t)=1$ and switch to the time variable
$\eta$ introduced in (\ref{eta}). 
We now discuss the reduction
of the general system (\ref{EP}) for the cases of the background
spherical collapse solution, linear perturbations and second-order
perturbations.

\subsection{Background evolution}
\label{app:Apr3}

The saddle-point solution is spherically symmetric. We characterize it
by its nonlinear profile in the Eulerian space
$\big(\hat\delta(\eta,r),\hat\Theta(\eta,r),\hat\Psi(\eta,r)\big)$. From
(\ref{EP}) we obtain that it obeys the equations,
\bseq
\label{Beqs}
\begin{align}
\label{Beqs1}
&\d_\eta\hat\delta-\d_r\hat\Psi\d_r\hat\delta-(1+\hat\delta)\hat\Theta=0\;,\\
\label{Beqs2}
&\d_\eta\hat\Theta+\frac{1}{2}\hat\Theta-\frac{3}{2}\hat\delta
-\d_r\hat\Psi\d_r\hat\Theta-\hat\Theta^2+\frac{2}{r^2}(\d_r\hat\Psi)^2
+\frac{4}{r}\d_r^2\hat\Psi\d_r\hat\Psi=0\;,\\
\label{Beqs3}
&\d_r(r^2\d_r\hat\Psi)=r^2\hat\Theta\;.
\end{align}
\eseq
This system can be solved by the standard methods performing the map
to the Lagrangian space which leads to the relations (\ref{sphmap})
between nonlinearly and linearly evolved densities. The explicit
expressions for various quantities characterizing the saddle-point
background profile can be found in \cite{Ivanov:2018lcg}.

In this paper we often use the trajectory $\hat r(\eta)$ of the shell
of particles which arrive at $r_*$ by $\eta=0$. We can think of this
shell as the moving boundary of the cell which, in the absence of
shell crossing, contains a constant mass of matter. This implies two
relations,
\be
\label{rmuhat}
\dot{\hat r}=-\d_r\hat \Psi(\eta,\hat r)\;,\qquad\qquad
\dot{\hat\mu}=0\;,
\ee
where
\be
\label{muhatdef}
\hat\mu(\eta)\equiv\int_0^{\hat r(\eta)}r^2dr\,\big(1+\hat\delta(\eta,r)\big)\;,
\ee
and dot stands for (total) derivative w.r.t. $\eta$. The first
relation further gives,
\be
\label{rhatddot}
\ddot{\hat
  r}=-\d_\eta\d_r\hat\Psi+\d_r\hat\Psi\d_r^2\hat\Psi\Big|_{\eta,\hat r}\;.
\ee
It is useful to obtain the equation for $\hat r$ in closed form. To
this end, we multiply Eq.~(\ref{Beqs2}) by $r^2$ and integrate it from
$0$ to $\hat r$. After several simplifications using Eq.~(\ref{Beqs3})
we obtain,
\be
\label{reqint}
\hat r^2\bigg(\d_\eta\d_r\hat\Psi-\d_r\hat\Psi\d_r^2\hat\Psi
+\frac{1}{2}\d_r\hat\Psi\bigg)\bigg|_{\eta,\hat r}
-\frac{3}{2}\bigg(\hat\mu-\frac{\hat r^3}{3}\bigg)=0\;.
\ee
In the brackets we recognize the first and second time-derivatives of
$\hat r$. Thus, we arrive at the equation,
\be
\label{rhateq}
\ddot{\hat r}+\frac{\dot{\hat r}}{2}-\frac{\hat
  r}{2}+\frac{3\hat\mu}{2\hat r^2}=0\;.
\ee
This is nothing but the Newton's law for a particle moving together
with the cell boundary. The trajectory interpolates between 
$\hat r(-\infty)=R_*$ and $\hat r(0)=r_*$.

\subsection{Linearized perturbations with $\ell>0$}
\label{app:Apr1}

Next we consider the evolution of linearized aspherical fluctuations
in the background of saddle point solution. 
We decompose all quantities into background and first-order aspherical
perturbations, $\delta=\hat\delta+\delta^{(1)}$, etc., and expand
the latter in spherical harmonics,
\be
\label{sphexp}
\delta^{(1)}({\bf x})=\sum_{\ell\geq 1}\sum_{m=-\ell}^\ell 
(-i)^\ell\, Y_{\ell m}({\bf x}/r)\,\delta^{(1)}_{\ell m}(r)\;,
\ee
In what follows we will omit the azimuthal ``quantum number'' $m$ since it does not appear explicitly in the equations.
To linear order in perturbations, the Euler--Poisson system written in
terms of the velocity potential takes the following form
\bseq
\label{EPppp}
\begin{align}
\label{EPppp1}
&\d_\eta{\delta}_\ell^{(1)}-\Theta_\ell^{(1)}-\d_r\hat\Psi\,\d_r\delta_\ell^{(1)}
-\hat\Theta\,\delta_\ell^{(1)}-\d_r\hat\delta\,\d_r\Psi_\ell^{(1)}
-\hat\delta\,\Theta_\ell^{(1)}=0\;,\\
\label{EPppp2}
&\d_\eta{\Theta}_\ell^{(1)}+\frac{1}{2}\Theta_\ell^{(1)}-\frac{3}{2}\delta_\ell^{(1)}
-\d_r\hat\Psi\,\d_r\Theta_\ell^{(1)}
-\d_r\hat\Theta\,\d_r\Psi_\ell^{(1)}
-2\d_r^2\hat\Psi\,\Theta_\ell^{(1)}\notag\\
&\qquad\qquad\qquad+2\bigg(\d_r^2\hat\Psi-\frac{\d_r\hat\Psi}{r}\bigg)
\,\bigg(\frac{2}{r}\d_r\Psi_\ell^{(1)}-\frac{\ell(\ell+1)}{r^2}\Psi_\ell^{(1)}\bigg)=0\;,\\
\label{EPppp3}
&\d_r^2\Psi_\ell^{(1)}+\frac{2}{r}\d_r\Psi_\ell^{(1)}
-\frac{\ell(\ell+1)}{r^2}\Psi_\ell^{(1)}=\Theta_\ell^{(1)}\;.
\end{align}
\eseq
This is a system of (1+1)-dimensional partial
differential equations with respect to
$(\delta_\ell^{(1)},\Theta_\ell^{(1)},\Psi_\ell^{(1)})$.  

In setting up the initial conditions we proceed as follows. At early
times the saddle-point background vanishes and the solution of the
above equations reduces to 
\[
\delta_\ell^{(1)}(\eta,r)\to \e^\eta\, \delta_{L,\,\ell}^{(1)}(r)\;,
\]
where $\delta_{L,\,\ell}^{(1)}$ is a fluctuation of the linear density
field. We need an orthogonal basis for such functions on the half-line
which are properly normalized with respect to the radial integration measure,
\be
\label{normlmk}
\int_0^\infty dr\,r^2\,\delta_{L,\,\ell}^{(1)}(r;k)\,\delta_{L,\,\ell}^{(1)}(r;k')
=(2\pi)^3k^{-2}\delta_D(k-k')\;.
\ee
The expression on the r.h.s. is the radial delta-function compatible with the mo\-men\-tum-space measure $\int [dk]$, Eq.~(\ref{kintradial}). 
A convenient basis with this property is provided by the spherical Bessel functions,
\be
\label{linit1}
\delta^{(1)}_{L,\ell}(r;k)=4\pi\,j_\ell(kr)\;.
\ee
The velocity potentials are expanded in spherical harmonics
in the same way as the density, and their initial conditions follow
from the adiabaticity of the linear mode,
\be
\label{linit2}
\Theta_{L,\,\ell}^{(1)}(r;k)=\delta_{L,\,\ell}^{(1)}(r;k)~,~~~~~
\Psi_{L,\,\ell}^{(1)}(r;k)=-\delta_{L,\,\ell}^{(1)}(r;k)/k^2\;.
\ee  
In setting up the initial condition for $\Psi_{L,\,\ell}^{(1)}$ we have used that Bessel functions are eigenstates of the radial part of the Laplace operator, see Eq. \eqref{Besseleq}.

\subsection{Quadratic perturbations in the monopole sector}
\label{app:Apr2}

Here we analyze the evolution of 
second-order monopole perturbation $\delta^{(2)}_0(\eta,r;k_I,k_J)$ induced by
two aspherical fluctuations of a given multipole $\ell$ and momenta $k_I$, $k_J$.
Expanding the Euler--Poisson equations to the quadratic order and
averaging over the angles one obtains 
\bseq
\label{EP2nd}
\begin{align}
\label{EP2nd1}
&\dot\delta_0^{(2)}-\Theta_0^{(2)}-\d_r\hat\Psi\,\d_r\delta_0^{(2)}
-\hat\Theta\,\delta_0^{(2)}-\d_r\hat\delta\,\d_r\Psi_0^{(2)}
-\hat\delta\,\Theta_0^{(2)}=\Xi_{\delta}\;,\\
\label{EP2nd2}
&\dot{\Theta}_0^{(2)}+\frac{1}{2}\Theta_0^{(2)}-\frac{3}{2}\delta_0^{(2)}
-\!\d_r\hat\Psi\,\d_r\Theta_0^{(2)}
-\!\d_r\hat\Theta\,\d_r\Psi_0^{(2)}\\
\nonumber
& ~~~~~~~~~~~~~~~~~~
-2\d_r^2\hat\Psi\,\Theta_0^{(2)}
+\frac{4}{r}\bigg(\d_r^2\hat\Psi-\frac{\d_r\hat\Psi}{r}\bigg)
\,\d_r\Psi_0^{(2)}=\Xi_{\Theta}\,,\\
\label{EP2nd3}
&\d_r^2\Psi_0^{(2)}+\frac{2}{r}\d_r\Psi_0^{(2)}=\Theta_0^{(2)}\;,
\end{align}
\eseq
where 
\be
\label{XiUps}
\Xi_\delta=\frac{1}{r^2}\d_r(r^2\Upsilon_{\delta,IJ})~,\qquad
\Xi_\Theta=\frac{1}{r^2}\d_r(r^2\Upsilon_{\Theta,IJ})\;,
\ee
and $\Upsilon_\delta$ and $\Upsilon_\Theta$ are defined as
\bseq
\label{Ups}
\begin{align}
\label{Upsd}
\Upsilon_{\delta,IJ}(\eta,r)=&\frac{1}{8\pi}\,
\delta^{(1)}_{\ell,I}\d_r\Psi^{(1)}_{\ell,J}
+(I\leftrightarrow J)
\;,\\
\Upsilon_{\Theta,IJ}(\eta,r)
=&\frac{1}{8\pi}\,\bigg[\Theta_{\ell,I}^{(1)}\d_r\Psi_{\ell,J}^{(1)}
-\frac{2}{r}\d_r\Psi_{\ell,I}^{(1)}\d_r\Psi_{\ell,J}^{(1)}
+\frac{2\ell(\ell+1)}{r^2}\Psi_{\ell,I}^{(1)}\d_r\Psi_{\ell,J}^{(1)}\notag\\
&\qquad-\frac{\ell(\ell+1)}{r^3}\Psi_{\ell,I}^{(1)}\Psi_{\ell,J}^{(1)}\bigg]
+(I\leftrightarrow J)\;.
\label{UpsT}
\end{align}
\eseq
To analyze this system, 
it is convenient to switch to the
total monopole density combining the background and perturbations, 
$\delta_0=\hat\delta+\delta^{(2)}_0$, and similarly for $\Theta_0$ and
$\Psi_0$. Summing (\ref{EP2nd}) with the background Eqs.~(\ref{Beqs})
we find the system for $(\delta_0,\Theta_0,\Psi_0)$ which has the same
form as (\ref{Beqs}), but with the sources $\Xi_\delta$, $\Xi_\Theta$
in the continuity and Euler equations, respectively.  

The rest of the analysis proceeds similarly to
appendix~\ref{app:Apr3}. We introduce the moving radius of the cell
$r_{\rm cell}(\eta;k_I,k_J)$ which obeys
\be
\label{RPsi}
\dot r_{\rm cell}=-\d_r\Psi_0(\eta, r_{\rm cell})\;
\ee 
and arrives at $r_{\rm cell}=r_*$ at $\eta=0$. Then we multiply the
continuity equation by $r^2$ and integrate from $0$ to $r_{\rm
  cell}(\eta)$. This gives
\be
\label{masschange1}
\dot \mu_{\rm cell}
=r^2_{\rm cell}\Upsilon_{\delta,IJ}(\eta,r_{\rm cell})\;,
\ee
where
\be
\label{massdef}
\mu_{\rm cell}(\eta)\equiv
\int_0^{r_{\rm cell}(\eta)}r^2dr\;\big(1+\delta_0(\eta,r)\big)
\ee
is the mass inside the cell. We see that it is no longer constant due
to perturbations. Finally, from the Euler equation we obtain the
equation of motion for the cell boundary, similar to (\ref{rhateq}),
but now with a non-zero r.h.s.,
\be
\label{Rchange}
\ddot r_{\rm cell}+\frac{\dot r_{\rm cell}}{2}
-\frac{r_{\rm cell}}{2}+\frac{3\mu_{\rm cell}}{2r_{\rm cell}^2}
=-\Upsilon_{\Theta,IJ}(\eta,r_{\rm cell})\;.
\ee
The last step is to decompose the cell radius and the mass back into
the background values and perturbations,
\bseq
\begin{align}
\label{R2def}
&r_{\rm cell}=\hat r+r^{(2)}_{IJ}\;,\\
\label{massdecomp}
&\mu_{\rm cell}=\hat\mu+A(\eta)r^{(2)}_{IJ}+\mu^{(2)}_{IJ}\;,
\end{align}
\eseq
where $A(\eta)$ is defined as
\be
\label{A}
A(\eta)=\hat r^2(\eta)\big[1+\hat\delta\big(\eta,\hat
r(\eta)\big)\big]\;,
\ee
and
\be
\label{mass*def}
\mu^{(2)}_{IJ}(\eta)\equiv\int_0^{\hat r(\eta)}r^2
dr\,\delta_0^{(2)}(\eta,r;k_I,k_J)\;. 
\ee
Linearizing Eqs.~(\ref{masschange1}), (\ref{Rchange}) with respect to
$\mu^{(2)}_{IJ}$, $r^{(2)}_{IJ}$ and using that $\hat \mu$, $\hat r$ obey
the background equations, we arrive at two linear equations,
\bseq
\label{mu2r2}
\begin{align}
	\label{mu2r21}
	&\dot \mu^{(2)}+A(\eta)\,\dot r^{(2)}
	+\frac{dA}{d\eta}\,r^{(2)} 
	=\hat r^2\Upsilon_\delta(\eta,\hat r)\;\\
	\label{mu2r22}
	&\ddot r^{(2)}+\frac{1}{2}\dot r^{(2)}
	+B(\eta)\, r^{(2)}
	+\frac{3}{2\hat r^2}\mu^{(2)}
	=-\Upsilon_\Theta(\eta,\hat r)\;,
\end{align}
\eseq
where we have omitted the indices $I,J$ to simplify notations and $B(\eta)$ is defined as
\be
\label{B}
B(\eta)=1+\frac{3}{2}\hat\delta\big(\eta,\hat r(\eta)\big)
	-\frac{R_*^3}{\hat r^3(\eta)}\;.
\ee
Note that the the sources on the r.h.s. of 
Eqs.~\eqref{mu2r2} are evaluated at the unperturbed cell boundary $(\eta,\hat
r(\eta))$.
The coefficient functions $A(\eta)$ and $B(\eta)$ depend only on the local value of the background density contrast on the same boundary.

The equations (\ref{mu2r2}) must be solved with the final condition
$r^{(2)}_{IJ}(0)=0$ and the initial conditions
\be
\label{init}
\mu^{(2)}_{IJ}\propto \e^{2\eta}~,\qquad 
\dot r^{(2)}_{IJ}
+\frac{d\ln A}{d\eta}\,r^{(2)}_{IJ} \propto \e^{2\eta}~,\qquad
\text{at}~\eta\to-\infty\;. 
\ee
The first condition expresses the fact that the $\mu^{(2)}_{IJ}$ is proportional to the
second-order density contrast $\delta_0^{(2)}$ by definition \eqref{mass*def}.
The latter is required for the consistency of Eq. \eqref{mu2r21}.

\subsection{Redefining sources for second-order perturbations}
\label{app:Apr4}

Here we show that the source $\Upsilon_\delta$
in 
the perturbed continuity equation (\ref{mu2r21}) can be set to zero
by a proper redefinition of the moving cell radius. 
This is consistent with the EFT of LSS where 
only the Euler equation
is modified by the introduction of the effective stress tensor,
whereas the continuity equations remains unchanged.

Let us redefine the cell radius by replacing 
\be
\label{epsilon0}
r^{(2)}_{IJ}(\eta)\mapsto r^{(2)}_{IJ}(\eta)+\epsilon_{IJ}(\eta)\;,
\ee
where $\epsilon_{IJ}(\eta)$ is chosen to obey the equation
\be
\label{epsilon}
	A(\eta)\,\dot \epsilon_{IJ}
	+\frac{dA}{d\eta}\,\epsilon_{IJ}
	=\hat r^2\Upsilon_{\delta,IJ}(\eta,\hat r)\;.
\ee
Physically, this corresponds to choosing the moving cell boundary in
such a way that the mass inside the cell --- including background and
perturbations --- stays constant. 
As a result, the r.h.s. of Eq.~(\ref{mu2r21}) vanishes, $\widetilde \Upsilon_\delta=0$.

Substituting (\ref{epsilon0}),
(\ref{epsilon}) into the Euler equation (\ref{mu2r22}) we obtain,
\be
\label{r2new}
	\ddot r^{(2)}_{IJ}+\frac{1}{2}\dot r^{(2)}_{IJ}
	+B(\eta)\, r^{(2)}_{IJ}
	+\frac{3}{2\hat r^2}\mu^{(2)}_{IJ}+B_1(\eta)\,\epsilon_{IJ}
	=-\widetilde\Upsilon_{\Theta,IJ}(\eta,\hat r)\;,
\ee
where
\bseq
\begin{align}
\label{B1}
&B_1=\l\frac{d\ln A}{d\eta}\r^2-\frac{d^2\ln A}{d\eta^2}
-\frac{1}{2}\frac{d\ln A}{d\eta}+B\;,\\
\label{SourceTheta}
&\widetilde\Upsilon_\Theta=\Upsilon_\Theta
+\l\frac{1}{2}-\frac{d\ln A}{d\eta}\r\frac{\Upsilon_\delta}{1+\hat\delta}
+\frac{d}{d\eta}\l\frac{\Upsilon_\delta}{1+\hat\delta}\r\;. 
\end{align}
\eseq
Note that in the latter expression the background density $\hat\delta$
is evaluated at the unperturbed cell boundary $(\eta,\hat
r(\eta))$. 
Let us show the function $B_1(\eta)$ vanishes
on the background equations of motion.
Recalling the definition (\ref{A}) we have,
\be
\label{etalog}
\frac{d\ln A}{d\eta}=\frac{2\dot{\hat r}}{\hat
  r}+\frac{\d_\eta\hat\delta+\dot{\hat
    r}\d_r\hat\delta}{1+\hat\delta}\bigg|_{\eta,\hat r}
=\d_r^2\hat\Psi(\eta,\hat r)\;,
\ee
where in the second equality we have used Eqs.~(\ref{rmuhat}),
(\ref{Beqs1}), (\ref{Beqs3}). Similarly, from Eqs.~(\ref{rhateq}),
(\ref{Beqs2}), (\ref{Beqs3}) we find,
\be
\label{dr2psi}
\frac{d^2\ln A}{d\eta^2}=
\big(\d_r^2\hat\Psi\big)^2-\frac{1}{2}\d_r^2\hat\Psi
+1+\frac{3}{2}\hat\delta-\frac{R_*^3}{\hat r^3}\bigg|_{\eta,\hat r}\;.
\ee
Collecting these expressions along with the definition \eqref{B} indeed yields $B_1(\eta)=0$.
Thus, the Euler equation takes the form (\ref{mu2r22}) with the new source 
$\widetilde\Upsilon_\Theta$ given by
Eq.~(\ref{SourceTheta}).  

Let us now simplify $\widetilde\Upsilon_\Theta$.
Using the equations obeyed by the first order perturbations and the
background, we can get rid of explicit time derivatives in
$\widetilde\Upsilon_\Theta$. 
Starting from $\Upsilon_\delta$ defined in
Eq.~(\ref{Upsd}) and using Eqs.~(\ref{EPppp}), we have
\[
\begin{split}
\frac{d}{d\eta}\bigg(\frac{\Upsilon_\delta}{1+\hat\delta}\bigg)
=\frac{1}{4\pi(1+\hat\delta)}\bigg[&(\d_\eta\delta_\ell^{(1)}-
\d_r\hat\Psi\d_r\delta_\ell^{(1)})\d_r\Psi_\ell^{(1)}
+\delta_\ell^{(1)}(\d_\eta\d_r\Psi_\ell^{(1)}-\d_r\hat\Psi\d_r^2\Psi_\ell^{(1)})
\\&-\frac{\d_\eta\hat\delta-\d_r\hat\Psi\d_r\hat\delta}{1+\hat\delta}
\delta_\ell^{(1)}\d_r\Psi_\ell^{(1)}\bigg]\;,
\end{split}
\]
where we have suppressed the indices $I$, $J$ to avoid clutter of
notations. We now use Eqs.~(\ref{EPppp1}), (\ref{Beqs1}) to simplify,
\[
\frac{d}{d\eta}\bigg(\frac{\Upsilon_\delta}{1+\hat\delta}\bigg)
=\frac{1}{4\pi}\bigg[\Theta_\ell^{(1)}\d_r\Psi_\ell^{(1)}
+\frac{\d_r\hat\delta}{1+\hat\delta}(\d_r\Psi_\ell^{(1)})^2
+\delta_\ell^{(1)}(\d_\eta\d_r\Psi_\ell^{(1)}-\d_r\hat\Psi\d_r^2\Psi_\ell^{(1)})\bigg]\;.
\]
Finally, the radial part of the linearized Eq.~(\ref{EP2}) before
elimination of the Newton potential reads,
\be
\label{PsiPhi}
-\d_\eta\d_r\Psi_\ell^{(1)}-\frac{\d_r\Psi_\ell^{(1)}}{2}
+\d_r\hat\Psi\d_r^2\Psi_\ell^{(1)}+\d_r^2\hat\Psi\d_r\Psi_\ell^{(1)}
+\frac{\d_r\Phi_\ell^{(1)}}{\H^2}=0\;.
\ee
We collect these expressions into Eq.~(\ref{SourceTheta}) and
eliminate $\Theta_\ell^{(1)}$, $\delta_\ell^{(1)}$ using
(\ref{EPppp3}) and the Poisson Eq.~(\ref{EP3}). 
It is convenient to cast the result into the
following suggestive form,
\be
\label{Ups2}
\widetilde\Upsilon_{\Theta,IJ}(\eta)=\frac{1}{\H^2(1+\hat\delta)}
\bigg[
\d_r\tau_{\paral,IJ}
+\frac{2}{r}\tau_{\paral,IJ}-\frac{2}{r}\tau_{\perp,IJ}
\bigg]
\bigg|_{\eta,\hat r(\eta)}\;,
\ee
where 
\bseq
\label{Ssigmas}
\begin{gather}
\label{sigmas}
\tau_{\paral,IJ}=\frac{\H^2}{4\pi}(1+\hat\delta)\d_r\Psi^{(1)}_{\ell,I}
\d_r\Psi^{(1)}_{\ell,J}+
\frac{1}{12\pi \H^2}
\bigg[\d_r\Phi^{(1)}_{\ell,I}
\d_r\Phi^{(1)}_{\ell,J}-\frac{\ell(\ell+1)}{r^2}
\Phi^{(1)}_{\ell,I}
\Phi^{(1)}_{\ell,J}\bigg]\;,\\
\tau_{\perp,IJ}=\frac{\H^2\ell(\ell+1)}{8\pi r^2}(1+\hat\delta)
\Psi^{(1)}_{\ell,I}
\Psi^{(1)}_{\ell,J}
-\frac{1}{12\pi \H^2}\d_r\Phi^{(1)}_{\ell,I}
\d_r\Phi^{(1)}_{\ell,J}\;.
\end{gather}
\eseq
In Sec.~\ref{sec:couterterm} we interpret $\tau_{\paral,\perp}$ as the components
of the effective stress produced by the short modes.

\section{WKB expansion for UV modes}
\label{app:WKB}

\subsection{Equations and sources}
\label{app:WKB1}

The initial conditions (\ref{linit1}) for linear aspherical
perturbations 
are provided by the spherical Bessel functions which are 
exponentially suppressed at $kr<(\ell+1/2)$ for high multipoles $\ell\gg 1$. This implies that the
perturbations with $kr_*\ll \ell$ do not contribute into the
variation of the density inside the cell and the corresponding elements
of the response matrix are suppressed, 
$Q_\ell(k,k')\approx0$ for $k$ or
$k'$ much smaller than $\ell/r_*$. The dominant
contribution into the response matrix then comes from the modes with 
\be
\label{klineq}
k\gtrsim\ell/r_*\gg1/r_*\;.
\ee
These modes oscillate much faster than the background solution, so one
can apply the WKB approximation to find their evolution.  

Using the momentum as a large parameter, we write the leading WKB expressions,
\bseq
\label{WKBform}
\begin{align}
\label{WKBform1}
	&\delta_\ell^{(1)}=(\ell+1/2)^{-1}\,\updelta_{\ell}\,\e^{ikS_\ell-i\pi/4}+{\rm
          h.c.}~,~~\\  
\label{WKBform2}
	&\Theta_\ell^{(1)}=(\ell+1/2)^{-1}\,\upvartheta_{\ell
        }\,\e^{ikS_\ell-i\pi/4}+{\rm h.c.}~,~~\\  
\label{WKBform3}
	&\Psi_\ell^{(1)}=\!(\ell+1/2)^{-1}k^{-2}\,\uppsi_{\ell }\,
       \e^{ikS_\ell-i\pi/4}+{\rm h.c.}~,~~\\ 
	&\Phi_\ell^{(1)}=\!\tfrac{3}{2}\H^2
(\ell+1/2)^{-1}k^{-2}\,\upvarphi_{\ell }\,\e^{ikS_\ell-i\pi/4}+{\rm h.c.} 
\label{WKBform4}
\end{align}
\eseq
where $\updelta_{\ell},\upvartheta_\ell$ etc. are slowly varying
functions. Note that $\Psi_\ell^{(1)}$ and $\Phi_\ell^{(1)}$ are suppressed by two
powers of $k$ compared to $\delta_\ell^{(1)}$ and
$\Theta_\ell^{(1)}$. 
For convenience, we have also inserted the overall factors
$(\ell+1/2)^{-1}$, as well as complex phases in front of various
functions.\footnote{Note that these normalization conventions differ from
  Ref.~\cite{Ivanov:2018lcg}.}  
Then the initial conditions at $\eta\to-\infty$, 
set up by the
asymptotic expansion of spherical Bessel functions at large order
\cite{Ivanov:2018lcg}, are real and depend only on the ratio $\vk$
defined in Eq.~(\ref{vk}),  
\bseq
\label{WKBinit}
\begin{align}
\label{Slinit}
&S_\ell=\frac{1}{\vk}\bigg[\sqrt{(\vk r)^2-1}-\arccos\frac{1}{\vk
  r}\bigg]\;,\\
\label{dl1init}
&\updelta_{\ell}=\upvartheta_{\ell}=-\uppsi_{\ell}=-\upvarphi_{\ell }=
\frac{2\pi}{\sqrt{\vk r}\,[(\vk r)^2-1]^{1/4}}\cdot \e^\eta\;.
\end{align}
\eseq
These expressions apply at $r>1/\vk$. At smaller radii the Bessel
functions are exponentially suppressed, so we set all coefficients in
the WKB expansion to
zero. 

From the equation \eqref{EPppp3} for velocity potential we obtain 
\bseq\label{PsiPhiLO}
\be
\label{PsiLO}
\uppsi_{\ell}=-\frac{\upvartheta_{\ell}}{(S_\ell')^2+(\vk r)^{-2}}\;,
\ee
where we have denoted with prime the radial derivative $\d_r $.
Similarly, for the gravitational potential we have, 
\be
\label{PhiLO}
\upvarphi_{\ell}=-\frac{\updelta_{\ell}}{(S_\ell')^2+(\vk r)^{-2}}\;.
\ee
\eseq
Further, we substitute the form \eqref{WKBform} into 
Eqs.~\eqref{EPppp1}, \eqref{EPppp2}. At the leading order $O(k)$ both
equations reduce to 
\be
\label{flowcons}
\frac{d S_\ell}{d\eta}\bigg|_{\rm flow}=0\;.
\ee
where we have used the derivative along the background flow lines,
\be
\label{flow}
\frac{d}{d\eta}\bigg|_{\rm
  flow}=\frac{\d}{\d\eta}-\hat\Psi'\frac{\d}{\d r}\;.
\ee
We conclude that $S_\ell$ is conserved along the flow and we can write at
all times,
\be
\label{Salleta}
S_\ell=\frac{1}{\vk}\bigg[\sqrt{(\vk R)^2-1}-\arccos\frac{1}{\vk
  R}\bigg]
\qquad\Longleftrightarrow\qquad
\frac{\d S_\ell}{\d R}=\frac{\sqrt{(\vk R)^2-1}}{\vk R}\;,
\ee
where $R$ is the Lagrangian radial coordinate labeling
the background flow lines.

From the order $O(1)$ of Eqs.~\eqref{EPppp1}, \eqref{EPppp2} we
obtain the equations for the first-order WKB coefficients 
\bseq
\label{WKBNLO}
\begin{align}
\label{WKBNLO1}
&\frac{d\updelta_{\ell}}{d\eta}\bigg|_{\rm flow}-\hat\Theta\updelta_{\ell}
-(1+\hat\delta)\upvartheta_{\ell}=0\;,
\\
\label{WKBNLO2}
&\frac{d\upvartheta_{\ell}}{d\eta}\bigg|_{\rm flow}-\frac{3}{2}\updelta_{\ell}
+\bigg[\frac{1}{2}
-\frac{2(\vk r S_\ell')^2\hat\Psi''}{1+(\vk r S_\ell')^2}
-\frac{2\hat\Psi'}{r(1+(\varkappa r S_\ell')^2)}\bigg]\upvartheta_{\ell}=0\;.
\end{align}
\eseq
We notice that Eqs.~\eqref{WKBNLO} do not contain
spatial derivatives of the functions $\updelta_{\ell}$,
$\upvartheta_{\ell}$, 
so they form a
system of ordinary differential equations for these functions along
the background flow lines.
We solve these equations numerically in the vicinity
of the flow line corresponding to the moving cell boundary $\hat
r(\eta)$, which in the Lagrangian coordinates is fixed at
$R=R_*$. 
This is sufficient for evaluating the sources
(\ref{Ups2}). 

Let us obtain the explicit expressions for the stress tensor components $\tau_{\paral,\perp}$ \eqref{Ssigmas}
in the WKB limit. 
Considering only the diagonal components with
$k_I=k_J=k$, the aggregated contribution of the UV modes
into the effective stress tensor reads
\begin{align}
\label{UpsCum}
&\sum(2\ell+1)
\int[dk]  P(k)\, \tau_\alpha(\eta,R;\ell,k)\notag\\
&=2\H^2 \int\limits_{k\gg R_*^{-1}} \frac{dk\,P(k)}{(2\pi)^3}\cdot
\int_{R^{-1}}^\infty \frac{d\vk}{\vk}\,\sum_{a=\kin,\pot}\chi_\alpha^{a}\big(\eta,R;\vk\big)\;,
\qquad \alpha=\paral,\perp\;,
\end{align}
where we have defined new kinetic and potential WKB sources $\chi_\alpha^{a}$, $a=\kin,\pot$.
Take, for example, the kinetic part of $\tau_{\paral}$ given by the first term in Eq.~(\ref{sigmas}). Using
Eq.~(\ref{WKBform3}), it takes the form
\be
\label{UpsPhiWKB}
\tau_{\paral}^\kin=
\frac{\H^2(1+\hat\delta)}{2\pi(\ell\!+\!1/2)^2
  k^2}
(S_\ell')^2\uppsi_\ell^2\big(1-\cos(2ikS_\ell-i\pi/4)\big)\;.
\ee
The second term in brackets quickly oscillates as function of $k$ and
averages away in the cumulative quantities involving integrals over
$k$. 
Neglecting it,
we arrive at the expression for the rescaled component of the stress,
\bseq
\label{chis}
\be
\label{chis1}
\chi^\kin_{\paral}=\frac{1+\hat\delta}{2\pi} (S_\ell')^2\uppsi_\ell^2\;.
\ee
Similar calculation for other components yields,
\begin{align}
\label{chis2}
&\chi^\kin_\perp=\frac{1+\hat\delta}{4\pi}\frac{\uppsi_\ell^2}{(\vk
  r)^2}\;,\\
\label{chis3}
&\chi^\pot_{\paral}=\frac{3}{8\pi} \bigg((S_\ell')^2
-\frac{1}{(\vk r)^2}\bigg)\upvarphi_\ell^2\;,\\
\label{chis4}
&\chi_\perp^\pot=-\frac{3}{8\pi}(S_\ell'^2)\upvarphi_\ell^2\;.
\end{align}
\eseq
Note that according to the initial conditions (\ref{WKBinit}), the WKB
functions entering here vanish at early times if $\vk r<1$. Since 
the WKB evolution is ultralocal along the flow, we conclude that this
property is preserved in terms of the Lagrangian radius $R$. Namely, 
the stress tensor components (\ref{chis}) are non-vanishing only if
$\vk > 1/R$, which implies the lower limit of the $\vk$-integration in Eq.~(\ref{UpsCum}).

Finally, we can collect the stress tensor components into the rescaled sources (\ref{upschi}) used in the counterterm model,  
\bseq
\label{UpsF}
\begin{align}
\label{UpsDeF}
\upsilon^\kin&=\frac{1}{2\pi}\bigg[2(S_\ell')^2\uppsi_{\ell}\uppsi_{\ell}'
+\bigg(2S_\ell'S_\ell''+\frac{2(S_\ell')^2}{r}-\frac{1}{\vk^2 r^3}
+\frac{\hat\delta'}{1+\hat\delta}
(S_\ell')^2\bigg)\uppsi_{\ell}^2
\bigg]\bigg|_{\eta,\hat r}\;,\\
\upsilon^\pot&=\frac{3}{4\pi(1+\hat{\delta})}\bigg[
\bigg((S_\ell')^2-\frac{1}{(\vk r)^{2}}\bigg)\upvarphi_{\ell}\upvarphi_{\ell}'
+\bigg(S_\ell'S_\ell''+\frac{2(S_\ell')^2}{r}\bigg)\upvarphi_{\ell}^2\Big]
\Big|_{\eta,\hat r}
\;,
\label{UpsThF}
\end{align}
\eseq

\subsection{Evaluation of the $\varkappa$-integrals}
\label{app:WKB2}

The WKB expressions for the effective stress $\chi^a_\alpha$ are
singular at $\vk R\to 1$. 
This singularity stems from the initial conditions
(\ref{dl1init}) and, since the dynamical equations (\ref{WKBNLO}) are
ultralocal, propagates to arbitrary time. It is spurious and arises from
the breakdown of the WKB approximation at the ``turning point'' $\vk
R=1$. Indeed, the standard condition for the validity of the WKB
expansion is the smallness of the second derivative of the phase,
\be
\label{WKBcond}
S_\ell''/(S_\ell')^2\ll k\;.
\ee
Substituting here the result (\ref{Salleta})
and using that $\d_r R$ is an order-one quantity, we conclude that the
WKB approximation 
requires 
\be
\label{WKBvk}
|\vk R-1| \gg (\ell+1/2)^{-2/3}\;.
\ee
Strictly speaking, we can use the WKB expressions 
only at $\vk>(1+\epsilon)R^{-1}$, where $\epsilon\gg
\ell^{-2/3}$. 

However, when computing the contribution of the high-$\ell$ modes into
the fluid prefactor and the counterterm
we need to integrate over all $\vk$ for which the
sources are non-zero. 
We can try to naively extend the WKB expressions in the integrals down
to $\epsilon=0$ hoping that the error we make in this way is
small. This indeed works for the integrals of the type appearing in the
counterterm stress tensor (\ref{eftpartau}) since the singularity in
$\chi_\alpha^a$ has the form $(\vk R-1)^{-1/2}$ and is
integrable. So, in principle, one can first compute
$\tau_\alpha^{a,\ctr}$ along several flow lines and obtain the
counterterm stress as function of $\eta$ and $r$ in the vicinity of
the cell boundary. Then the radial derivative entering the counterterm
sources (\ref{Upsctr}) can also be evaluated numerically. This method
is, however, inefficient since it requires numerical evaluation of the
$\vk$-integral at several points in $R$ and at every time slice.
Instead, we use the expression (\ref{eftpar}) for the counterterm
sources and notice that we can split it into contributions of
different $\vk$, solve Eqs.~(\ref{murctr}) for each $\vk$ separately,
and take the $\vk$-integral only once in the end. The same approach is used
to calculate the high-$\ell$ fluid prefactor (\ref{AprUV}).  

However, here we encounter a problem since the integrand contains a
singularity $(\vk R_*-1)^{-3/2}$ and a naive $\vk$-integration
diverges. This singularity can be traced to the same singularity in
$\upsilon^\pot(\eta;\vk)$ which, in turn, comes from the derivative
$\d_r\chi^\pot_{\paral}(\eta,R;\vk)$.\footnote{It is easy to see that
  $\upsilon^\kin$ does not lead to any troubles since the strongest
  singularity in it is $(\vk R_*-1)^{-1/2}$.} 
The divergence is spurious and arises because of interchanging the order of
$r$-differentiation and $\vk$-integration. To resolve it, we
have to treat this step more carefully.

In order to make the discussion general, we consider an arbitrary
function $\chi(\!R;\!\vk)$, such that 
\be
\label{chising}
\chi(R;\vk)\simeq \frac{\chi_{\frac{1}{2}}}{(\vk R-1)^{1/2}}\;,\qquad
\text{at}~\vk R\to 1\;.
\ee
We have suppressed a possible time dependence of $\chi$ since it is
irrelevant for the argument. Then we have a chain of relations,
\be
\label{qint}
\d_r\int_{R^{-1}}^\infty \frac{d\vk}{\vk}\,\chi
=\lim_{\epsilon\to 0}\d_r\int_{(1+\epsilon)R^{-1}}^\infty
\frac{d\vk}{\vk}\,\chi
=\lim_{\epsilon\to 0}
\bigg[\int_{(1+\epsilon)R^{-1}}^\infty \frac{d\vk}{\vk}\, 
\d_r\chi-\frac{2}{\sqrt\epsilon}[\d_r\chi]_{\frac{3}{2}}
\bigg]\;,
\ee
where
\be
\label{CWKB}
[\d_r\chi]_{\frac{3}{2}}=-\frac{\d_r R}{2R}\chi_{\frac{1}{2}}
=
\lim_{\epsilon\to 0} \big[\epsilon^{3/2} 
\d_r\chi\big(R;(1+\epsilon)R^{-1}\big)\big]\;
\ee
is the coefficient of the $(\vk R-1)^{-3/2}$ singularity in
$\d_r\chi$. Thus, we see that a proper treatment of the
$\vk$-integral consists in regulating it at the lower end,
substituting a divergent boundary term and restoring the lower limit
afterwards. We denote the integrals understood in the sense of this
regularization procedure with a dash-integral sign.  

In practice, both terms in (\ref{qint}) are evaluated numerically at
small, but finite $\epsilon$. The difference between the
straightforward application of this formula and the exact result is
then $O(\sqrt\epsilon)$ which can be significant. The same is true
even for convergent integrals, like those of $\upsilon^\kin$ or
$\chi^a_{\paral}$ appearing in (\ref{UpsCum}).
To improve
precision, we evaluate this $O(\sqrt\epsilon)$ correction computing
the Taylor expansion of the functions $\upsilon^a$ and $\chi^a_\alpha$ at
$\vk\to R_*^{-1}$. 

We start by introducing dimensionless variables\footnote{These variable should not be confused with the use of notations $x$, $y$ in other sections of the paper.} 
\be
\label{xy}
x=\vk R_*-1~,~~~~~ y=R/R_*-1\;,
\ee
and write
\be
\label{qfsmallx}
\upsilon(x)=\upsilon_{\frac{3}{2}}\, x^{-3/2}
+\upsilon_{\frac{1}{2}}\, x^{-1/2} +O(\sqrt{x})\;,
\ee
where $\upsilon$ stands for any of the functions $\upsilon^a$ or
$\chi^a_\alpha|_{y=0}$. Note that for all of them, except $\upsilon^\pot$, the
first term in this expansion will be zero. 
The
integral takes the form,\footnote{This formula can be easily
  generalized to the case when $\upsilon$ is multiplied by an
  arbitrary function regular at $x=0$, see \cite{Ivanov:2018lcg} for explicit
  expressions.} 
\be
\label{dashint}
\dashint_{R_*^{-1}}^\infty\frac{d\vk}{\vk}\,\upsilon
=-\frac{2\upsilon_{\frac{3}{2}}}{\sqrt{\epsilon}}+
2(\upsilon_{\frac{1}{2}}-\upsilon_{\frac{3}{2}})\sqrt{\epsilon}+
\int_\epsilon^\infty \frac{dx}{1+x}\,\upsilon(x)
+O(\epsilon^{3/2})\;.
\ee
To evaluate the counterterm prefactor, we choose a grid $x_I$,
solve the system (\ref{murctr}) separately with the sources given by
$\upsilon(x_I)$, $\upsilon_{\frac{3}{2}}$, $\upsilon_{\frac{1}{2}}$,
and combine the results at the final time slice using
Eq.~(\ref{dashint}). 

We still need to
find the coefficients $\upsilon_{\frac{3}{2}}$,
$\upsilon_{\frac{1}{2}}$. We make use of the ultralocality of the
WKB equations (\ref{WKBNLO}) which implies that in a small vicinity of
the cell boundary defined by ${y\ll x\ll 1}$ their solution has the form,  
\bseq
\label{dTl1}
\begin{align}
\updelta_{\ell}(\eta,y;x)=\big(\a_0(\eta)+\a_1(\eta)x+\a_2(\eta)y
+O(x^2,xy)\big)\,\updelta_{\ell}^{\rm sing}(y;x)\;,\\
\upvartheta_{\ell}(\eta,y;x)=\big(\b_0(\eta)+\b_1(\eta)x+\b_2(\eta)y
+O(x^2,xy)\big)\,\updelta_{\ell}^{\rm sing}(y;x)\;.
\end{align}
\eseq
Here
\be
\label{deltal10}
\begin{split}
\updelta_{\ell}^{\rm sing}(y;x)\equiv
\frac{2\pi}{\sqrt{\vk R}\,[(\vk R)^2-1]^{1/4}}\approx\frac{2\pi}{(2x)^{1/4}}
\bigg[1-\frac{5}{8}x-y\bigg(\frac{1}{4x}+\frac{23}{32}\bigg)\bigg]\;
\end{split}
\ee
encapsulates the singular initial data, whereas the 
regular evolution kernels $\alpha_0(\eta)$, $\alpha_1(\eta)$, etc. can
be found by integrating (\ref{WKBNLO}) at $|x|,|y|\ll 1$ with 
$x$- and $y$-independent initial conditions 
$\updelta_{\ell}=\upvartheta_{\ell}=\e^{\eta}$ at $\eta\to-\infty$. 
Further, we have
\be
\label{S1S2}
S_\ell'\big|_{y=0}=\frac{\d R}{\d r}\Big|_{\hat
  r}\cdot\sqrt{2x} +O(x^{3/2})\;,~~~~
S_\ell''\big|_{y=0}=\frac{1}{R_{*}}
\bigg(\frac{\d R}{\d r}\Big|_{\hat
  r}\bigg)^2\cdot\frac{1}{\sqrt{2x}} +O(x^{1/2})\;.
\ee
Using these relations and Eqs.~(\ref{PsiLO}), (\ref{PhiLO}) we find
\bseq
\label{Phil1*}
\begin{align}
\label{Phil11}
 \upvarphi_{\ell}\big|_{y=0}=&-\frac{2\pi \hat r^2}{
  R_*^2}
\cdot\frac{1}{(2x)^{1/4}}\bigg[\a_0+x\,
\bigg(\frac{11\a_0}{8}-2\a_0\bigg(\frac{\d\ln R}{\d\ln
  r}\Big|_{\hat r}\bigg)^2
+\a_1\bigg)\bigg]\;,\\
\upvarphi_{\ell}'\big|_{y=0}=&\frac{2\pi \hat r^2}{
  R_*^3}\frac{\d R}{\d r}\Big|_{\hat r}\cdot\frac{1}{(2x)^{5/4}}
\bigg[\frac{\a_0}{2}\notag\\
&\qquad+x\,\bigg(\frac{39}{16}\a_0
+3\a_0\bigg(\frac{\d\ln R}{\d\ln r}\Big|_{\hat r}\bigg)^2
-4\a_0\bigg(\frac{\d\ln R}{\d\ln r}\Big|_{\hat r}\bigg)^{-1}
+\frac{\a_1}{2}-2\a_2\bigg)\bigg]\;,
\label{Phil12}
\end{align}
\eseq
and similarly for $\uppsi_{\ell},\uppsi_{\ell}'\big|_{y=0}$ with the
replacement $\a_i\mapsto \b_i$, $i=0,1,2$. Finally, substitution into
Eqs.~(\ref{UpsF}) yields, 
\bseq
\label{allbt}
\begin{align}
&\chi^\kin_{\paral,\frac{3}{2}}= \chi^\kin_{\paral,\frac{1}{2}}= 
\chi^\kin_{\perp,\frac{3}{2}}= \chi^\pot_{\paral,\frac{3}{2}}= 
\chi^\pot_{\perp,\frac{3}{2}}=\chi^\pot_{\perp,\frac{1}{2}}=  
\upsilon^\kin_{\frac{3}{2}}=0\;,\\
&\chi^\kin_{\perp,\frac{1}{2}}
=\frac{\pi(1+\hat\delta)\hat r^2}{\sqrt{2}R_*^2}\beta_0^2\;,\\
&\chi^\pot_{\paral,\frac{1}{2}}=-\frac{3\pi\hat r^2}{2\sqrt{2}R_*^2}\alpha_0^2\;,\\
&\upsilon^\kin_{{1\over 2}}=\frac{\sqrt{2}\pi \hat r}{R_*^2}
\bigg[\bigg(\frac{\d\ln R}{\d\ln r}\Big|_{\hat r}\bigg)^3-1\bigg]\b_0^2\;,\\
&\upsilon^\pot_{{3\over 2}}=\frac{3\pi \hat r^2}{4\sqrt{2} R_*^3(1+\hat\delta)}
\frac{\d R}{\d r}\Big|_{\hat r}\a_0^2\;,\\
&\upsilon^\pot_{{1\over 2}}=\frac{3\pi \hat r^2}{\sqrt{2}R_*^3(1+\hat\delta)}
\frac{\d R}{\d r}\bigg|_{\hat r}\bigg[\a_0^2
\bigg(\frac{17}{16}\!+\!\frac{3}{2}\bigg(\frac{\d\ln R}{\d\ln
	r}\Big|_{\hat r}\bigg)^2
\!\!-\!2\bigg(\frac{\d\ln R}{\d\ln
	r}\Big|_{\hat r}\bigg)^{-1}\bigg)
\!+\!\frac{\a_0\a_1}{2}\!-\!\a_0\a_2\bigg].
\end{align}
\eseq

\section{Alternative estimates of shell-crossing scale}
\label{app:ksc}

Here we explore two alternative definitions of the shell-crossing
scale, in addition to that presented in Sec.~\ref{sec:ksc}. We start
with the linear perturbation matrix (\ref{dispmatr}). Combining
Eqs.~(\ref{traject1}) and (\ref{traject2}) we get an equation for its
general element,
\be
\label{displgen}
\frac{d\foc_i^j}{d\eta}=\d_i\d_j\d_k\hat\Psi\,x_k^{(1)}
+\d_j\d_k\hat\Psi\,\foc_i^k
-\d_i\d_k\hat\Psi\,\foc_j^k
+\d_i\d_j\Psi^{(1)}\;,
\ee
where all quantities are evaluated on the unperturbed trajectory $(\eta,\hat\x)$.
The variance of $\foc_i^j$ is dominated by short modes implying that
the derivatives acting on perturbations are enhanced compared to
derivatives acting on the background. This allows us to neglect the
first term on the r.h.s. of (\ref{displgen}). Further, by symmetrizing
the equation in the indices $i$ and $j$ we get rid of the second and
third terms. In particular, the trace over $i$ and $j$ gives
Eq.~(\ref{displ}) from the main text.

\begin{figure}
	\centering
	\includegraphics[width=0.50\linewidth]{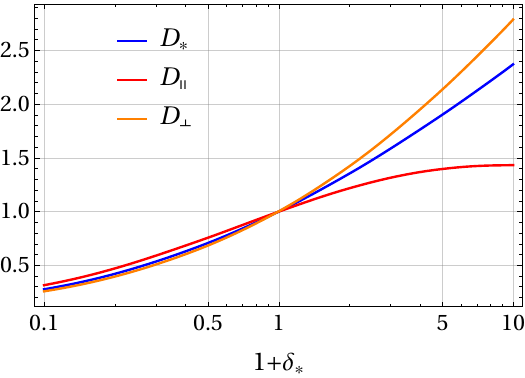}
	\caption{The radial and transverse growth factors, evaluated
          at the cell boundary, as functions
          of $\delta_*$ at $\eta=0$, together with the cumulative
          growth factor (\ref{Dstar}).}
	\label{fig:ksc2}
\end{figure}

Since the background solution is spherically symmetric, it is natural
to split $\x^{(1)}$
into the radial (parallel)
and transverse (perpendicular) parts. For the radial part we have
\be
\label{displpar}
\frac{d\foc_{\paral}}{d\eta}\equiv-\frac{d}{d\eta}\d_r r^{(1)}=\d_r^2\Psi^{(1)}\;.
\ee 
Computing its variance with the help of the WKB
expressions from appendix~\ref{app:WKB1} we obtain,
\be
\label{varpar}
\left\langle\big(\foc_{\paral}\big)^2\right\rangle_{k_{\rm max}}
\approx\frac{1}{5}\cdot 4\pi\int^{k_{\rm max}}[dk]P(k)
\cdot\big(D_{\paral}(\eta,R)\big)^2\;,
\ee
where
\be
\label{Dpar}
D_{\paral}(\eta,R)=\left(5\int_{1/R}^\infty 
\frac{d\vk}{(2\pi)^2\vk}\left|\int_{-\infty}^\eta d
\eta'\,(\d_r S_\ell)^2\,\uppsi_{\ell}\Big|_{\eta',R;\vk}\right|^2\right)^{1/2}
\ee
is the radial growth factor, which we normalized to coincide with
$\e^\eta$ at $\delta_*=0$. Similarly, for the transverse part:
\be
\label{displperp}
\frac{d\foc_\perp}{d\eta}\equiv-\frac{d}{d\eta}\d_\mu x_\mu^{(1)}=\frac{\Delta_\Omega\Psi^{(1)}}{r^2}\;,
\ee 
where $\mu=\theta,\phi$ are angular directions, and $\Delta_\Omega$
stands for the angular part of the Laplacian. The variance reads,
\be
\label{varperp}
\left\langle\big(\foc_\perp\big)^2\right\rangle_{k_{\rm max}}
\approx\frac{8}{15}\cdot 4\pi\int^{k_{\rm max}}[dk]P(k)
\cdot\big(D_\perp(\eta,R)\big)^2\;
\ee
with the transverse growth factor
\be
\label{Dperp}
D_\perp(\eta,R)=\left(\frac{15}{8}\int_{1/R}^\infty 
\frac{d\vk}{(2\pi)^2\vk^5}\left|\int_{-\infty}^\eta 
d\eta'\,\frac{\uppsi_{\ell}(\eta',R;\vk)}{r^2(\eta',R)}\right|^2\right)^{1/2}\;.
\ee
The two new growth factors, evaluated at the cell boundary, 
are compared with $D_*(\eta)$ in
Fig.~\ref{fig:ksc2}. We see that they follow the same
qualitative trend: They are larger than $1$ for overdensities and
decrease below $1$ in the underdense regions. The difference between
different growth factors is more significant at
$\delta_*>0$ with the hierarchy
\be
\label{DDs}
D_{\paral}(\eta,R_*)<D_*(\eta)<D_\perp(\eta,R_*)
\qquad\text{for}~~\delta_*>0\;.
\ee 
On the other hand, for underdensities all three growth factors almost
coincide and the ordering is reversed.

\begin{figure}[t]
	\centering
	\includegraphics[width=0.50\linewidth]{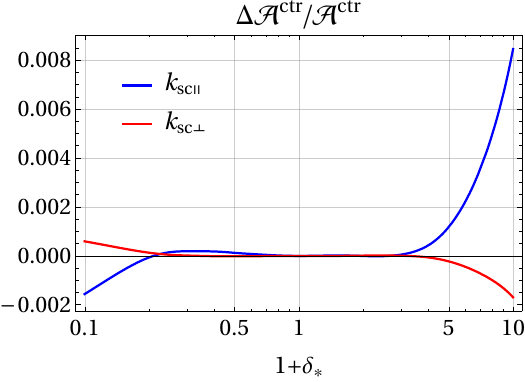}
	\caption{Difference between the counterterm prefactor computed with alternative estimates of the shell-crossing scale and 
 the baseline model. The curves are obtained from fitting
          the same N-body data as in the main text. The cell radius is
        $r_*=10\,\Mpch$ and redshift $z=0$.}
	\label{fig:TE}
\end{figure}

The new growth factors characterize shell-crossing in the radial and
transverse directions, so we introduce
\be
\label{ksc2}
k_{{\rm sc}\,\paral}\propto D_{\paral}^{-m/2}~,
~~~~~~k_{{\rm sc}\,\perp}\propto D_\perp^{-m/2}\;.
\ee
We consider these estimates as two extremes bracketing reasonable
dependences of $\ksc$ on the background and reflecting the
theoretical uncertainty of the counterterm.
To assess this uncertainty, we fit
the counterterm models with $D_*$ replaced by $D_{\paral}$ and
$D_\perp$ to the N-body data following the procedure described in
Sec.~\ref{sec:result}. The difference between the resulting
counterterms prefactors and the prefactor obtained with the baseline
model is plotted in Fig.~\ref{fig:TE} for the cell radius
$r_*=10\,\Mpch$ and redshift $z=0$. We observe
that the difference is very small and stays within $1\%$ even at the
tails. The picture for other cell radii and redshifts is similar.

\section{Sensitivity of the ``speed of sound'' to cosmology}
\label{app:gamma}

In Sec.~\ref{sec:result} we have measured 
the values of the EFT parameters in the power spectrum from the
Farpoint simulation, Eq.~(\ref{mdpl}).
They differ by some $\sim 30\%$ from the values previously reported in the
literature. While the error bars of our measurement are quite large
and encompass within $2\sigma$ the previous values,
it is worth understanding the origin of this discrepancy. We are going
to argue that it can be attributed to the difference between the
Planck cosmological parameters used in the Farpoint simulation and the
WMAP-based cosmologies used for previous measurements. For comparison
we take the cosmological parameters (cf. (\ref{Planckparam})),
\be
\label{WMAPparam}
\Omega_m=0.26~,~~~~\Omega_b=0.044~,~~~~h=0.72~,~~~~n_s=0.96~,~~~~\sigma_8=0.794\;. 
\ee 
These are the input parameters for the Horizon Run 2 (HR2) simulation
\cite{Kim:2011ab} which was used in \cite{Ivanov:2018lcg} to measure
the speed of sound 
\be
\label{gammaHR2}
\gamma_0^{\rm HR2} = (1.51\pm 0.07) (\Mpch)^2\;.
\ee
Previous estimates
\cite{Baldauf:2014qfa,Foreman:2015uva} for similar cosmology gave
the value $m\approx 8/3$ for the exponent of the power-law scaling of
$\gamma(z)$.

\begin{figure}[t]
	\centering
	\includegraphics[width=0.47\linewidth]{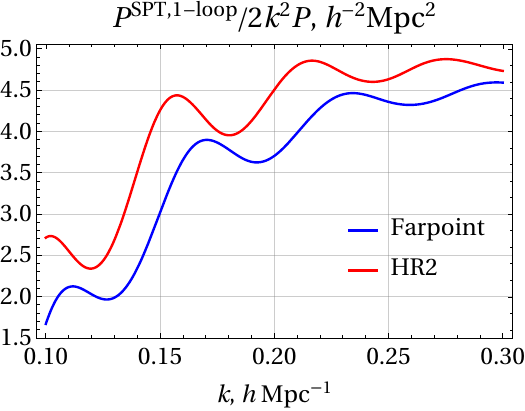}~~~~
	\includegraphics[width=0.49\linewidth]{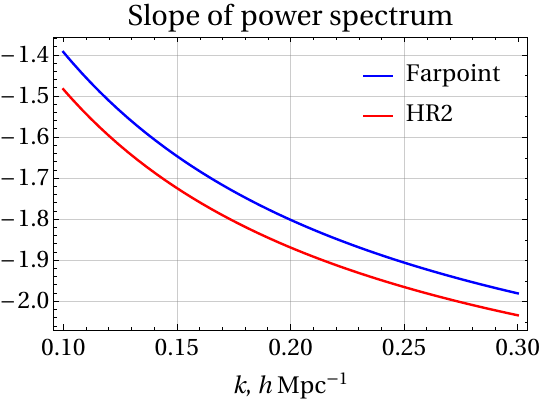}
	\caption{Ratio (\ref{ratgamma}) ({\it left}) and the 
slope of the linear power
          spectrum ({\it right}) in cosmologies used in the Farpoint and
          Horizon Run 2 simulations.
We find the slop from the broad-band part of the power
          spectrum approximated with the Eisenstein--Hu formula.} 
	\label{fig:Pk}
\end{figure}

We start by observing that the speed of sound enters into the template
for the nonlinear power
spectrum as part of the renormalized one-loop contribution,
\be
\label{PNLtemplate}
P_{\rm NL}(k)\big|_{z=0}=P(k)+P^\text{SPT,\,1-loop}(k)-2\gamma_0 k^2P(k)\;,
\ee  
where $P^\text{SPT,\,1-loop}$ is the one-loop correction computed in the
Standard Perturbation Theory (SPT) (i.e. without renormalization). Thus, we
can get a rough idea of a possible shift of $\gamma_0$ due to different
cosmologies without reference to N-body data by looking at the change in the SPT one-loop contribution. More precisely,
we want to compare the ratio 
\be
\label{ratgamma}
\frac{P^\text{SPT,\,1-loop}(k)}{2k^2 P(k)}
\ee
at $k\in [0.1\div 0.3]\, h/\Mpc$ --- the range of momenta typically
used in the $\gamma_0$ measurements.\footnote{Recall that we include
  theoretical uncertainty which makes the measurement of $\gamma_0$
  weakly sensitive to $k$ above $0.1\,h/\Mpc$.}
In the left plot of Fig.~\ref{fig:Pk} we show this ratio in the HR2
and Farpoint cosmologies. We emphasize that to calculate it we use
only the linear power spectrum for the two cosmologies which we
process with \texttt{CLASS-PT} \cite{Chudaykin:2020aoj} to evaluate
one-loop. We see that, modulo the baryon acoustic oscillations (BAO),
the ratio differs on average by $\sim 0.5 (\Mpch)^2$. This is of the
same magnitude as the change in $\gamma_0$ found from simulations. 

An alternative estimate can be obtained by considering the
UV-sensitive part of $P^\text{SPT,\,1-loop}$. In this part $\gamma_0$ is
known to renormalize the so-called velocity dispersion,
\be
\label{sigmav}
\sigma_v^2\equiv\frac{1}{6\pi^2}\int_{k_{\rm min}}^\infty dk\,P(k)\;,
\ee 
entering into the full one-loop expression in the combination 
\be
\label{combgamma}
\frac{61}{210}\sigma_v^2+\gamma_0\;.
\ee
The integral in (\ref{sigmav}) runs from $k_{\rm min}\simeq
0.1\,h/\Mpc$ to infinity. One expects the change in the two terms in (\ref{combgamma}) to
be comparable when the cosmological parameters are varied, which gives
\be
\label{gammadiff}
\gamma_0\big|_{\rm Farpoint}-\gamma_0\big|_{\rm HR2}
\sim \frac{61}{210} \Big(\sigma_v^2\big|_{\rm Farpoint}
-\sigma_v^2\big|_{\rm HR2}\Big)=0.22\;.
\ee 
This is again in the ballpark of the $\gamma_0$-shift between
cosmologies which we found from simulations.

The expected shift in $m$ can be estimated similarly. From the
scaling universe approximation we have
\be
\label{plmn}
m\sim 4/(n+3)\;,
\ee
where $n$ is the slope of the linear power spectrum. In the right
panel of Fig.~\ref{fig:Pk} we show the slope $d\ln[P(k)]/d\ln k$ in
the two compared cosmologies. To obtain it, we removed the BAO by
approximating the broad-band part of the 
linear power spectrum with the 
Eisenstein--Hu template \cite{Eisenstein:1997jh}. 
We see that the slope in Farpoint
cosmology is bigger, implying lower $m$. Substituting
the slope evaluated at $k=0.1\,h/\Mpc$ one obtains
\be
\label{mdiff}
m\big|_{\rm Farpoint}-m\big|_{\rm HR2}\sim -0.17\;.
\ee
This has the same sign, but somewhat smaller magnitude than the shift
found from the simulations. Still, it again lies in the right
ballpark. In this respect, it is worth observing that the slope of the
linear power spectrum varies significantly over the relevant range of
momenta, suggesting that the relation (\ref{plmn}) should not be trusted literally.
Rather than trying to connect $m$ to the
linear slope, it is more appropriate to think of it as an effective
power-law resulting from non-linear dynamics involving different
scales. The scaling universe intuition is still useful, but only qualitatively.

\section{Transients from Zeldovich initial conditions}
\label{app:ZA}

Zeldovich initial conditions (ZIC) are known to give rise to
long-lived transients in the N-body simulations that bias their
results \cite{Scoccimarro:1997gr,Crocce:2006ve}. In this appendix we
estimate this effect in the case of the PDF. We focus on the
exponential part of the PDF since it is expected to give the leading
contribution. We work in the EdS approximation and follow the method
of Ref.~\cite{Valageas:2001qe}.

ZIC modify the map (\ref{eq:scmap}) between the spherically symmetric
linear and non-linear density contrasts. To find this modification,
let us reconsider the equations for spherical collapse. Motion of a
spherical shell of matter is described by the energy conservation, 
\be 
\label{energy}
\frac{1}{2}\left(\frac{dy}{d\tau}\right)^2-\frac{G{\cal M}}{y}={\cal E}\,,
\ee
where $y$ is the physical radius of the shell, $\tau$ is the physical
time, ${\cal M}$ is the total mass inside the shell, and ${\cal E}$ is the total
shell's energy. We are going to see that underdense regions are
stronger affected by ZIC, so we choose ${\cal E}>0$. The solution
to this equation is written in the parametric form,
\bseq
\label{eq:ytau}
\begin{align}
\label{eq:y}
&y=\frac{G{\cal M}}{2{\cal E}}(\ch\theta-1)\;,\\
\label{eq:tau}
&\tau=\frac{G{\cal M}}{(2{\cal
    E})^{3/2}}(\sh\theta-\theta-\sh\theta_i+\theta_i)+\tau_i\;, 
\end{align}
\eseq
where $\theta_i$ is the value of the parameter $\theta$ at the initial
time $\tau_i$ corresponding to the start of the simulations. Both
$\theta_i$ and $\tau_i$ vanish if the initial conditions are set
in the exact growing mode. We are going to see that for ZIC they are
non-zero. 

Next we change the variables from $(\tau,y)$ to the scale factor $a$
and the comoving shell radius $r=y/a$. We use the relations,
\bseq
\label{aM}
\begin{align}
\label{a}
&a=\bigg(\frac{8\pi
  G}{3}\rho_ia_i^3\bigg)^{1/3}\bigg(\frac{3}{2}\tau\bigg)^{2/3}\;,\\
\label{M}
&{\cal M}=\frac{4\pi}{3}\rho_i a_i^3 R^3\;,
\end{align}
\eseq
where in the last formula we expressed the mass inside the shell
through its Lagrangian radius $R$; $\rho_i$ is the initial average
density of the universe and $a_i$ is the initial scale factor. This
casts the parametric solution into the form,
\bseq
\label{sphercol}
\begin{align}
\label{sphercol1}
&r=R\l\frac{2}{9}\r^{1/3}\frac{\ch\theta-1}{\left[\sh\theta-\theta-\sh\theta_i
+\theta_i+{\cal C}\right]^{2/3}}\;,\\
\label{sphercol2}
&a=a_i\,{\cal C}^{-2/3}
\left[\sh\theta-\theta-\sh\theta_i+\theta_i+{\cal C}\right]^{2/3}\;,
\end{align} 
\eseq
where we have introduced the notation
\be
\label{C}
{\cal C}=\tau_i\frac{(2{\cal E})^{3/2}}{G{\cal M}}=\frac{\sqrt{2}}{3}
\left[\frac{3{\cal E}}{2\pi G\rho_ia_i^2R^2}\right]^{3/2}\;.
\ee
By mass conservation, the non-linear density contrast inside the shell
is 
\be
\label{fdef}
1+\bar\delta(r,a)=\bigg(\frac{R}{r}\bigg)^{3}
=\frac{9\left(\sh\theta-\theta-\sh\theta_i
+\theta_i+{\cal C}\right)^2}{2(\ch\theta-1)^{3}}\;.
\ee
Our task now is to connect the integration constants $\theta_i$,
${\cal C}$ to the linear density field.

We recall that in Zeldovich approximation the Eulerian (${\bf x}$) and
Lagrangian (${\bf X}$) coordinates of a particle are related as 
\be
\label{xXZA}
{\bf x}={\bf X}-\nabla\Psi_L({\bf X})\;,
\ee
where $\Psi_L$ is the linear velocity potential. From the identity
$\Delta\Psi_L=\Theta_L=\delta_L$ we find for the case of spherical
symmetry
\be
\label{PsiLspher}
\d_R\Psi_L=\frac{R}{3} \bar\delta_L(R,a)\;,
\ee
where $\bar\delta_L(R,a)$ is the average {\em linear} density contrast
inside $R$ at the time set by $a$. Imposing the relation
(\ref{xXZA}) at the initial moment of the simulation we find, 
\bseq
\label{yvi}
\be
\label{yi}
r_i=R\l1-\frac{1}{3}\bar\delta_{L}(R,a_i)\r
\ee
Imposing the Zeldovich approximation on the initial velocity yields,
\be
\label{vi}
\frac{dr}{da}\bigg|_i=-\frac{R}{3}\frac{\bar\delta_{L}(R,a_i)}{a_i}\;.
\ee
\eseq
Substituting Eqs.~(\ref{sphercol}) into (\ref{yvi}) we obtain, 
\bseq
\label{thetaiC}
\begin{align}
\label{thetai}
&\ch\theta_i+1=2\l1-\frac{a_i}{3a}\bar\delta_{L}\r
\l1-\frac{2a_i}{3a}\bar\delta_{L}\r^2\;,\\
\label{Ec}
&{\cal C}=\frac{\sqrt{2}}{3}\l-\frac{10a_i}{3a}\bar\delta_{L}\r^{3/2}
\l1-\frac{8a_i}{15a}\bar\delta_{L}+\frac{4a_i^2}{45a^2}\bar\delta_{L}^2\r^{3/2}
\l1-\frac{a_i}{3a}\bar\delta_{L}\r^{-3/2}\;,
\end{align}
\eseq
where $\bar\delta_L$ is the shortcut for $\bar\delta_L(R,a)$ and we have used that linear density contrast scales linearly with the scale factor, $\bar\delta_L(R,a_i)=a_i\bar\delta_L(R,a)/a$. These
equations allow us to determine $\theta_i$ and ${\cal C}$ once
$\bar\delta_L$ is given. Equation~(\ref{sphercol2}) then expresses
$\theta$ as function of $a$,
\be
\label{thetaa}
\sh\theta-\theta=\sh\theta_i-\theta_i-{\cal
  C}+\l\frac{a}{a_i}\r^{3/2}{\cal C}\;. 
\ee
Finally, Eq.~(\ref{fdef}) gives the non-linear density. In this way,
we obtain the modified map,
\bseq
\label{ZICmaps}
\be
\label{fZIC}
\bar\delta(r,a)=f_{\rm ZIC}\big(\bar\delta_L(R,a);a_i/a\big)
\ee
and its inverse
\be
\label{FZIC}
\bar\delta_L(R,a)=F_{\rm ZIC}\big(\bar\delta(r,a);a_i/a\big)\;.
\ee
\eseq
The set of equations for overdensities is derived similarly and reads,
\bseq
\label{ZAfinalup}
\begin{align}
\label{ZAfinalup0}
&\cos\theta_i+1=2\l1-\frac{a_i}{3a}\bar\delta_{L}\r\l1-\frac{2a_i}{3a}\bar\delta_{L}\r^2\;,\\
\label{ZAfinalup2}
&{\cal C}=\frac{\sqrt{2}}{3}\l\frac{10a_i}{3a}\bar\delta_{L}\r^{3/2}
\l1-\frac{8a_i}{15a}\bar\delta_{L}+\frac{4a_i^2}{45a^2}\bar\delta_{L}^2\r^{3/2}
\l1-\frac{a_i}{3a}\bar\delta_{L}\r^{-3/2}\;,\\
\label{ZAfinalup3}
&\theta-\sin\theta=\theta_i-\sin\theta_i-{\cal
  C}+\l\frac{a}{a_i}\r^{3/2}{\cal C}\;,\\
\label{ZAfinalup4}
&1+\bar\delta(r,a)=\bigg(\frac{R}{r}\bigg)^{3}
=\frac{9\left(\theta-\sin\theta-\theta_i
+\sin\theta_i+{\cal C}\right)^2}{2(1-\cos\theta)^{3}}\;.
\end{align} 
\eseq 
Note that the maps (\ref{ZICmaps}) are redshift dependent. Note also
that in the limit $a_i\to 0$ corresponding to exactly adiabatic
initial conditions we have 
\be
\theta_i={\cal C}=0~,~~~~~\sh\theta-\theta=\frac{1}{6}\l-\frac{20}{3}\,\bar
\delta_L\r^{3/2} ~~{\rm or}~~~
\theta-\sin\theta=\frac{1}{6}\l \frac{20}{3}\,\bar
\delta_L\r^{3/2}
\ee
and we recover the usual redshift-independent EdS map (\ref{FGplusminus}), (\ref{EdSfF}). 

It is instructive to simplify the result
assuming $|a_i\bar\delta_L/a|\ll 1$. This is satisfied for the Farpoint
simulation in the considered range of densities. Focusing again on
underdensities, one can show that the combination
$(-\sh\theta_i+\theta_i+{\cal C})$ is at least of order ${\cal
  O}\big(|a_i\bar\delta_L/a|^{5/2}\big)$ and is negligible. The leading
correction then comes from the shift of ${\cal C}$,
\be
\label{Ccorr}
{\cal C}=\frac{\sqrt{2}}{3}\l-\frac{10a_i}{3a}\bar\delta_L\r^{3/2}
\l1-\frac{3a_i}{10a}\bar\delta_L\r\;, 
\ee
which substituted into (\ref{thetaa}) gives,
\be
\bar\delta_L\l1-\frac{a_i}{5a}\bar\delta_L\r=F(\bar\delta)\;.
\ee
Expressing $\bar\delta_L$, we find
\be
\label{FZICsimpl}
F_{\rm ZIC}(\bar\delta;a_i/a)=F(\bar\delta)\l1+\frac{a_i}{5a}F(\bar\delta)\r\;.
\ee
This final expression is also valid for overdensities. Note that the
transients decay slowly --- only as the first power of the inverse
scale factor. Thus, they survive even at large hierarchy between $a_i$
and $a$. Since the mapping function $F$ enters into the PDF
exponentially, even a per cent shift in it leads to large systematic
bias of the PDF. All formulas derived above can be applied to
$\Lambda$CDM cosmology upon replacing the scale factor $a$ by the
growth factor $g(z)$. 

\begin{figure}
	\centering
	\includegraphics[width=0.48\linewidth]{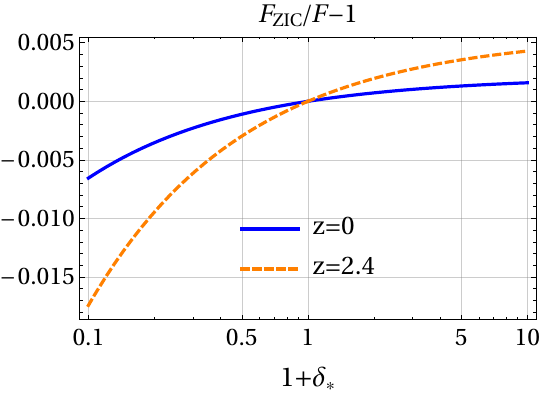}
\caption{Relative shift of the mapping function of spherical
  collapse assuming Zeldovich initial conditions (ZIC) at $z_i=200$
  with respect to the exact dynamics.}
	\label{fig:sys_map}
\end{figure}

We evaluate the shift of the spherical collapse map for the Farpoint
simulation \cite{HACC:2021sgt} which starts from ZIC at $z_i=200$. The
result is shown in Fig.~\ref{fig:sys_map}.\footnote{The plot is
  obtained using the full expressions (\ref{thetaiC}), (\ref{thetaa}),
  (\ref{fdef}), (\ref{ZAfinalup}). It is well described by the  
  approximate formula (\ref{FZICsimpl}). } 
We see that the
relative shift is larger for underdensities and reaches $0.7\%$
($1.7\%$) at $1+\delta_*=0.1$ for $z=0$ ($z=2.4$).  

We can now estimate the effect of transients on the PDF by taking the
ratio of the exponential parts with ZIC and in the exact case,
\be
\label{sys_res}
\frac{\P_{\rm ZIC}}{\P}\simeq\exp
\left\{-\frac{1}{2g^2(z)\sigma^2_{R_*}}\left[F_{\rm
      ZIC}^2-F^2\right]\right\}
\approx \exp
\left\{-\frac{g_i}{5g^3(z)\sigma^2_{R_*}}F^3(\delta_*)\right\}\;.
\ee
This factor is close to $1$ in the central region of the PDF, but
quickly deviates from unity at the tails. It is particularly strong
for underdensities where the function $F(\delta_*)$ is large in
absolute value (see the left panel of Fig.~\ref{fig:FEdS}). Also, as
expected, the effect is larger for higher redshifts. 

\begin{figure}
	\centering
	\includegraphics[width=0.48\linewidth]{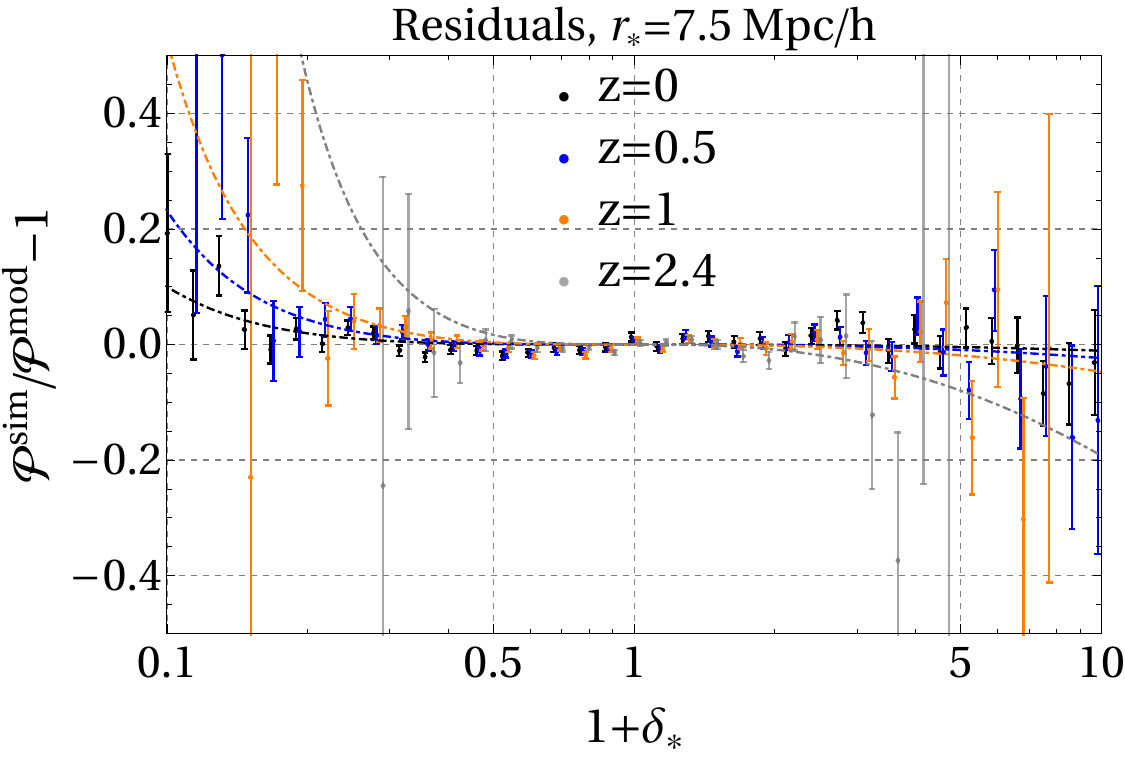}~~
	\includegraphics[width=0.48\linewidth]{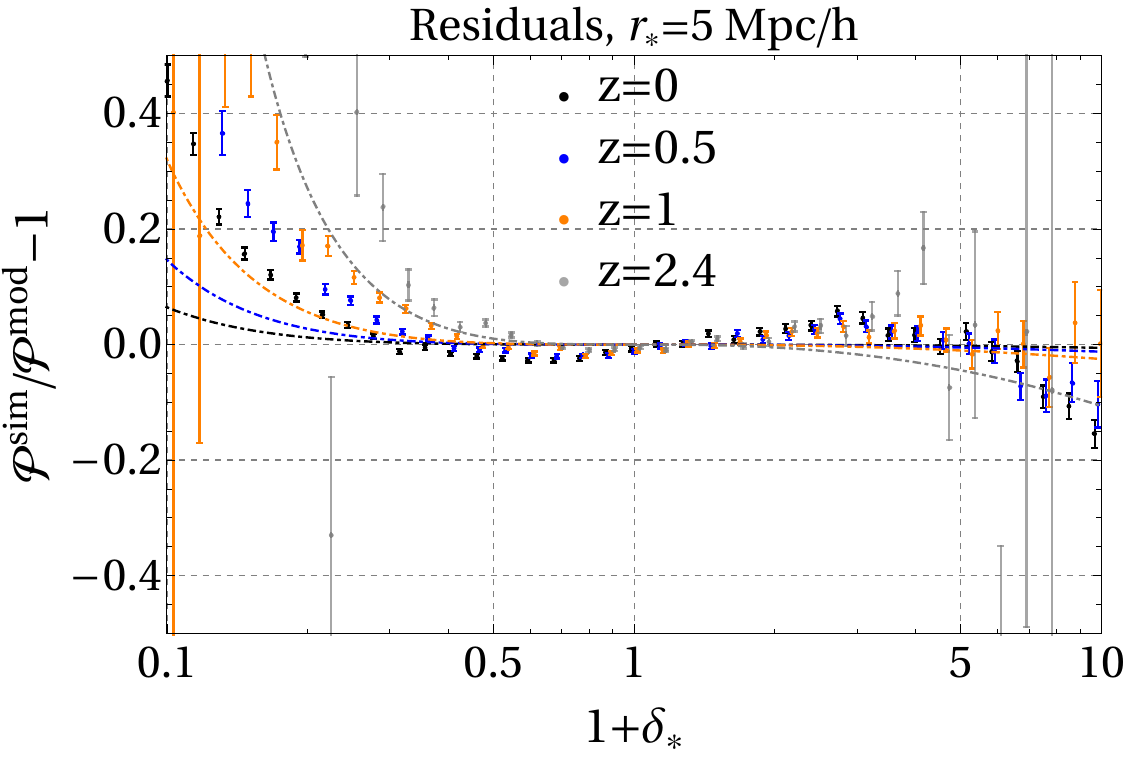}
\caption{The systematic bias in the PDF due to
  Zeldovich initial conditions (dotted lines). We also show the 
residuals between N-body data and the theoretical model of the main
text (cf. lower panels of Fig.~\ref{fig:far_2}). Note that for 
$r_*=5\,\Mpch$ (right panel) the model has been constructed by fitting
the data in the range $(1+\delta_*)>(1+\delta_*)_{\rm min}$ with 
$(1+\delta_*)_{\rm min}$ listed in Eq.~(\ref{deltamin}).} 
	\label{fig:sys_map2}
\end{figure}

In Fig.~\ref{fig:sys_map2} we compare the correction (\ref{sys_res})
to the residuals between the N-body PDF and the theoretical model
constructed in the main text. The latter does not include the effect
of transients and thus we expect a systematic difference between the
model and the data. The residuals indeed appear to follow the trend
implied by Eq.~(\ref{sys_res}). However, for cells of radius
$r_*=7.5\,\Mpch$ (left panel) or larger this systematic error is
burried under the statistical uncertainty, so we neglect it in our
analysis. On the other hand, for $r_*=5\,\Mpch$ (right panel) the
effect of ZIC is statistically significant at strong
underdensities. This is consistent with the findings of \cite{Uhlemann:2019gni}.
We use in the analysis only the part of the data for
which the systematic bias (\ref{sys_res}) does not exceed the
statistical error. This leads to the cuts
$(1+\delta_*)>(1+\delta_*)_{\rm min}$, where
\be
\label{deltamin}
(1+\delta_*)_{\rm min}=\{0.28,\,0.36,\,0.44,\,0.6\}~~~~~\text{for}~~
z=\{0,\,0.5,\,1.0,\,2.4\}\;.
\ee 

A comment is in order. For cells with $r_*=5\,\Mpch$ the estimate
(\ref{sys_res}) correctly captures the qualitative trend of the
residuals. However, it cannot fully account for their magnitude,
underpredicting them by a factor of a few at strong underdensities. It
is important to investigate the origin of this additional
discrepancy. There are at least three logical possibilities:
large corrections from ZIC to the PDF prefactor; large higher-order
corrections to the PDF model; or unknown systematics in the N-body
data. We leave this question for future. 
In this respect it is worth noting that using 2LPT initial
conditions, instead of ZIC, strongly suppresses the effects of transients
in the simulations~\cite{Crocce:2006ve,Uhlemann:2019gni}.

\section{Reduced models for counterterm prefactor}
\label{app:comp}

In Sec.~\ref{sec:ctr} we have formulated the model for counterterm prefactor with three free fitting parameters: the scaling index $m$ and the kinetic and potential counterterm amplitudes $\zeta^\kin$, $\zeta^\pot$. In Sec.~\ref{sec:result} we have shown that this model well describes the data of N-body simulations. Here we explore the possibility of fitting PDF with a reduced number of parameters by imposing constraints on $(m,\zeta^\kin,\zeta^\pot)$. We also comment on the relation of our model to Model 2 of Ref.~\cite{Ivanov:2018lcg}.

\begin{figure}
	\centering
	\includegraphics[width=0.48\linewidth]{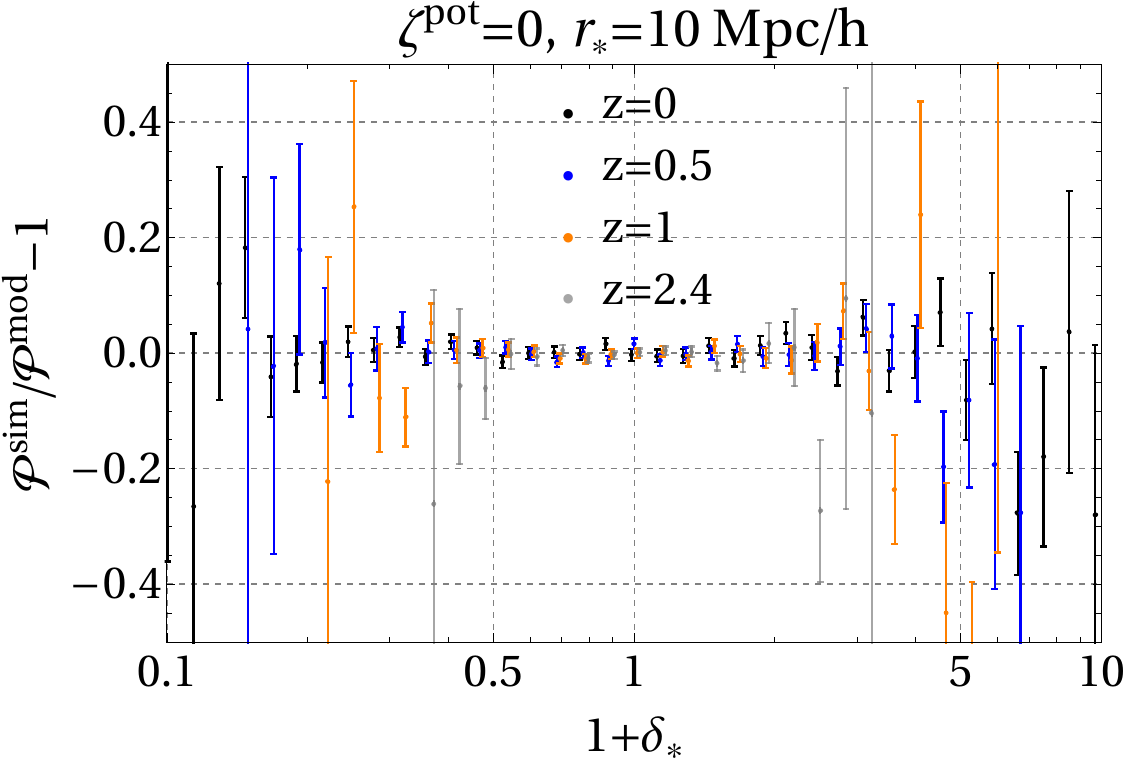}~~
	\includegraphics[width=0.48\linewidth]{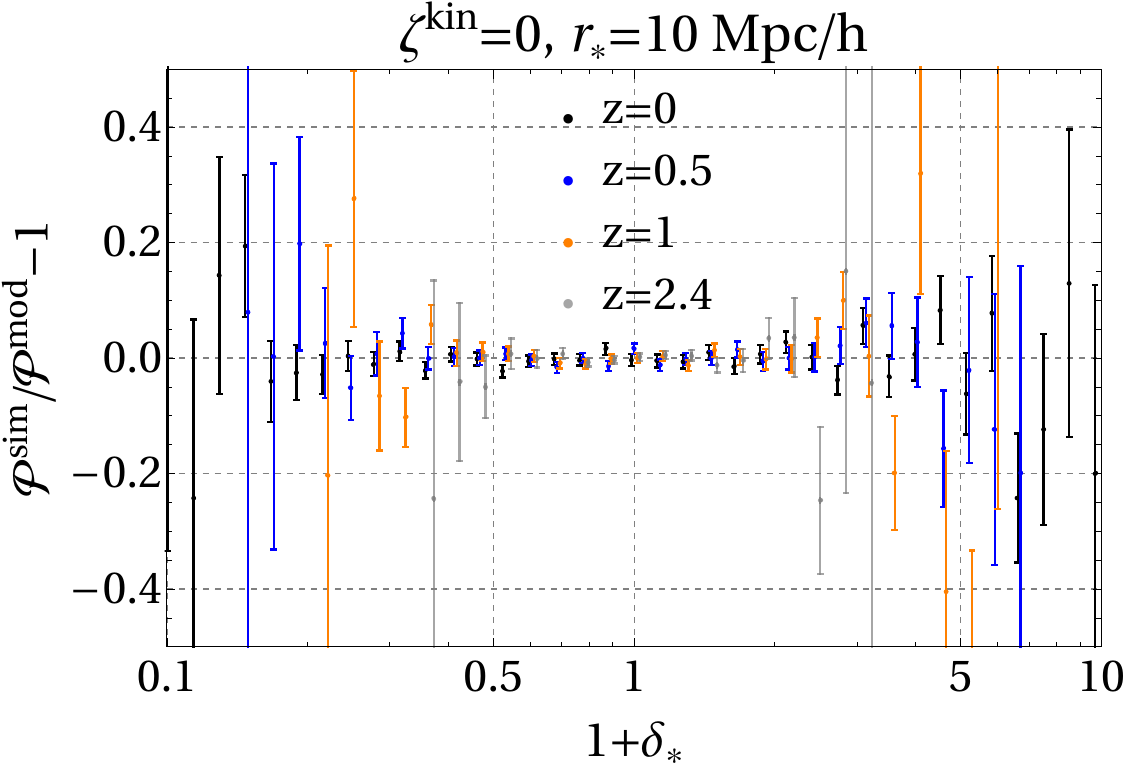}
	\caption{Residuals between the PDF extracted
          from N-body data and the reduced theoretical model with $\zeta^\pot=0$ ({\it left}) and $\zeta^\kin=0$ ({\it right})
          for cell radius $r_*=10\Mpc/h$.  }
	\label{fig:fitres2}
\end{figure}

As a first test, we perform the fitting procedure when setting either the kinetic or potential counterterm amplitude to zero.
The residuals of the fits for the cell radius $r_*=10\Mpc/h$ are shown in Fig. \ref{fig:fitres2}.
We found that the both models capture the behaviour of the PDF fairly well.
The good quality of the fit is confirmed by the values of $\chi^2$ listed in Tab.~\ref{tab:large2}.
\begin{table}[ht]
\begin{center}
\begin{tabular}{|c|c|c|c|c|c|}
\hline 
Model & $m$ & $\zeta^\kin$ & $\zeta^\pot$ & $\chi^2/N_{\rm
  dof}$ & $\gamma_0$\\
\hline 
Baseline &  $2.14\pm 0.10$ & $0.48\pm0.29$ & $-2.03\pm0.69$  & $108/110$ &
$1.73\pm0.08$\\ 
$\zeta^\pot=0$ &  $2.42\pm 0.05$ & $1.42\pm0.06$ & $-$ & $118/111$ &
$1.95\pm 0.06$\\
$\zeta^\kin=0$ &  $1.97\pm 0.04$ & $-$ & $-3.23\pm0.11$  & $111/111$ &
$1.63\pm0.05$\\ 
$\zeta^\kin=\zeta^\pot$ &  $2.63\pm 0.07$ & $2.68\pm0.18$ & $2.68\pm0.18$ & $185/111$ &
$2.22\pm0.08$\\ 
\hline
\end{tabular}
\end{center}
\caption{Best fit parameters of the counterterm models, $\chi^2$ per number of degrees of freedom, and the derived effective speed of sound $\gamma_0$.
\label{tab:large2}
}
\end{table}
However, the predictions of model parameters deviate significantly.
In particular, the value of $\zeta^\kin$ extracted from the $\zeta^\pot=0$ fit differs from the result using the full theoretical PDF model by more than $3\sigma$.
Also, the values of $m$ and $\gamma_0$ are systematically higher for this fit than 
in the baseline analysis. 
The fit $\zeta^\kin=0$ is closer to the baseline model but gives lower values of $m$, $\gamma_0$ which are mildly in tension with the ranges extracted from the power spectrum (see Eq.~(\ref{mdpl})).  

As a second test, we consider the equal counterterm amplitudes, $\zeta^{\kin}=\zeta^{\pot}$.
This choice takes place if one renormalizes the stress tensor for short-scale contributions as the whole without splitting the latter into the kinetic and potential parts. 
The results listed in Tab.~\ref{tab:large2} reflect that this reduced model fails to reproduce the data with
required precision.
In particular, it leads to a severe $4.4\sigma$ tension with the N-body data.
It is thus impossible to describe the counterterm stress with a single coefficient.
This result confirms that the kinetic and potential counterterms are physically distinct. We do not see any firm reason to discard any of them and recommend keeping $\zeta^\kin$, $\zeta^\pot$ as independent free parameters.

\begin{figure}
	\centering
    \includegraphics[width=0.49\linewidth]{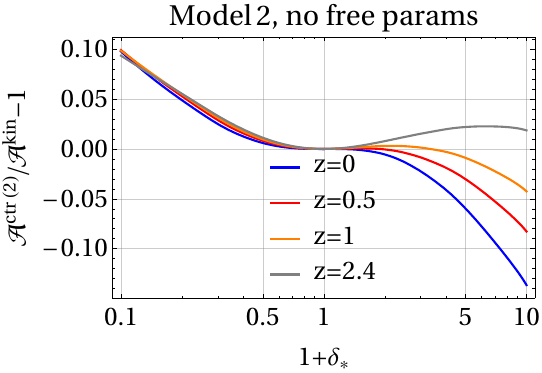}
	\includegraphics[width=0.49\linewidth]{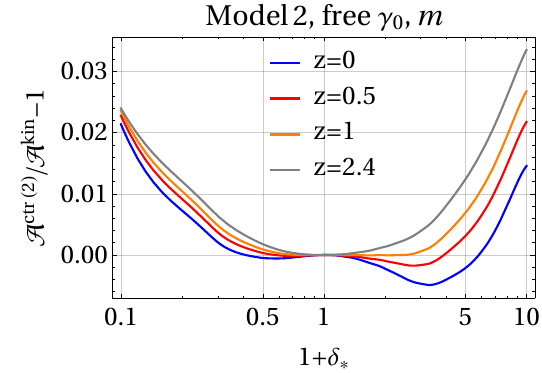}
	\caption{The difference between the counterterm prefactor (\ref{Actr2}) from Model 2 of \cite{Ivanov:2018lcg} and reduced theoretical model of this paper with $\zeta^\pot=0$ for cell radius $r_*=10\Mpch$ at various redshifts. {\it Left:}
the parameters of 
 Model 2 are fixed to $m=8/3$, $\gamma_0=1.95$. {\it Right:} $(m,\gamma_0)$ 
  are fitted from the PDF data with correlated Gaussian priors inferred from the power spectrum fit.
          }
	\label{fig:fitmod2}
\end{figure}

Incidentally, the model with $\zeta^\pot=0$ happens to give a counterterm prefactor similar to Model 2 of  
Ref.~\cite{Ivanov:2018lcg}.
The latter was proposed as an ad hoc way to take into account the dependence of the shell-crossing scale on the mean density of the cell. In more detail, the counterterm prefactor of Model 2 reads, 
\be
\label{Actr2}
{\mathcal A}^{\ctr(2)}=\exp\left(-\frac{315\gamma_0}{122\pi}
  \Big(\frac{g(z)}{F'(\delta_*)(1+\delta_*)}\Big)^{m-2}\times
  2\hat{\lambda}\int \frac{d\vk}{\vk}q(\vk)\right) \,, 
\ee
where $q(\vk)$ is the WKB function appearing in the short-mode contribution to the prefactor, Eq.~(\ref{AprUV}). 
Following~\cite{Ivanov:2018lcg}, Model 2 uses as input the value $\gamma_0$ from the fit of the power spectrum and $m=8/3$, and thus does not introduce any new fitting parameters.
The ratio between ${\cal A}^{\ctr(2)}$ computed in this way and the best-fit counterterm prefactor with $\zeta^\pot=0$ is shown on the left panel of Fig.~\ref{fig:fitmod2}. We see that the difference between two models is only about $10\%$ at the tails. As a result, the Model 2 without any free parameters describes the data fairly well, with only a weak
$1.8\sigma$ tension,
as follows from the value of 
$\chi^2$ listed in the upper row of 
Table~\ref{tab:compmod2}.
\begin{table}[ht]
\begin{center}
\begin{tabular}{|c|c|c|c|}
\hline 
Model 2 &
$m$ &
$\gamma_0$ &
$\chi^2/N_{\rm dof}$ \\
\hline
No free params &
$8/3$ &
$1.95$ &
$134/112$ \\ 
Free $\gamma_0$ &
$8/3$ &
$1.97\pm 0.06$ &
$134/112$ \\  
Free $\gamma_0$, $m$ &  
$2.48\pm 0.04$ &
$2.03\pm0.06$ & 
$120/111$ \\ 
\hline
\end{tabular}
\end{center}
\caption{Parameters of the counterterm prefactor \eqref{Actr2} from Model 2 of \cite{Ivanov:2018lcg} and the quality of the fit to N-body data. 
Three cases are considered: fixed parameters (first row), one free fitting parameter $\gamma_0$ (second row), and two free parameters $\gamma_0$, $m$ (third row).
\label{tab:compmod2}
}
\end{table}
We do not have any physical explanation of this quantitative success. It would be interesting to understand if it persists for more general cosmologies, such as, e.g., scaling universes, or is a pure accident of $\Lambda$CDM.

In the spirit of EFT approach to PDF, we can allow the parameters of Model~2 to be determined from the PDF data themselves.  A
natural extension is to set $\gamma_0$ as a free parameter.
We fit $\gamma_0$ to the
aspherical prefactor extracted from simulations using the procedure
described in Sec.~\ref{sec:result}.
We impose the Gaussian prior on $\gamma_0$ from the fit of the power spectrum
(\ref{mdpl}).
The results are presented in the second row of Tab.~\ref{tab:compmod2}.
The quality of the fit does not change and the value of $\gamma_0$ is entirely consistent with the result inferred from the power spectrum.

Moving one step further we let both parameters $(m, \gamma_0)$ to vary freely in the fit. 
We impose correlated Gaussian priors following from the fit of
the power spectrum (\ref{mdpl}).
The results listed in Tab.~\ref{tab:compmod2} reflect that the quality of the fit improves compared to that in the $m=8/3$ scenarios.
In particular, Model~2 with free $(m, \gamma_0)$ is in $1.1\sigma$ agreement with the data.\footnote{For comparison, the baseline theoretical model of this paper shows $0.6\sigma$ agreement.}
On the right of Fig.~\ref{fig:fitmod2} we plot the relative difference between the best-fit Model 2 \eqref{Actr2} 
and the $\zeta^\pot=0$ result for different redshifts.
We find that the Model 2 with free $(m, \gamma_0)$ reproduces the kinetic counterterm prefactor with a few per cent precision.

\bibliographystyle{JHEP}
\bibliography{short.bib}

\end{document}